\documentclass[a4paper,fleqn,usenatbib]{mnras}

\usepackage{mathptmx}
\usepackage[T1]{fontenc}
\usepackage{ae,aecompl}
\usepackage{graphicx}	
\usepackage{amsmath}	
\usepackage{amssymb}	
\usepackage{verbatim} 
\usepackage[english]{babel}
\usepackage[dvipsnames]{xcolor}
\usepackage{natbib}
\usepackage{graphicx}
\usepackage{txfonts}
\usepackage[normalem]{ulem}
\usepackage{float}
\usepackage[caption = false]{subfig}
\usepackage{array}
\usepackage{multirow}
\usepackage{pifont}

 \title[Exploring MPs Effect on Stellar IMF Estimates]{Multiple populations within globular clusters in Early-type galaxies \\ {\Large Exploring their effect on stellar initial mass function estimates}}   

\author[W. Chantereau et al.]{
W. Chantereau,$^{1}$\thanks{E-mail: W.Chantereau@ljmu.ac.uk}
C. Usher,$^{1}$
N. Bastian$^{1}$
\\
$^{1}$Astrophysics Research Institute, Liverpool John Moores University, 146 Brownlow Hill, Liverpool L3 5RF, UK
}

\date{Accepted XXX. Received YYY; in original form ZZZ}

\pubyear{2018}

\begin{document}
\label{firstpage}
\pagerange{\pageref{firstpage}--\pageref{lastpage}}
\maketitle

\begin{abstract}
It is now well-established that most (if not all) ancient globular clusters host multiple populations, that are characterised by distinct chemical features such as helium abundance variations along with N-C and Na-O anti-correlations, at fixed [Fe/H]. These very distinct chemical features are similar to what is found in the centres of the massive early-type galaxies and may influence measurements of the global properties of the galaxies. Additionally, recent results have suggested that \textit{M}/\textit{L} variations found in the centres of massive early-type galaxies might be due to a bottom-heavy stellar initial mass function.  We present an analysis of the effects of globular cluster-like multiple populations on the integrated properties of early-type galaxies. In particular, we focus on spectral features in the integrated optical spectrum and the global mass-to-light ratio that have been used to infer variations in the stellar initial mass function. To achieve this we develop appropriate stellar population synthesis models and take into account, for the first time, an initial-final mass relation which takes into consideration a varying He abundance. We conclude that while the multiple populations may be present in massive early-type galaxies, they are likely not responsible for the observed variations in the mass-to-light ratio and IMF sensitive line strengths. Finally, we estimate the fraction of stars with multiple populations chemistry that come from disrupted globular clusters within massive ellipticals and find that they may explain some of the observed chemical patterns in the centres of these galaxies.
\end{abstract}

        

        
\section{Introduction}

Massive early-type galaxies (ETGs) have long been known to show non-solar abundance patterns \citep[e.g.][]{Worthey92, Trager00, Gallazzi06} with the observed increase in [$\alpha$/Fe] with galaxy mass being interpreted as an inverse relationship between mass and star formation time scale \citep[e.g.][]{Matteucci94, Thomas05}.   
More recent works \citep[e.g.][]{Graves08, Johansson12, Conroy14, Worthey14} have allowed the abundances of a range of elements to be measured from integrated galaxy light and shown that the abundances of elements such as C, N and Na also increase with galaxy mass.

However, several lines of evidence show that the stellar populations of massive ETGs are not as simple as expected for populations that formed in a relatively short burst.
Many ETGs show brighter far UV-luminosities than would be expected for an old, metal rich stellar population \citep[e.g.][]{Code79,Burstein88,Greggio90,Dorman95,Brown97,OConnell99,Brown00, Yi11}.
The colours of this `UV-upturn' are correlated to the stellar mass-to-light ratio (\textit{M}/\textit{L}) and mass of the host galaxy \citep{Burstein88,Greggio90,OConnell99,Donas07,Dalessandro12,Zaritsky15}.
Recent works have also shown interesting radial abundance gradients in the centres of at least some ETGs.
\citet{vanDokkum17} found strong positive [O/Fe] and negative [Na/Fe] radial gradients but no or weak age or heavier [$\alpha$/Fe] gradients in 6 ETGs while \citet{Sarzi17} found a strong negative [Na/Fe] gradient in the centre of M87, in addition to the reported IMF variations. These abundance trends do not appear to be the result of a classical galactic chemical enrichment, such as $\alpha$-enhancement \citep[see e.g.][]{vanDokkum17}. Thus the origin of these gradients is still uncertain \citep{Schiavon07}. 

\cite{Cappellari12} found that \textit{M}/\textit{L} strongly varies in the $r$-band among ETGs and is correlated to the galaxy's velocity dispersion. They interpreted this as an IMF variation towards a bottom-heavy distribution (low-mass dwarf stars) for the most massive ETGs (those with the highest velocity dispersion). Evidence for a bottom heavy IMF in the centre of galaxies also comes from detailed studies of the spectra of massive ETGs \citep[e.g.,][]{vanDokkum10,Conroy12,LaBarbera17} where absorption lines distinctive of dwarf stars are stronger and absorption lines indicative of giant stars are weaker than in stellar population synthesis models with Milky Way-like IMFs. Interpretation of these observations of IMF sensitive spectral features is complicated by their (strong) dependence on chemical abundance \citep[e.g.,][]{Conroy12_counting} in massive ETGs.

Meanwhile, thanks to  advances in instrumentation including multiplexed high resolution spectrographs and high resolution optical and ultraviolet imaging from space, ground breaking observations have shaken the long-standing paradigm of globular clusters (GCs) being composed of a simple stellar population \citep[e.g.,][]{Catelan10,Gratton12,Charbonnel16_EES}. Photometric observations highlight a large variety of features in colour-magnitude diagrams not consistent with simple stellar populations \citep[see e.g.,][]{Piotto09, Milone17} while spectroscopic studies provide conspicuous evidence of significant star-to-star variations of the light elements abundances. It appears that there are (at least) two main stellar populations within GCs, the first (1P) consists of stars that show the same chemical abundance patterns as field stars (at a given [Fe/H]). The second population (2P) shows distinct patterns, namely enhanced He, N, Na, and Al accompanied by depletion of C, O and sometimes Mg \citep[e.g.][]{Gratton12}. These abundance anomalies have been found in (nearly) all of the ancient GCs studied to date, including extremely metal poor and metal rich (near solar) GCs in the Milky Way and nearby galaxies \citep[e.g.][]{Bastian18}.

GCs are mainly composed of a monometallic (i.e., they lack spreads in heavy elements such as Fe and Ca) and a coeval population of stars, therefore these abundance variations are not expected to be the results of standard stellar or galactic chemical evolution. These observed chemical features are rather thought to be the direct signatures of hydrogen-burning at high temperature, a process that is not at work in main sequence low-mass stars displaying these peculiar chemical patterns \citep{Denisenkov90}. However, the origin of the abundance variations, or even if they are limited to globular clusters, is currently unknown \citep[see the recent review by][]{Bastian18}. 
We note that MPs are observed in massive star clusters as young as 2 Gyr \citep{Martocchia18} and in GCs with near solar metallicities \citep{Tang17, Schiavon17_inner_gcs} showing that MP requires neither special conditions in the early universe nor low metallicities. \medskip

There are recent hints that the presence of MPs may not be a phenomenon restricted to GCs. For instance, a small fraction ($\sim 2$ \%) of field stars in the halo and the inner Galaxy bear chemical features typical of these MPs. \citep{Spiesman92,Carretta10,Martell10,Ramirez12,Martell16,Schiavon17}. At galactic scales, \cite{Strader13} found an ultra-compact dwarf thought to be the stripped core of a galaxy enhanced in N and Na that could indicate the presence of similar self-enrichment to what is present in GCs. The chemical composition of MPs, especially the He-enhancement, has a great impact on their stellar evolution. Thus the presence of these He rich stars has important effects on the observed properties of GCs \citep[e.g.,][]{D'Antona10,Cassisi14,Chantereau16,Charbonnel16,Tailo16}, hence it is timely to developed studies on their effects on galactic scales with appropriate stellar population synthesis models. \medskip

The peculiar chemical features observed in massive ETGs at a given age and metallicity (Na and N enhancement, \citealt{Schiavon07,Conroy14,vanDokkum17}, and O-depletion to a certain extent, \citealt{vanDokkum17}) are reminiscent of the MP chemical patterns found in GCs.
Parallels can also be drawn between the UV-upturn in ETGs and the bright UV luminosities of GCs with hot horizontal branches due to He enhanced stars. 
In addition, at a given age and metallicity, He rich stars of MPs are less massive, have different luminosities, and display larger final masses than their He normal counterpart. Thus, we might expect the He rich population to have a different \textit{M}/\textit{L} than the chemically `normal' population. It would then be interesting to investigate the integrated properties of MPs, and investigate if He rich stars of GCs would be a viable alternative explanation for the observed \textit{M}/\textit{L} and spectral index variations between galaxies.

As the origin of the multiple population phenomenon is currently unknown there are two possible mechanisms for which they might appear in massive early-type galaxies.  The first is that the chemically peculiar stars are formed in globular clusters which are subsequently disrupted due to the tidal field of the host galaxy.  The second is that multiple populations are not limited to globular clusters and can form within galaxies under specific conditions (i.e. high stellar densities or pressures). In the present work we explore the former option. We note that massive early-type galaxies posses populous GC systems \citep[e.g.][]{Harris13} and that their GC systems display broad metallicity distributions with substantial numbers of GCs at solar and higher metallicity \citep[e.g.][]{Peng06, Usher12}. Note also that our current understanding of star formation indicates that intense star formation (high metallicity) is accompanied by the formation of GCs \citep{Kruijssen15, Pfeffer17}.

The goals of this paper are twofold.  First, we explore the extent to which disrupted GCs are expected to contribute to the stellar populations of massive ETGs. For this we exploit recent advances in our understanding of star/cluster formation within galaxies as well as how the stellar population properties (i.e. the extent of MPs within them) depend on the cluster properties (i.e. mass).  Secondly, we explore the effects of the peculiar abundance pattern of MPs (i.e. enhanced He, N, Na, Al and depleted C and O) on the integrated properties of the populations, focusing on the \textit{M}/\textit{L} ratio, UV flux and spectral indices.  To do this we create tailored integrated light stellar population models with MPs and compare them to both other models in the literature as well as to selected observations of ETGs.  The underlying question motivating the current work is to what extent can the observed MP phenomenon explain the observed features of massive ETGs?

This paper is organised as follows. In Section~\ref{Disrupted} we investigate the fraction of second population stars from GCs that might contribute to the peculiar abundance patterns observed in the core of ETGs. Then in Section~\ref{SPS}, we present the initial-final mass relations and the stellar population synthesis models adapted to the MPs framework. In  Section~\ref{MLratio} we investigate the effects of the typical initial chemical compositions of MPs, initial mass functions and ages on the resulting \textit{M}/\textit{L}.  We present the effects of MPs on the spectral indices in Section~\ref{index_strengths} and we investigate their effect in the central regions of ETGs (Section~\ref{application}). We finally discuss and summarise the results in Sections~\ref{Discussion} and \ref{Conclusion} respectively.

\section{Contribution to the Field of Disrupted Globular Clusters}\label{Disrupted}

We do not know if MPs can form directly within ETGs as their formation mechanism within GCs (and elsewhere) is currently unknown.
However, disrupting GCs are expected to contribute heavily to the stellar populations in the central regions of ETGs since intense star formation leads to the formation of large numbers of massive star clusters \citep[e.g.,][]{Whitmore99,Bastian08,Kruijssen14} and since the tidal fields in the centres of massive galaxies effectively disrupt star clusters \citep[e.g.,][]{Goudfrooij04, Pfeffer17}. 
In this section we estimate the contribution of disrupted GCs to the field of the core of ETGs.
The basic question that we hope to address later in the paper with this toy model is whether disrupted GCs could potentially be the origin of the `anomalous chemistry' seen in ETGs. \medskip

For the initial estimate we assume that all formed GCs in the ETGs core have been disrupted, which agrees with the depressed numbers of GCs seen projected near the centres of ETGs relative to the outer regions of ETGs \citep[e.g.,][]{McLaughlin99, Forbes06, Peng08}. For the GC mass function, we adopt a power-law with index of $-2$ and an upper limit of $10^8$~M$_{\sun}$, comparable to the most massive observed young massive clusters \citep[e.g.,][]{Maraston04, Bastian06}. While there is evidence for an environmentally dependent upper mass truncations \citep[e.g.,][]{Gieles06, Larsen09, Bastian12, Johnson17}, we expect a high upper limit in the extreme high pressure environments in the inner regions of ETGs and do not expect much variations from galaxy to galaxy. Similarly, we expect the cluster formation efficiency ($\Gamma$) to be relatively high in such high pressure environments \citep[e.g.,][]{Kruijssen15, Pfeffer17}, between 0.5 and 1, meaning that half to all stars form in clusters. Though in our toy model we assume all stars have the same age and metallicity, we would expect the age and metallicity distribution of stars that formed in GCs in the centres of ETGs to mirror that of field stars.

Recent surveys have found that the fraction of second population stars (enriched stars, $F_{\rm enriched}$) with GCs increases with increasing cluster mass ranging from $\sim50$\% at $10^5$~M$_{\sun}$ to near $85$\% at $10^6$~M$_{\sun}$ \citep{Milone17}. We use these measurements and an initial-current GC mass relationship \citep[Fig.~7,][]{Kruijssen15} to derive a linear relation (in logarithmic mass) between initial cluster mass and enriched fraction between a faction of zero at $1.7 \times 10^{5}$~M$_{\sun}$ and one at $3.2 \times 10^{6}$~M$_{\sun}$. Below an initial mass of $1.7 \times 10^{5}$~M$_{\sun}$ we set the fraction to 0; above $3.2 \times 10^{6}$~M$_{\sun}$ to 1. Note that unlike some MP formation scenarios we do not adopt extreme mass loss of the young GCs, i.e. that they were 10-30$\times$ more massive than at present. \medskip

Adopting a lower cluster mass limit of 10$^2$~M$_{\sun}$, the fraction of enriched stars in the central regions of ETGs is then given by:

\begin{equation} \label{equation:f_enriched}
F_{\rm ETG, enriched} =  \frac{\Gamma}{M_{\rm total}} \times \int_{10^2}^{10^{8}} M^{-2} \times F_{\rm enriched}(M) dM
\end{equation}

Adopting $\Gamma=0.75$ and the \citet{Milone17} relation for the relation between the fraction of enriched stars and the initial cluster mass results in an estimate of $F_{\rm ETG, enriched} = 0.35$, i.e. a third of the stars in the central regions of ETGs are expected to show the anomalous chemistry observed in GCs. However, not only does the enriched fraction varies with cluster mass, the {\em degree} of enrichment is also a function of cluster mass, with higher mass clusters hosting larger spreads in He, Na and N \citep{Carretta10,Schiavon13,Milone15_NGC6266,Milone17}. In the majority of GCs, however, most of the stars do not display the extreme abundances.  If we use NGC~2808 as a test case, $\sim13$\% of the stars in this cluster belong to the extreme population, i.e. that with the largest He variations ($\Delta Y=0.15$)\footnote{We use the He content as a discriminant between the multiple populations as it is the element which has the largest impact on the evolution of these stellar populations.}, and a further 15\% of the stars are mildly enriched ($\Delta Y=0.07$) with the rest not showing an appreciable variation in He. To estimate the possible He spread in ETGs due to disrupted clusters we use the relation of \citet{Milone15_NGC6266} between cluster mass and the maximum He spread ($f_{\rm max, \Delta Y}$) and then distribute the stars in a similar way as in NGC~2808 with $13$\% having the maximum $\Delta Y$, $15$\% half the maximum $\Delta Y$ and the rest having $\Delta Y=0$. \medskip

\begin{figure}
   \centering
   \includegraphics[width=0.45\textwidth]{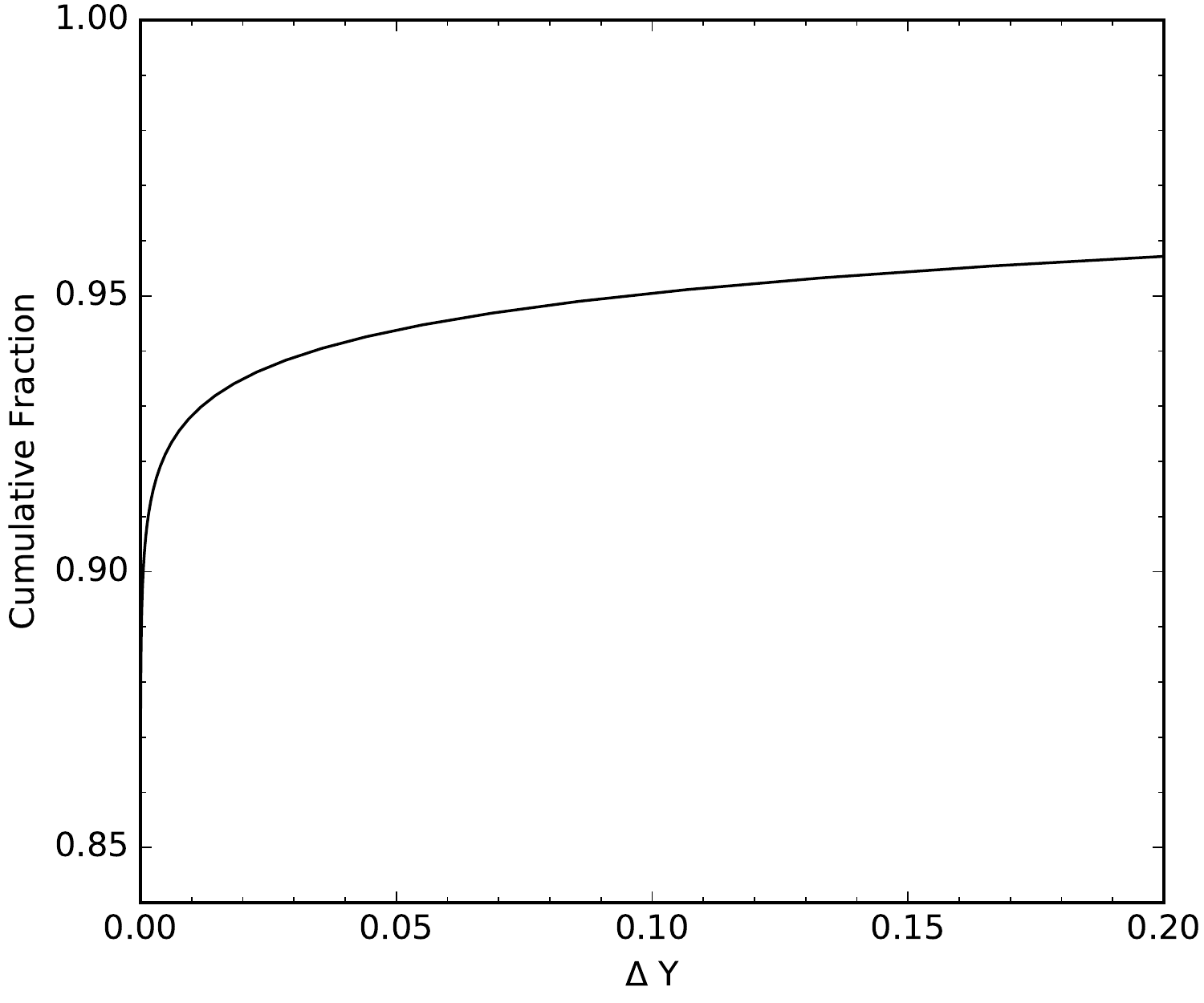}   
    \caption{Cumulative fraction of helium spread ($\Delta Y$) for our ETG model. The fraction is based on the relation between GC luminosity and He spread of \citet{Milone15_NGC6266} and the assumption that $13$\% of the stars have the maximum $\Delta Y$, $15$\% have half the maximum $\Delta Y$ and the rest having $\Delta Y=0$ together with the cluster mass function in Equation~\ref{equation:f_enriched}.}
    \label{Figure:He_distribution}
\end{figure}

The resulting distribution of $\Delta Y$ is shown in Figure~\ref{Figure:He_distribution}. For the given set of assumptions, we find that 93\% of the stars of the galaxy would be expected to have $\Delta Y < 0.01$ and 95\% of the stars have $\Delta Y < 0.07$.  However, there is a relatively extended tail with $5$\% of the stars having $\Delta Y>0.1$. We note that our calculations are sensitive to the maximum initial GC mass, the initial-current GC mass relationship, and the assumption that present day fraction of second population stars is the same as at formation. \medskip

In summary, we find that disrupted GCs may contribute a substantial amount ($\sim 35$\%) of stars with second population chemistry (enriched in N and Na while depleted in C and O) to the field population in the centres of ETGs.
Disrupted GCs also contribute a small but significant fraction ($\sim 5$ \%) of strongly He enhanced stars.

\section{Stellar population synthesis models}\label{SPS}
In this section we describe the stellar population synthesis models and initial-final mass relations used in this work.
We emphasise that our models are intended to study the qualitative effects of He enhancement and of the CNONa abundance patterns present in GCs and are not constructed to fully reproduce the observed spectral energy distributions of GCs or galaxies.
In the Appendix we compare our models to those of \citet{Conroy12_counting} as well as observed spectra of metal rich GCs in the Milky Way and find qualitative agreement.
We calculated stellar population synthesis models for a solar He content ($Y_\mathrm{ini} = 0.26$, also referred as He normal content at $Z_{\sun}$) and for an enhanced He content ($Y_\mathrm{ini} = 0.4$) at solar metallicity and an age of 12.6 Gyr \citep[typical ETG age and metallicity, e.g.,][]{Gallazzi05,Conroy12,Conroy14}. We kept the metallicity mass fraction constant \citep[$Z = 0.0134$,][]{Asplund09} between the solar He and He rich models as observed in most GCs. 
For both normal and He rich abundances we calculate models with a scaled solar composition and with a chemical composition typical of the second population GC stars ([C/Fe] $=$ [O/Fe] $= -0.6$, [N/Fe] $= +1.0$, [Na/Fe] $= +1.0$). This abundance pattern was made to reproduce observed GC abundances due to multiple populations \citep[e.g.,][]{Carretta09} with the constraint that the C+N+O and Ne+Na abundances are kept constant \citep[as observed in most GCs, e.g.,][]{Brown91,Dickens91,Norris95,Smith96,Ivans99,Cohen05,Yong15}. \medskip 

We follow \citet{Conroy12_counting} and used the 2012 version of the \citet{Dotter07} isochrones for the main sequence (MS) through to the tip of the red giant branch (RGB).  For post RGB evolution we used the \citet{Bertelli08} isochrones matched in age, helium content and metallicity to the \citet{Dotter07} models. However, the models from the two database do not predict the same initial mass at the tip of the RGB. To prevent a discontinuity in the initial masses of the combined isochrones, we subtract the difference ($\sim 0.01$ M$_{\sun}$) from the initial masses of \citet{Bertelli08} isochrones to match the \citet{Dotter07} isochrones \citep[as done in][]{Conroy12_counting}. Our choice of isochones was dictated by the need to vary He and to have the models extend to low stellar masses. For each isochrone we selected semi-uniformly $\sim100$ points in $\log(L)$-$\log(T_{eff})$ space covering all stages of stellar evolution from the lowest mass MS star to the end of the AGB phase. We selected more points from evolutionary phases such as the main sequence turn-off (MSTO) where there are sharp variations in the $\log(L)$-$\log(T_{eff})$ space. We refined our selection such that no point contributes more than 2\% to the integrated light assuming a \citet{Kroupa01} IMF. We used the same isochones for the N, Na enhanced, C, O depleted and solar He models (referred as CNONa models hereafter, see Table~\ref{models_nomenclature}) as for the solar composition models since stellar evolution is not affected by abundance variations of C, N, O and Na if the C+N+O abundance is kept constant at fixed metallicity \citep[as shown in][]{Salaris06,Coehlo11,Sbordone11}.

\subsection{Initial-final mass relations}\label{IRMR}

\begin{figure}
   \centering
   \includegraphics[width=0.45\textwidth]{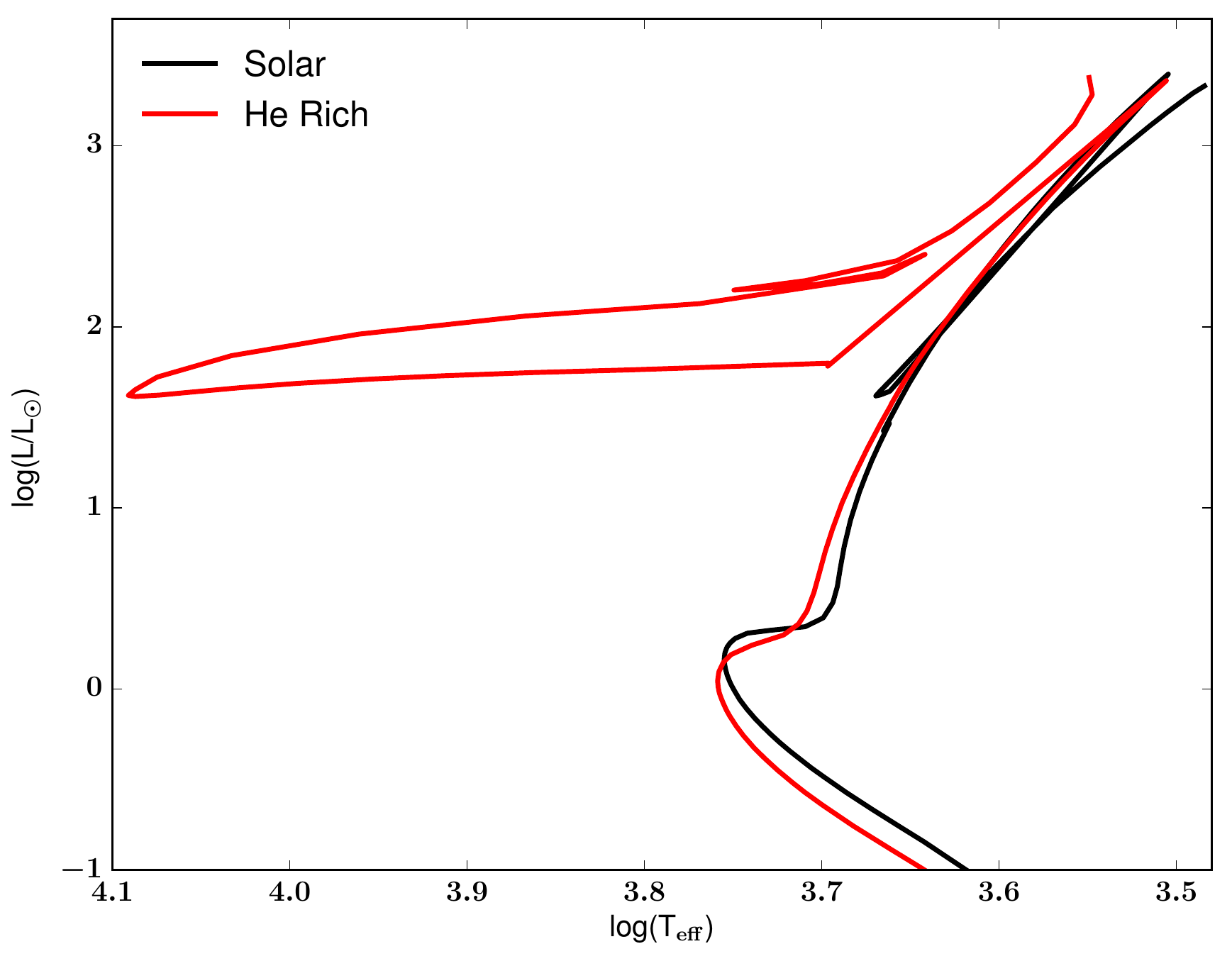}   
    \caption{Isochrones at $Z_{\sun}$ and 12.6~Gyr for He normal and He rich populations (black and red respectively).
    The He rich isochrone has a significantly hotter horizontal branch as well as a slightly hotter main sequence, red giant branch and asymptotic giant branch.}
    \label{Figure:isochrone}
\end{figure}

We investigate in this section the initial-final mass relation for our He normal and He rich models (cf. isochrones in Fig.~\ref{Figure:isochrone}). As previously discussed, the inclusion of the other abundance variations (such as C, N, O, Na) in our case does not influence the stellar evolution and in turn does not affect the initial-final mass relation. Thus, in this section we only refer to the He normal and He rich populations, but the results are transposable to the He normal and He rich populations with CNONa variations (referred as He rich CNONa models hereafter). From the isochrones available, we can only take as final mass for stars ending their life as white-dwarf (WD) the core mass at the start of the thermally-pulsing asymptotic giant branch. In this case, the final mass is slightly underestimated \citep{Chantereau17}, but since we are interested in differential effects between two model populations treated self-consistently, this approximation have only a negligible effect on the results. \medskip

\begin{figure*}
   \centering
   \includegraphics[width=0.75\textwidth]{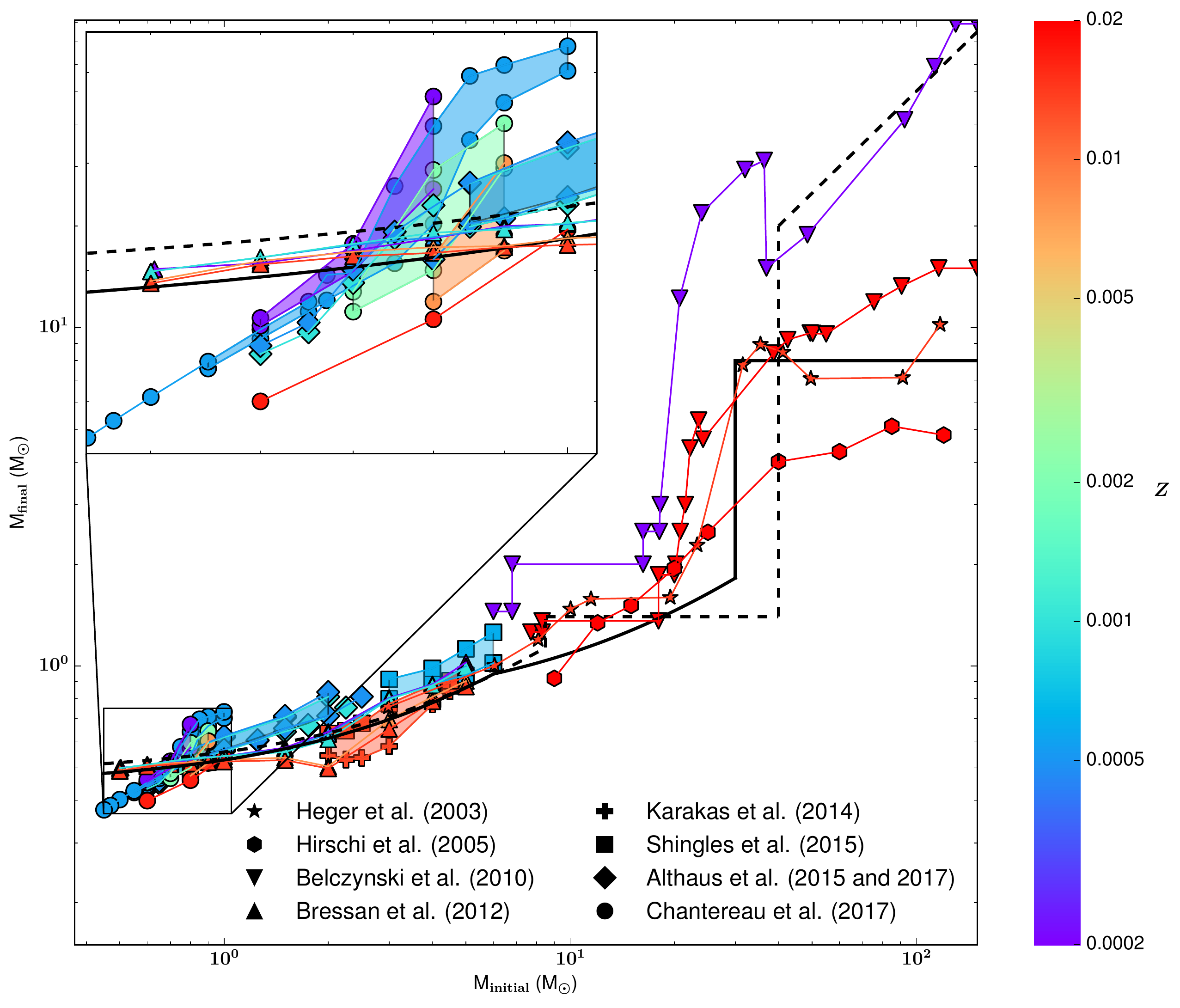}   
    \caption{Theoretical predictions from the literature for initial-final mass relations at several metallicities (colour-coded). The relation from \protect\cite{Renzini93} represented by a black dashed line is a standard prescription used in stellar population synthesis \protect\citep[e.g.,][]{Maraston98,Conroy09}. The GALEV evolutionary synthesis model \protect\citep{Kotulla09} is represented by a continuous black line. The filled areas correspond to differences between He normal and He rich ($Y_\mathrm{ini}$ = 0.4) predictions.}
    \label{Figure:IRMR_Z}
\end{figure*}

\begin{figure}
   \centering
   \includegraphics[width=0.45\textwidth]{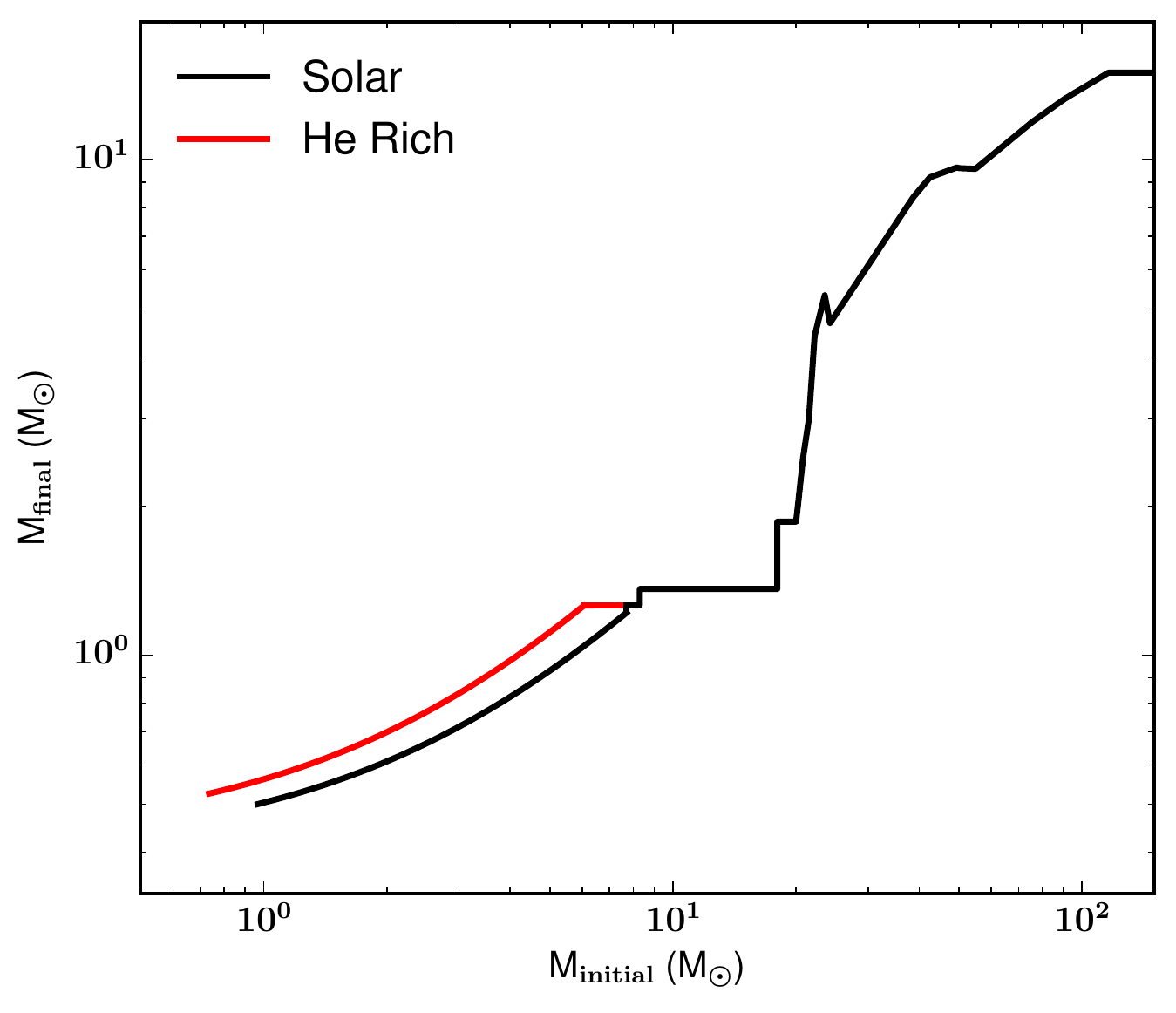}   
    \caption{Initial-final mass relationship as a function of $M_\mathrm{ini}$ for He normal and He rich stars (black and red respectively) at $Z_{\sun}$ and 12.6~Gyr.}
    \label{Figure:IRMR}
\end{figure}

For an initial helium content of 0.4, \cite{Chantereau17} predict an average increase of the final mass of low-mass stars at low metallicity by $\sim$10\% (with respect to the final mass of He normal stars). This result is fully consistent with an independent study from \cite{Althaus17} using different mass-loss prescriptions, even though final masses are very sensitive to these prescriptions (see Fig.~\ref{Figure:IRMR_Z}). This increase is rather constant as a function of the metallicity. Thus, since there are no predictions of the final mass of low mass stars at $Z_{\sun}$ for $Y_\mathrm{ini} = 0.4$, we assume here an average increase of the final mass of He rich stars of $\sim$10\% for $M_\mathrm{ini} < 2.0$~M$_{\sun}$. At $Z_{\sun}$, \cite{Karakas14} found that 2.0-4.5~M$_{\sun}$ He rich stars have an average core mass at the first thermal pulse $\sim$20\% higher than their He normal counterpart. We assume the same increase for the final mass for stars with $M_\mathrm{ini}$ between 2 and 5~M$_{\sun}$. We then use a linear fit derived by least square fitting of these final masses to determine an initial-final mass relationship for both populations with $M_\mathrm{ini}$ between 0.5 and 7.0~M$_{\sun}$  (Fig.~\ref{Figure:IRMR}). \medskip

More massive stars end their life as neutron stars (NS) or black holes \citep[BH, $M_\mathrm{ini}$>7.7 and 20~M$_{\sun}$ respectively,][]{Belczynski10}, however for $Y_\mathrm{ini} = 0.4$, there are no available initial-final mass relations. But since these stars contribute only to a minority of all the mass locked in the cluster's remnants, it is reasonable to assume a similar initial-final mass relation for He rich and He normal stars for $M_\mathrm{ini}>7.7~$M$_{\sun}$. Note that if we assume that NSs formed through electron-capture supernovae have the same minimum mass for both populations, then the minimum initial mass needed to form these NSs from He rich progenitors is dramatically shifted down to $\sim6.1~$M$_{\sun}$ (since the final mass of He rich stars is greater than its He normal counterparts for a given initial mass). Thus, the He enhancement will increase the formation rate of NSs through electron-capture supernovae (assuming an identical IMF), as pointed out in \cite{Shingles15}.

\subsection{Stellar spectra and model atmosphere} 

We computed stellar model atmospheres using \textsc{ATLAS12} \citep{Kurucz70,Kurucz05} and synthesised stellar spectra using \textsc{SYNTHE} \citep{Kurucz79,Kurucz81}.
Both of these codes assume one-dimensional, static and plane parallel atmospheres in local thermodynamic equilibrium.
We used the same versions of these codes as in \citet{Martocchia17} and refer the reader there for further details of our stellar atmosphere calculations. We synthesised the spectra over the 2000 to 10 000 \AA{} wavelength range at a resolution of $R = 200$ 000 before using a Gaussian kernel to smooth it to $R = 5000$ for future analysis. \medskip

We calculated the spectral energy distribution for each chemical composition by splitting the mass range between 0.08 M$_{\sun}$ and the initial stellar mass of the AGB-tip (see e.g. \citealt{Wachter02,Kamath12} for the terminology; 0.967 and 0.735 M$_{\sun}$ for the $Y_\mathrm{ini} = 0.26$ and the $Y_\mathrm{ini} = 0.4$ models respectively) into bins centred on each of the stellar spectral models. We then summed together the model stellar spectral energy distributions weighted by the initial stellar mass in each bin according to a \citet{Kroupa01} IMF to produce the model spectral energy distribution for the population. We also calculated bottom heavy (dwarf star rich) solar composition models using power-law IMFs with a slope of $\alpha = -3.0$, $\alpha = -2.7$ and $\alpha = -2.35$ \citep{Salpeter55}. To compare our models with non-scaled solar chemistry with the effects of age and metallicity, we also calculated a model with lower metallicity ([Fe/H] = -0.30) as well as a model with a younger age (8.9 Gyr) at solar composition. \medskip

Here we have chosen for the second population stars a $Y_\mathrm{ini}$ as high as 0.4, to cover the extreme value found in a few Galactic GCs \citep[NGC~2808, $\omega$~Cen, NGC 6388, and to a certain extent NGC~2419,][]{Busso07,King12,DiCriscienzo15,Milone15}. However ETGs likely possess a range of He abundances rather than the two extreme values we consider, thus we have also interpolated from these two values a mixture of He normal models with 5, 10, 20 and 50\% of He rich CNONa models. We also create a population more representative of the core of ETGs (see Section~\ref{Disrupted}) made up of $\sim 35$\% second population stars by combining our models. We model the He rich population by representing the 5\% fraction of stars with $\Delta Y > 0.1$ with our $Y = 0.4$ model (He rich CNONa model). Of the remaining 30\% of second population stars, we assume that half have extreme chemistry but solar He (CNONa model), and the other half have intermediate chemistry which we represent with a 50-50 mix of our solar model and the CNONa model. Thus, we approximate the expected chemistry of the centre of a massive ETG with a mixture that is 5\% He rich CNONa, 22.5\% CNONa and 72.5\% solar composition. 

\begin{figure*}
   \centering
   \includegraphics[width=0.95\textwidth]{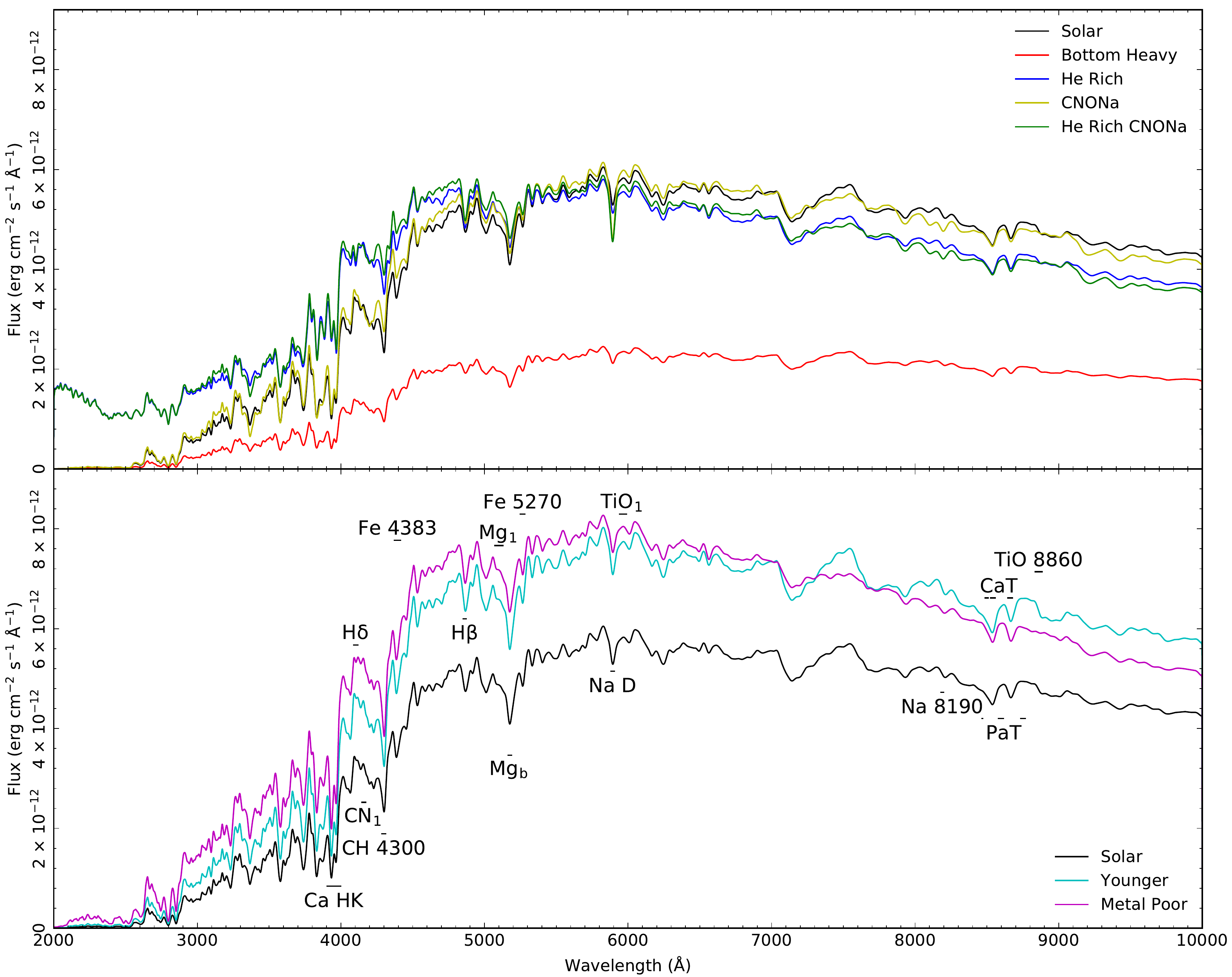}
   \caption{Model spectral energy distributions. \emph{Top panel:} Effects of He, CNONa and the IMF. The solar composition, Kroupa IMF model is shown as a black line, the solar composition, bottom heavy IMF ($\alpha = -3.0$) model as a red line, the He rich ($Y_\mathrm{ini} = 0.4$) model as a blue line, the enhanced NNa, depleted CO model as a yellow line and the HeNNa enhanced, depleted CO model as a green line. The bottom heavy IMF model is significantly fainter than the Kroupa IMF model but has a similar shaped spectral energy distribution. The He rich models are bluer than the He normal models, especially shortwards of 3000 \AA{}. The CNONa models are broadly similar to the models with the same He abundances but differ in the strengths of Na and molecular features. \emph{Bottom panel:} Effects of age and metallicity. The solar metallicity, 12.6~Gyr model is shown as a black line, the younger 8.9 Gyr model as a cyan line and the subsolar metallicity ([Fe/H] $= -0.3$) model as a magenta line. Both the metal poor and the younger models are brighter and bluer than the reference model and have weaker metal lines and molecular bands as well as stronger Balmer lines. Each spectrum has been smoothed to a spectral resolution of $R = 200$ for the purposes of display and represents the flux from 1~M$_{\sun}$ of initial stellar mass.}
   \label{Figure:physical_sed_all}
\end{figure*}

\subsection{Magnitudes and spectral indices}
\label{mags_indices}

To quantify large scale differences between the spectral energy distribution of our different models we calculated synthetic photometry by convolving each of the stellar population spectral energy distributions with the Hubble Space Telescope Wide Field Camera 3 (WFC3) filter curves\footnote{\url{http://www.stsci.edu/hst/wfc3/ins_performance/throughputs/Throughput_Tables}}.
We used the photometric zeropoints provided by the instrument website \footnote{\url{http://www.stsci.edu/hst/wfc3/analysis/uvis_zpts}} to calculate magnitudes in the Vega system (F225W, F275W, F336W, F438W, F475W, F555W, F606W, F625W, F775W, F814W, F850LP). \medskip

To study how a range of spectral features varies between models, we also calculated a range of Lick indices \citep{Worthey94} and indices used in IMF studies \citep[e.g.,][]{Maraston03,Conroy12_counting}. We calculated indices both sensitive to the IMF (e.g. Na D, Na 8190 and CaT) and insensitive (e.g., H$\beta$ and Fe 5270).  
Before calculating line strength indices, we broadened our model spectra to a velocity dispersion of 250 km s$^{-1}$. We then used the index definitions displayed in Table~\ref{table:index_definitions} and the formula provided in \citet{Cenarro01} to calculate the index strengths. As we are only interested in the differences between models, we do not attempt to place our Lick indices on the Lick system.

\begin{table*}
\caption{Stellar population models}
\begin{tabular}{l c c c c c}
\hline
Name & Helium & CNONa & [Fe/H] & Age & IMF \\
 & & & & [Gyr] & \\
(1) & (2) & (3) & (4) & (5) & (6) \\ \hline
He normal / Solar & $Y_\mathrm{ini} = 0.26$ & 0 \% & $\hphantom{-}0.0$ & 12.6 & Kroupa \\
He rich & $Y_\mathrm{ini} = 0.40$ & 0 \% & $\hphantom{-}0.0$ & 12.6 & Kroupa \\ 
He rich CNONa & $Y_\mathrm{ini} = 0.40$ & 100 \% & $\hphantom{-}0.0$ & 12.6 & Kroupa \\ 
CNONa & $Y_\mathrm{ini} = 0.26$ & 100 \% & $\hphantom{-}0.0$ & 12.6 & Kroupa \\ 
Bottom Heavy & $Y_\mathrm{ini} = 0.26$ & 0 \% & $\hphantom{-}0.0$ & 12.6 & $\alpha = -3.0$ \\ 
$\alpha = -2.7$ IMF & $Y_\mathrm{ini} = 0.26$ & 0 \% & $\hphantom{-}0.0$ & 12.6 & $\alpha = -2.7$ \\ 
Salpeter IMF & $Y_\mathrm{ini} = 0.26$ & 0 \% & $\hphantom{-}0.0$ & 12.6 & $\alpha = -2.35$ \\ 
Younger & $Y_\mathrm{ini} = 0.26$ & 0 \% & $\hphantom{-}0.0$ & $\hphantom{0}8.9$ & Kroupa \\  
Metal Poor & $Y_\mathrm{ini} = 0.26$ & 0 \% & $-0.3$ & 12.6 & Kroupa \\ 
He rich CNONa 5\% & 95 \% $Y_\mathrm{ini} = 0.26$ + $\hphantom{0}5$ \% $Y_\mathrm{ini} = 0.40$ & 5 \% & $\hphantom{-}0.0$ & 12.6 & Kroupa \\
He rich CNONa 10\% & 90 \% $Y_\mathrm{ini} = 0.26$ + 10 \% $Y_\mathrm{ini} = 0.40$ & 10 \% & $\hphantom{-}0.0$ & 12.6 & Kroupa \\
He rich CNONa 20\% & 80 \% $Y_\mathrm{ini} = 0.26$ + 20 \% $Y_\mathrm{ini} = 0.40$ & 20 \% & $\hphantom{-}0.0$ & 12.6 & Kroupa \\
He rich CNONa 50\% & 50 \% $Y_\mathrm{ini} = 0.26$ + 50 \% $Y_\mathrm{ini} = 0.40$ & 50 \% & $\hphantom{-}0.0$ & 12.6 & Kroupa \\
ETG mix & 95 \% $Y_\mathrm{ini} = 0.26$ + $\hphantom{0}5$ \% $Y_\mathrm{ini} = 0.40$ & 27.5 \% & $\hphantom{-}0.0$ & 12.6 & Kroupa \\ \hline
\end{tabular}

\medskip
\emph{Notes}. Column (1): Model name. Column (2): Initial helium mass fraction. Column (3): Percentage of N, Na enhanced, and C, O depleted stars. Column (4): Metallicity. Column (5): Age in Gyr. Column (6) IMF. Kroupa means a \citep{Kroupa01} IMF while $\alpha = -3.0$, $\alpha = -2.7$ and $\alpha = -2.35$ are power-law IMFs with the given slopes.
\label{models_nomenclature}
\end{table*}

\begin{table}
\caption{Spectroscopic Index definitions}
\begin{tabular}{l c c c}
\hline
Index & Feature & Continuum & Notes \\
& [\AA] & [\AA] & \\
(1) & (2) & (3) & (4) \\ \hline
Ca HK    & 3898.4 - 4002.4 & 3805.4 - 3832.7 & S05  \\
         &                 & 4019.6 - 4051.3 &      \\
H$\delta$ & 4083.5 - 4122.3 & 4041.6 - 4079.8 & Lick \\
         &                 & 4128.5 - 4161.0 &      \\
CN$_{1}$ & 4142.1 - 4177.1 & 4080.1 - 4117.6 & Lick \\
         &                 & 4244.1 - 4284.1 &      \\  
CH 4300  & 4281.4 - 4316.4 & 4266.4 - 4282.6 & Lick \\
         &                 & 4318.9 - 4335.1 &      \\
Fe 4383  & 4369.1 - 4420.4 & 4359.1 - 4370.4 & Lick \\
         &                 & 4442.9 - 4455.4 &      \\         
H$\beta$ & 4847.9 - 4876.6 & 4827.9 - 4847.9 & Lick \\
         &                 & 4876.6 - 4891.6 &      \\
Mg$_{1}$ & 5069.1 - 5134.1 & 4895.1 - 4957.6 & Lick \\
         &                 & 5301.1 - 5366.1 &      \\
Mg$_{b}$ & 5160.1 - 5192.6 & 5142.6 - 5161.4 & Lick \\
         &                 & 5191.4 - 5206.4 &      \\
Fe 5270  & 5245.7 - 5285.7 & 5233.2 - 5248.2 & Lick \\
         &                 & 5285.7 - 5318.2 &      \\
Na D     & 5876.9 - 5909.4 & 5860.6 - 5875.6 & Lick \\
         &                 & 5922.1 - 5948.1 &      \\
TiO$_{1}$ & 5936.6 - 5994.1 & 5816.6 - 5849.1 & Lick \\
         &                 & 6038.6 - 6103.6 &      \\
Na 8190  & 8174.7 - 8202.7 & 8167.7 - 8174.7 & C12  \\
         &                 & 8202.7 - 8212.7 &      \\
CaT      & 8484.0 - 8513.0 & 8474.0 - 8484.0 & C01  \\
         & 8522.0 - 8562.0 & 8563.0 - 8577.0 &      \\
         & 8642.0 - 8682.0 & 8619.0 - 8642.0 &      \\
         &                 & 8700.0 - 8725.0 &      \\
         &                 & 8776.0 - 8792.0 &      \\
PaT      & 8461.0 - 8474.0 & 8474.0 - 8484.0 & C01  \\
         & 8577.0 - 8619.0 & 8563.0 - 8577.0 &      \\
         & 8730.0 - 8772.0 & 8619.0 - 8642.0 &      \\
         &                 & 8700.0 - 8725.0 &      \\
         &                 & 8776.0 - 8792.0 &      \\         
TiO 8860 &                 & 8832.6 - 8852.6 & C12  \\
         &                 & 8867.6 - 8887.6 &      \\

\hline
\end{tabular}

\medskip
\emph{Notes:} Column (1): Index name. Column (2): Feature passband(s). Column (3): Continuum passbands. Column (4): C01: \citet{Cenarro01}, C12: \citet{Conroy12_counting}, Lick: Lick index \citep{Worthey94,Worthey97} S05: \citet{Serven05}. The TiO 8860 index was defined by \citet{Conroy12_counting} as the ratio of fluxes between blue and red passbands.
\label{table:index_definitions}
\end{table}

\subsection{Caveats to the modelling}\label{modeling_caveats}

In our stellar population synthesis models, we did not take into account variations of Mg and Al initial abundances. This is motivated by the fact that the Mg-Al anti-correlation is not an ubiquitous chemical feature in GCs \citep[see e.g.,][]{Carretta09_MgAl}. In addition \cite{Cassisi13} showed that the inclusion of Mg and Al abundance variations in stellar models and synthetic spectra has only a small effect on bolometric corrections in standard filters. We also assume the same surface abundances at all stages of stellar evolution. Thus, we do not consider the effects of several processes including diffusion, convective mixing and radiative levitation that modify the surface chemistry of a star as it evolves. But since we are interested here in differences between He normal and He rich models, it will not largely affect our results under the rough assumption that these processes have similar effects on both populations. However this assumption may be violated for the horizontal branch stars in our He rich models which are at the onset ($\sim$12 000 K) of the effects of radiative levitation \citep[e.g.,][]{Behr03, Pace06}. The simple models we present in this study do not account for blue stragglers, post-AGB stars, and binaries, but their effects are briefly discussed in Section~\ref{Discussion}.  \medskip

Since we are mainly interested in the changes in the spectral energy distribution due to chemical composition, we are less sensitive to the limitations of calculating synthetic spectra (incomplete line lists, etc.) that are especially important for cool stars. We only model stellar photospheres; however chromospheres can play an important role for some spectral features and the ultraviolet luminosity in cool stars \citep{Wedemeyer17}. While the effects of non-local thermal equilibrium can be significant for some spectra features \citep[e.g.][]{Asplund05, Short15}, the effects of non-local thermal equilibrium on the response to abundance changes are not significant \citep{Conroy18}. In addition, the solar abundance patterns adopted by the isochrones are slightly different from the ones we use for our stellar atmosphere calculations. Finally, we would like to note that stellar atmosphere modelling of cool stars is less accurate than for stars with higher temperature, which is a recurrent problem in stellar population synthesis studies. However, since here we are interested in differences between two populations with different initial chemical compositions, the effects on the results are mitigated.

\section{Mass-to-light ratio}\label{MLratio}

First, we determine the stellar mass for populations with different chemical compositions, initial mass functions, and ages. We assume that gas lost during stellar evolution - stellar winds and supernovae ejecta - is expelled from the stellar population. At $Z_{\sun}$, 12.6~Gyr, and for a normal initial helium content, all the stars with $M_\mathrm{ini}\geq 0.967~$M$_{\sun}$ are present in the form of remnants whereas for a higher initial helium content of 0.4 this transition is shifted down to $M_\mathrm{ini}$ = 0.735~M$_{\sun}$. This is due to the well-known fact that He rich stars have shorter lifetimes than their He normal counterparts \citep[e.g.,][]{Salaris06,Pietrinferni09,Sbordone11,Valcarce12,Chantereau15,Althaus17}. The total mass locked in remnants from He rich progenitors is then higher by $\sim$41\% with respect to the He normal population (+0.062~M$_{\sun}$ when the initial mass of the population is normalised to 1~M$_{\sun}$). The inclusion of the mass-loss during advanced phases does not significantly change the total mass of stars since there are only 3\% and 3.5\% of post-MS stars (in mass) for $Y_\mathrm{ini}$ = 0.26 and 0.4 respectively. Thus, the fact that the mass-loss dependence on the initial helium content is currently unknown does not affect our conclusions for the total stellar mass. The total mass of stars still in the nuclear burning phase of the He rich population is then lower by $\sim$14\% with respect to the He normal population (-0.055~M$_{\sun}$ when normalised to 1~M$_{\sun}$), which counterbalanced the mass locked into remnants. We can then conclude that at 12.6~Gyr and at $Z_{\sun}$, the mass of both He normal and He rich populations are very similar.

For a younger He normal population (8.9~Gyr), only stars with $M_\mathrm{ini}\leq 1.035~$M$_{\sun}$ are still in the nuclear burning phase, which is relatively close to our old He normal population. Thus, an age difference of 3.7~Gyr has only a negligible effect on the total mass of the population. At a lower metallicity of [Fe/H] $= -0.3$, He normal stars which are still in the nuclear burning phase have a maximum initial mass very similar to the He normal case at $Z_{\sun}$. As shown in Fig.~\ref{Figure:IRMR_Z}, since  the initial-final mass relation does not change between $Z_{\sun}$ and [Fe/H] = -0.3 ($Z$ = 0.0069), we keep the same relation used before. Therefore, the total mass of the He normal population is the same as in the solar metallicity case. Finally we also compute the mass for a range of bottom heavy IMFs ($\alpha = -3.0$, $-2.7$ and $-2.35$) at $Z_{\sun}$ and 12.6~Gyr. As the slope of the IMF steepens, a higher fraction of the initial stellar mass is still in nuclear active stars. Thus, the normalised mass (to 1~M$_{\sun}$) of the population approaches unity. \medskip

Our model spectral energy distributions are displayed in Fig.~\ref{Figure:physical_sed_all}. The shape of the solar composition bottom heavy IMF model ($\alpha = -3$) is almost identical to the solar composition Milky Way-like IMF one \citep{Kroupa01} but the bottom heavy model is much fainter for the same initial mass. The He rich model is dramatically brighter than the He normal model bluewards of 2600 \AA{}, brighter at wavelengths bluer than $\sim 5000$ \AA{}, but fainter redwards. The CNONa models have similar shaped spectral energy distributions to solar models (same He abundance) but differ in the strengths of several spectral features including molecular bands and strong Na lines. Finally both the metal poor and the younger models are brighter and bluer than our old, solar metallicity model but not as bright as the He rich models in the ultraviolet.

\begin{figure}
\centering
 \includegraphics[width=0.45\textwidth]{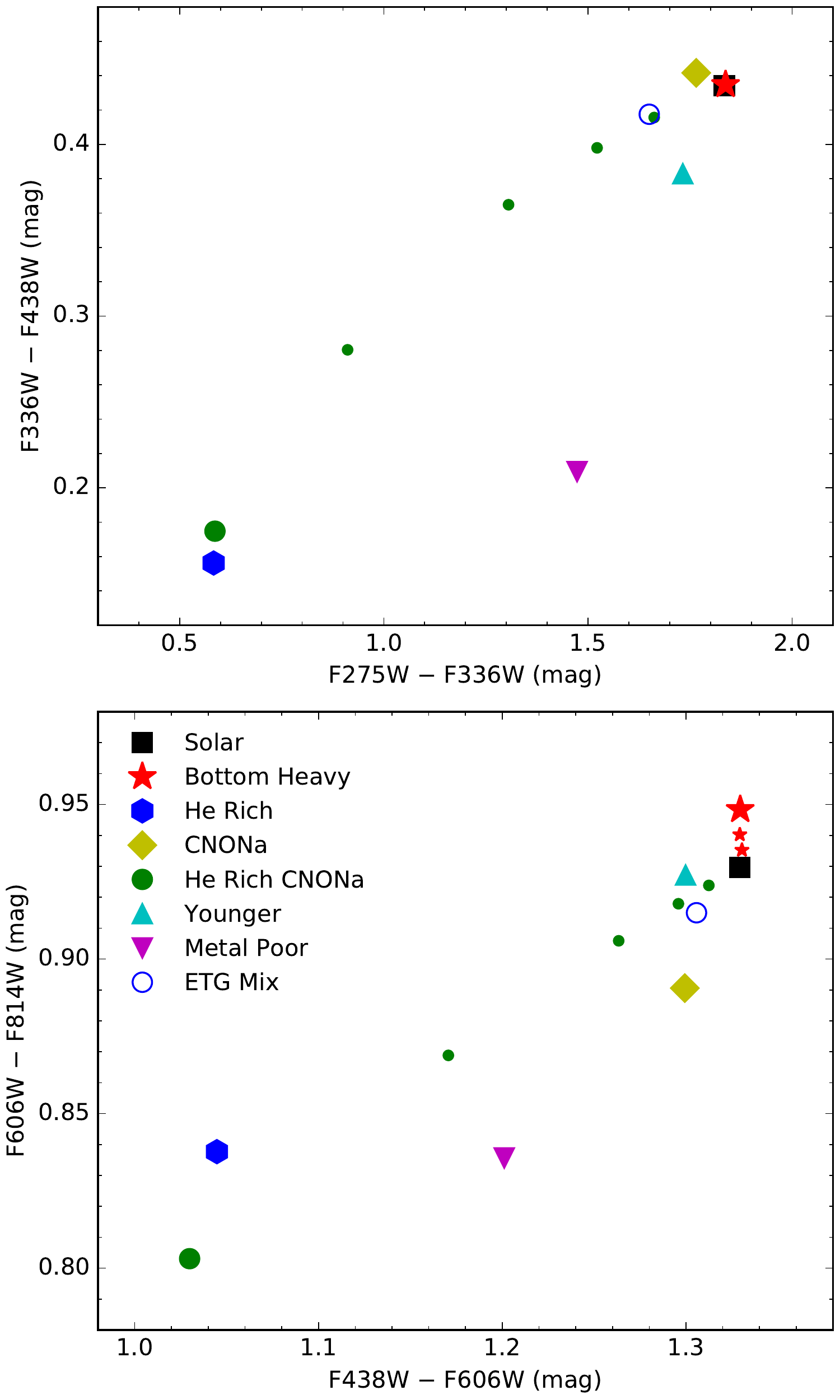}
 \caption{Colour-colour plots for our models in the HST WFC3 passbands. The solid black square is the solar composition model, the red star is the bottom heavy IMF model ($\alpha = -3.0$), the blue hexagon is the He rich model, the yellow diamond is the CNONa abundance pattern model ([C/Fe] = [O/Fe] $= -0.6$, [N/Fe] $= +1.0$, [Ne/Fe] $= -0.07$, [Na/Fe] $= +1.0$), the green circle is the He rich CNONa pattern, the cyan upwards triangle is the younger model (8.9 Gyr) and the magenta downwards pointing triangle is the metal poor model ([Fe/H] $= -0.3$). The small green circles represent mixtures of the solar composition model with 5, 10, 20 or 50\% of the He rich CNONa abundance pattern. The small red stars between the solar composition, Kroupa IMF model and the bottom heavy IMF model are the Salpeter ($\alpha = -2.35$) and $\alpha = -2.7$ IMF models respectively. The hollow blue circle represents our ETG mixture model (22.5\% CNONa model, 5\%  He rich CNONa model, 72.5\% solar model). He rich models are bluer than solar He models across all colours as are the younger model and the metal poor model. The bottom heavy IMF model has similar colours to the Milky Way-like IMF solar composition model except in (F606W $-$ F814W) where it is slightly redder. The CNONa model shows slightly bluer colours with respect to the solar composition model except in (F336W $-$ F438W) where the CNONa model is slightly redder. We note that the F336W, F438W, F606W and F814W WFC3 filters approximate the $UBVI$ Johnson-Cousins filters.}
 \label{Figure:colours}
\end{figure}

Colour-colour plots for our models are displayed in Fig.~\ref{Figure:colours}. Changing from a Milky Way-like IMF to a bottom heavy IMF has little effect on the colours except (F606W $-$ F814W) which is slightly redder. As expected from the spectral energy distributions, enhanced He models have bluer colours compare to models with solar  metallicities. Younger ages and lower metallicities also lead to bluer colours, but not as blue as the He rich models. CNONa abundances lead to slightly bluer colours except in (F336W $-$ F438W) which is marginally redder than the solar composition model. \medskip

\begin{table*}
\begin{tabular}{l c c c c c c c c c c c}
\hline
Models & Mass & F225W & F336W (U/u) & F438W (B) & F475W (g) & F555W (V) & F625W (r) & F775W (i) & F814W (I) & F850LP (z) \\ \hline \hline
Solar & 0.568 & 17.73 & 9.88 & 6.50 & 5.36 & 4.67 & 3.98 & 3.07 & 2.89 & 2.45 \\ \hline
He rich & 0.560 & 0.86 & 6.02 & 5.12 & 4.80 & 4.54 & 4.17 & 3.39 & 3.22 & 2.78 \\ \hline
He rich & \multirow{2}{*}{0.560} & \multirow{2}{*}{0.86} & \multirow{2}{*}{5.93} & \multirow{2}{*}{4.96} & \multirow{2}{*}{4.66} & \multirow{2}{*}{4.45} & \multirow{2}{*}{4.12} & \multirow{2}{*}{3.43} & \multirow{2}{*}{3.27} & \multirow{2}{*}{2.85} \\ 
CNONa & \\ \hline
CNONa & 0.568 & 16.12 & 9.48 & 6.20 & 5.16 & 4.55 & 3.92 & 3.11 & 2.94 & 2.51 \\ \hline
Bottom & \multirow{2}{*}{0.944} & \multirow{2}{*}{72.93} & \multirow{2}{*}{40.56} & \multirow{2}{*}{26.68} & \multirow{2}{*}{22.06} & \multirow{2}{*}{19.19} & \multirow{2}{*}{16.32} & \multirow{2}{*}{12.43} & \multirow{2}{*}{11.67} & \multirow{2}{*}{9.82} \\ 
Heavy & \\ \hline
Younger & 0.575 & 10.48 & 6.98 & 4.81 & 4.02 & 3.53 & 3.04 & 2.35 & 2.21 & 1.86 \\ \hline 
Metal Poor & 0.568 & 4.59 & 5.13 & 4.15 & 3.62 & 3.27 & 2.90 & 2.37 & 2.27 & 2.00 \\ \hline
He rich & \multirow{2}{*}{0.567} & \multirow{2}{*}{9.01} & \multirow{2}{*}{9.56} & \multirow{2}{*}{6.40} & \multirow{2}{*}{5.32} & \multirow{2}{*}{4.66} & \multirow{2}{*}{3.99} & \multirow{2}{*}{3.09} & \multirow{2}{*}{2.91} & \multirow{2}{*}{2.47} \\
CNONa 5\% & \\ \hline
He rich  & \multirow{2}{*}{0.567} & \multirow{2}{*}{6.03} & \multirow{2}{*}{9.27} & \multirow{2}{*}{6.31} & \multirow{2}{*}{5.28} & \multirow{2}{*}{4.65} & \multirow{2}{*}{3.99} & \multirow{2}{*}{3.11} & \multirow{2}{*}{2.93} & \multirow{2}{*}{2.49} \\
CNONa 10\% & \\ \hline
He rich  & \multirow{2}{*}{0.566} & \multirow{2}{*}{3.63} & \multirow{2}{*}{8.73} & \multirow{2}{*}{6.12} & \multirow{2}{*}{5.21} & \multirow{2}{*}{4.62} & \multirow{2}{*}{4.01} & \multirow{2}{*}{3.14} & \multirow{2}{*}{2.96} & \multirow{2}{*}{2.52} \\
CNONa 20\% & \\ \hline
He rich & \multirow{2}{*}{0.564} & \multirow{2}{*}{1.65} & \multirow{2}{*}{7.43} & \multirow{2}{*}{5.63} & \multirow{2}{*}{4.99} & \multirow{2}{*}{4.56} & \multirow{2}{*}{4.05} & \multirow{2}{*}{3.24} & \multirow{2}{*}{3.07} & \multirow{2}{*}{2.63} \\
CNONa 50\% & \\ \hline
\end{tabular}
\caption[]{Mass locked in the remnants and the stars (normalized to 1~$M_\odot$) and \textit{M}/\textit{L} in different WFC3 filters while the closest filters in the SDSS and Johnson-Cousins systems are displayed in brackets.}  \label{table:ML}
\end{table*}

\begin{figure*}
   \centering
   \includegraphics[width=0.95\textwidth]{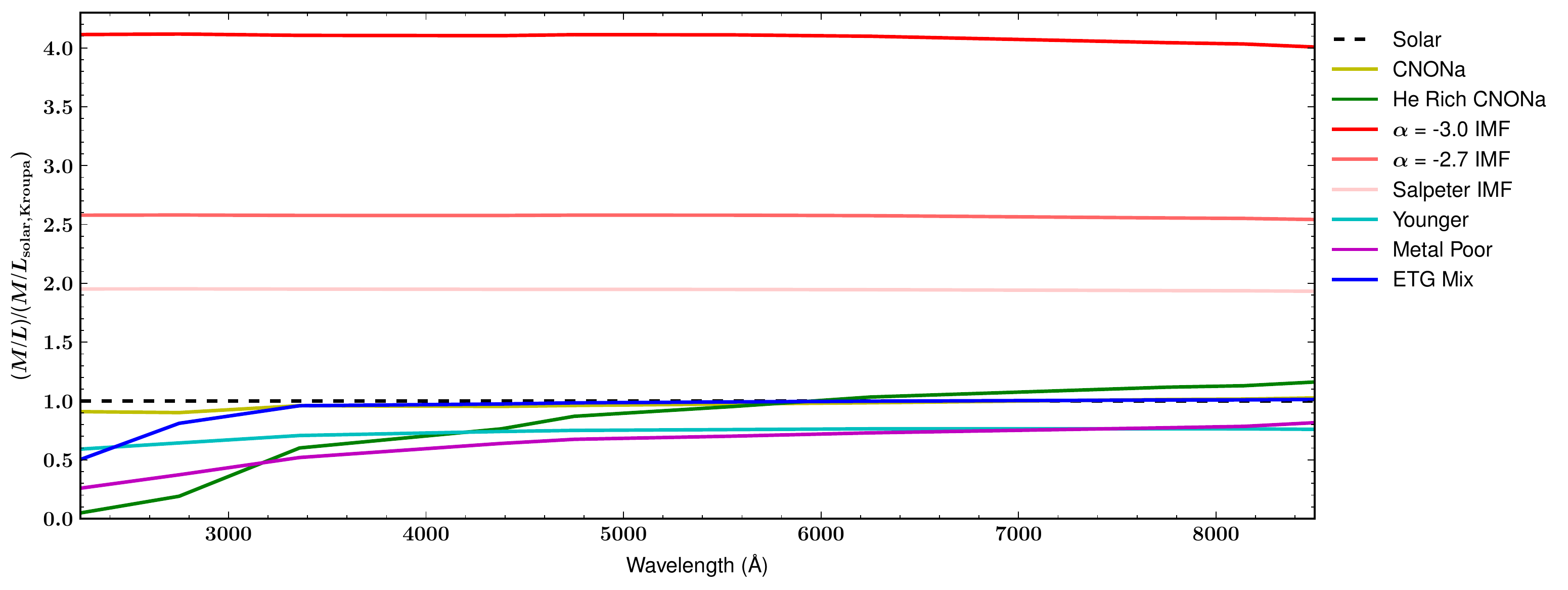}   
    \caption{Mass-to-light ratio as a function of wavelength. \textit{M}/\textit{L} for the solar composition, Milky Way-like IMF model (black), the bottom heavy IMF model (red), the CNONa model (yellow); the He rich CNONa model (green), the younger (8.9~Gyr) model (cyan), the metal poor ([Fe/H] $= -0.3$) model (magenta), and finally the ETG mixture model (blue). The He rich solar model (not displayed here) is largely similar to the He rich CNONa model. The bottom heavy IMF ($\alpha = -3.0$) has a higher \textit{M}/\textit{L} with respect to the Milky Way IMF model with a similar ratio ($\sim 4$) between the two models at all wavelengths. The He rich CNONa model has a significantly lower \textit{M}/\textit{L} at the bluest wavelengths than the solar model but slightly higher \textit{M}/\textit{L} at the reddest wavelengths. Increasing N and Na while decreasing C and O has little effect on the \textit{M}/\textit{L} at either solar or high He. Both the younger and the metal poor models show lower \textit{M}/\textit{L} with respect to the old, solar metallicity model at all wavelengths but show relatively lower \textit{M}/\textit{L} values in the bluer bands.}
    \label{Figure:ML}
\end{figure*}

\begin{figure*}
   \centering
   \includegraphics[width=0.95\textwidth]{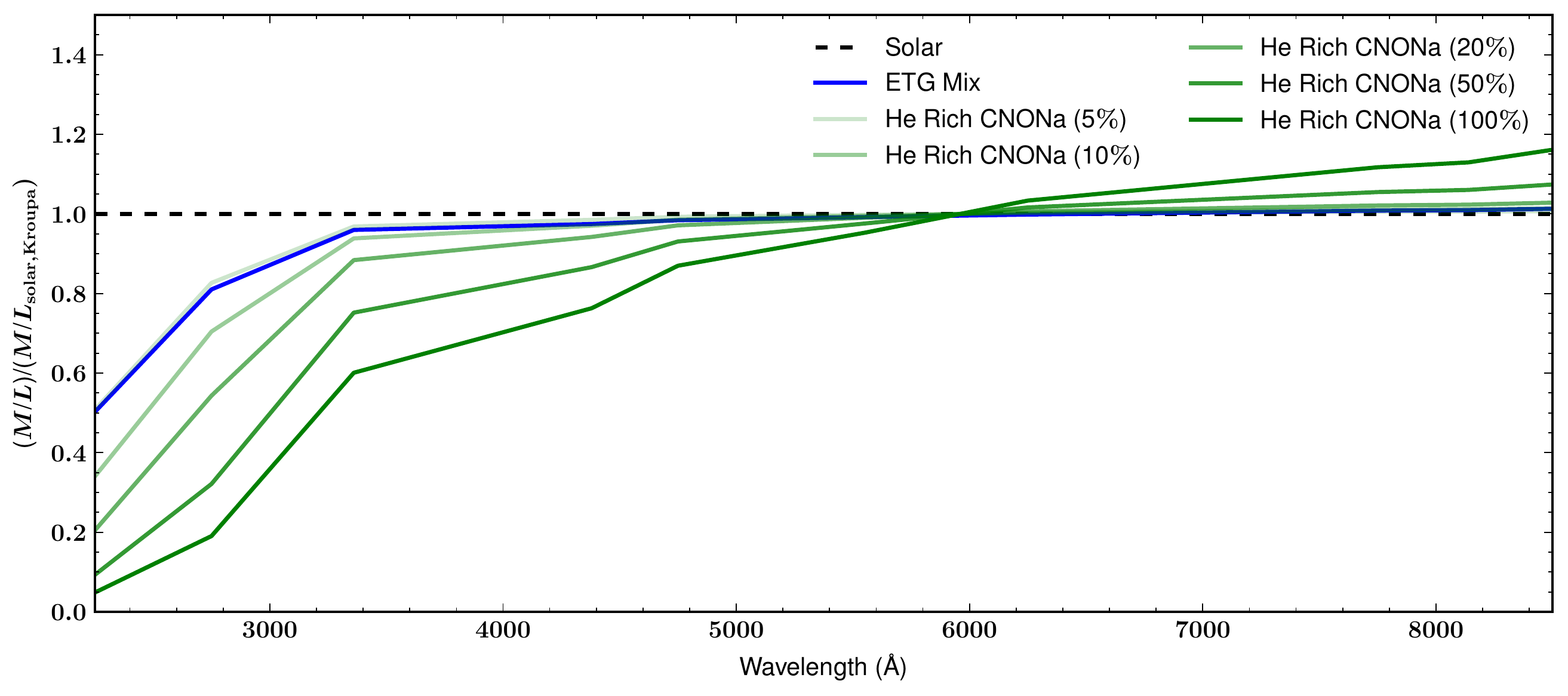}   
    \caption{Mass-to-light ratio as a function of wavelength. The \textit{M}/\textit{L} for the solar composition model is plotted in black, for the He rich CNONa model in green and for the various mixtures of the He rich CNONa model and the solar model as lighter shades of green. Unsurprisingly, the mixture models are intermediate to the two base models. The \textit{M}/\textit{L} for the ETG mixture model is plotted in blue.}
    \label{Figure:ML_2}
\end{figure*}

The mass-to-light ratios for our different models are displayed in table~\ref{table:ML}, Fig.~\ref{Figure:ML} and \ref{Figure:ML_2}. Since we have shown that He, CNONa, age (up to a 3.7~Gyr spread) and metallicity (up to $\Delta$[Fe/H] =  -0.3) do not dramatically affect the total stellar mass, \textit{M}/\textit{L} mainly depend on differences in the light contribution. The He rich population has a negligible effect on the \textit{M}/\textit{L} between $V$ (F606W) and $I$ (F814W) bands, which does not explain the variations found in the $r$-band by \cite{Cappellari12}. We also find that the differences in \textit{M}/\textit{L} between the case where only $Y_\mathrm{ini}$ changes (He rich model) with the case where all the initial abundances vary (He rich CNONa model) are negligible, highlighting the fact that as expected He is the main factor of \textit{M}/\textit{L} variations in the different filters. As expected \citep[e.g.,][]{Conroy09}, more bottom heavy IMFs lead to higher \textit{M}/\textit{L} in all bands. The ratio of the \textit{M}/\textit{L} of a model with a bottom heavy IMF to a population with the same chemistry and a Milky Way-like IMF is largely insensitive of wavelength. Therefore, here we can safely rule out the He rich population as the source of the \textit{M}/\textit{L} variations in $r$-band (F625W) in massive ETGs.

Both the younger and the more metal-poor models display similar behaviour to the He rich population in that their \textit{M}/\textit{L} decreases with shorter wavelengths.
However, unlike a He rich population, the younger and metal poor models show lower \textit{M}/\textit{L} with respect to the old, solar metallicity model in the $V$-band (F606W) and redwards. The 5, 10, and 20\% models are close to the He normal case in most of the bands. However, a conservative mixture such as 95\% of He normal stars with 5\% of He rich stars still leads to relatively large differences of \textit{M}/\textit{L} in the ultraviolet bands. The He rich population leads to a near-UV luminosity (in the F225W filter) $\sim$20 times higher than for the He normal population and $\sim$2 times higher if we take only 5\% of He rich stars into account. This is in good agreement with the statement that He rich populations in ETGs could play an important role in the UV-excess phenomenon \citep[e.g.,][]{Code79,Burstein88,Greggio90,Dorman95,Brown97,OConnell99,Brown00}.

\section{Spectral indices}\label{index_strengths}

\begin{figure*}
   \centering
   \includegraphics[width=504pt]{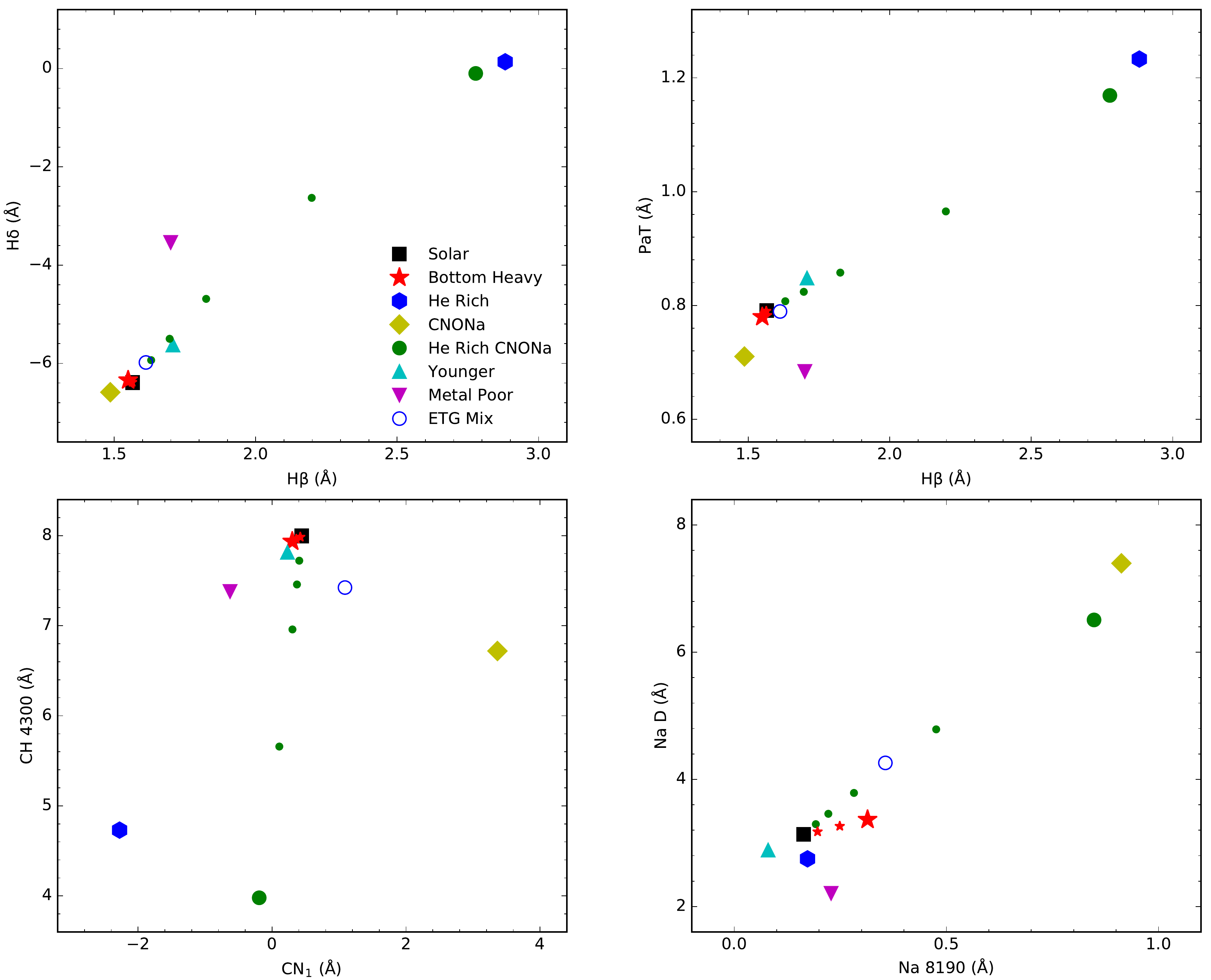}   
    \caption{Line index strengths.
    The symbol shapes and colours are as in Figure \ref{Figure:colours}.
    \emph{Top left:} Strength of the H$\delta$ line index versus the strength of the H$\beta$ line index.
    \emph{Top right:} Strength of PaT Paschen line index versus the strength of the H$\beta$ line index.	
	While the IMF has practically no effect on the strength of these H absorption lines, He abundance has a dramatic effect with a He rich population mimicking a significantly younger population.
	The CNONa model displays slightly weaker H lines with respect to the solar composition model.
    \emph{Bottom left:} Strength of the CH 4300 molecular index versus the strength of the CN$_{1}$ molecular index.
    While the IMF has virtually no effect on the strengths of the molecular bands, a He rich population has weaker CH and CN bands.
    The C poor and N rich second population abundance pattern models unsurprisingly display weaker CH bands and stronger CN bands.
    \emph{Bottom right:} Effects of He abundance on the Na D and Na 8190 line indices.
	While a bottom heavy IMF leads to stronger Na D and and Na 8190 absorption, enhanced He leads to weaker Na D lines but slightly stronger Na 8190 absorption.
	However, the effects of the IMF or He abundance on these Na indices are swamped by the effects of higher Na abundances in the second population abundance pattern models.}
    \label{Figure:indices_1}
\end{figure*}

\begin{figure*}
   \centering
   \includegraphics[width=504pt]{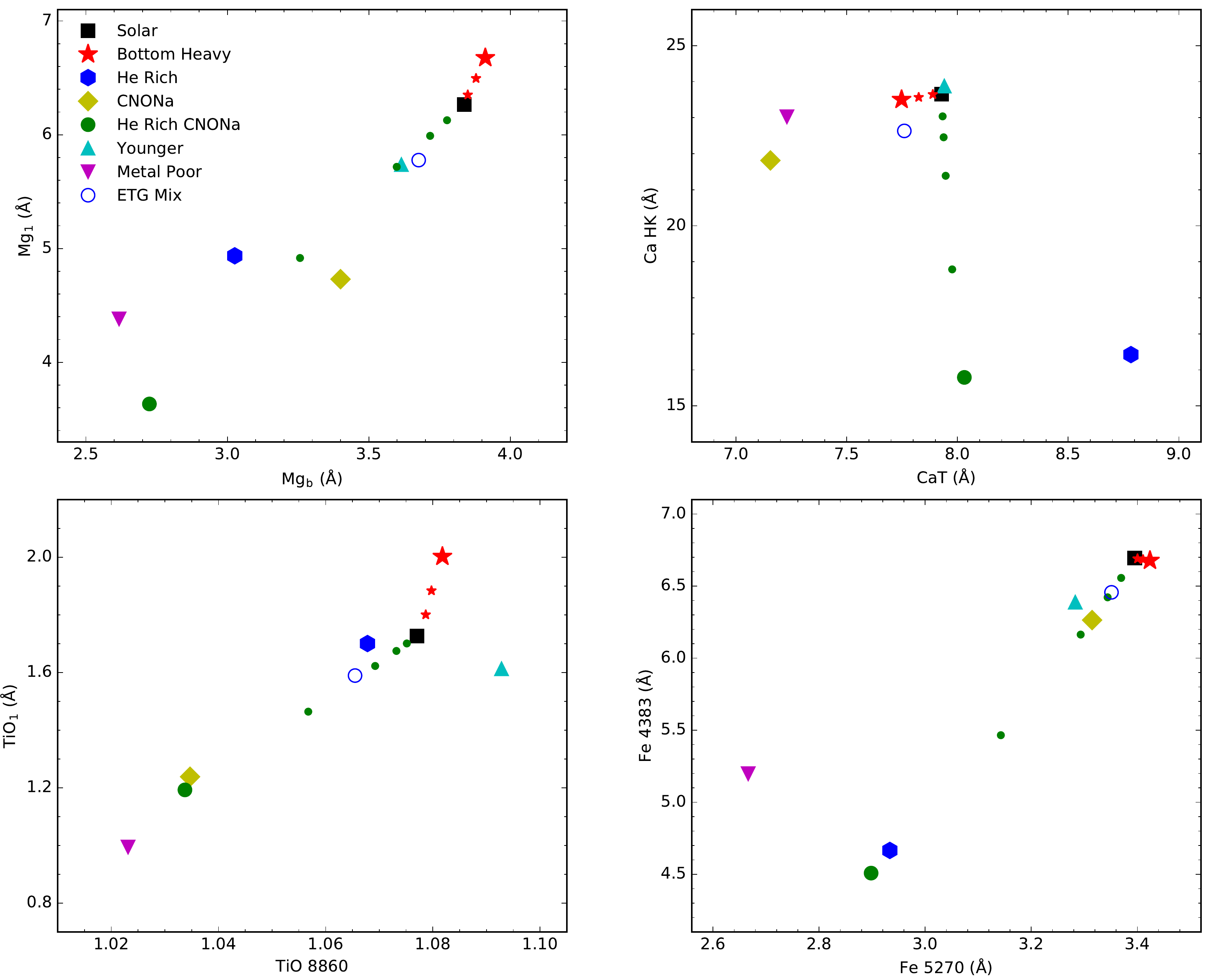}   
    \caption{Line index strengths.
    The symbol shapes and colours are as in Figure \ref{Figure:colours}.
  \emph{Top left:} Mg$_{1}$ MgH molecular band strength versus Mg$_{b}$ line index.
  While the bottom heavy IMF displays stronger Mg$_{1}$ and Mg$_{b}$ strengths, the He rich model shows weaker Mg$_{1}$, Mg$_{b}$.
  The CNONa mixture leads to weaker Mg index strengths.
    \emph{Top right:} Strength of the Ca H and K lines versus the strength of the calcium triplet (CaT).
	While the bottom heavy IMF model shows weaker CaT strength, similar H and K strengths with respect to the solar composition, Kroupa IMF model, the He rich model shows significantly weaker H and K absorption and stronger CaT absorption.
	The stronger CaT index strengths in the He rich model are due to contamination by the stronger Paschen line absorption in the He rich model (see Figure~\ref{Figure:indices_1}).
	The CNONa mixture leads to weaker Ca index strengths.
    \emph{Bottom left:} Effects of He abundance on the TiO$_{1}$ and TiO 8860 TiO molecular indices.
	Whereas the bottom heavy IMF model shows stronger TiO$_{1}$ and marginally stronger TiO 8860 band strengths, the He rich model shows weaker TiO 8860 absorption and no change in the TiO$_{1}$ index.
	The low O abundance second population abundance pattern models show weaker TiO band strengths with respect to the models with solar [O/Fe].
    \emph{Bottom right:} Effects of He on the Fe 4383 and Fe 5270 indices.
	While the IMF has little effect on the strength of these Fe absorption indices, increased He abundance produces weaker index strengths as does the CNONa mixture.
	The effect of enhanced He instead resembles the effects of a younger age.}
    \label{Figure:indices_2}
\end{figure*}

We plot the spectral index measurements for our models in Fig.~\ref{Figure:indices_1} and \ref{Figure:indices_2}. We note that the strength of a spectral index depends not just on the spectral feature that dominates the index, but on all atomic and molecular lines present in both the feature and continuum passbands. While spectral indices fail to capture all information available in the spectral energy distribution - they ignore the shape of the spectral feature being measured - they do provide insights into the effects of varying various stellar population parameters \citep{Conroy12_counting}. 

Our models confirm previous work on the effects of age, metallicity and the IMF on spectral indices.
In line with previous work \citep[e.g.,][]{Worthey94_comp,Thomas03,Bruzual03,Schiavon07}, young populations and more metal poor populations generally have weaker metal lines and stronger hydrogen Balmer lines with age having a stronger effect on the hydrogen lines and metallicity having a stronger effect on the metal lines and molecular bands.
The effects of a bottom heavy IMF on our modelled spectral indices are also broadly consistent with previous work \citep[e.g.][]{Conroy12_counting, LaBarbera13} with spectra features strong at low surface gravities, such as the CaT, are weaker in the bottom heavy models and spectral features stronger at high surface gravities, such as the Na doublets at 5890 and 8190 \AA{}, are stronger.
As has been previously shown, the IMF generally has little effect on spectral features in the blue portion of the optical wavelength range although some molecular bands such as the MgH band at 5100 \AA{} (which affects both the Mg$_{1}$ and Mg$_{b}$ indices) do show an IMF dependence.

The effect of He enhancement on Lick indices has also been considered in previous work \citep[e.g.,][]{Schiavon04, Chung13, Chung17}.
As in previous studies we see significantly stronger Balmer lines in the He rich models.
Enhanced He also leads to weaker metal lines, particularly at shorter wavelengths.
An exception to this trend is the CaT where the presence of the Pa 13, Pa 15 and Pa 16 hydrogen Paschen lines overlap in wavelength with the three CaT lines.
The enhanced He abundance leads to stronger Paschen absorption which causes a stronger CaT index strength although the strength of the CaT lines themselves is barely affected by the He enhancement.

These changes in line strength with He are mostly due to the effects of He on the horizontal branch as the main sequence and giant branches of the He rich model are only $\sim 200$ K hotter than the solar composition model while the horizontal branch of the He rich model is 7600 K hotter than the red clump of the solar He model (12300 K versus 4700 K).
We note the morphology of the horizontal branch is dependent on a host of factors besides He abundance including age and metallicity as well as the amount of mass loss on the RGB.
Thus, the effect of enhanced He varies with age and metallicity \citep[e.g.][]{Chung17}.
We note that a population with a hotter horizontal branch would likely have weaker H lines than our He rich model due to the increased ionisation of H with increasing temperature.
While to first order the effects of increased He mimic the effects of younger ages, more detailed comparisons do revel differences \citep{Schiavon04, Schiavon07}.

Unsurprisingly, the Na D and Na 8190 indices are significantly stronger in CNONa models.
Since N is the minority species, the CN$_{1}$ index is significantly stronger in the CNONa models while the C dominated CH 4300 index is weaker.
Although the strength of TiO bands is mainly set by the Ti abundance, the lower O abundance in the CNONa models leads to weaker TiO$_{1}$ and TiO 8860 bands.
We note the conversion of Ne into Na in the CNONa models has an important second order effect on many spectral features by raising the electron pressure as Na has a relatively low (5.1 eV) ionisation potential while Ne has a relatively high (21.6 eV) ionisation potential \citep{Conroy12_counting}.
Changes in the abundances of the CNO elements also play a secondary role in affecting index strengths by changing the strengths of molecular bands in both the feature passbands and in the continuum passbands.
An example of both these effects is the impact of NNa enhancement and CO depletion on the CaT index.
The higher electron pressure caused by the enhanced Na abundance lowers the abundance of singly ionised Ca.
This lowers the strength of the CaT lines themselves while the lower O reduces the strength of the TiO bands in the region of the CaT and the increased N increases the strengths of the CN bands in the same region.
Combined, these effects reduce the measured CaT strength.

To first order, the behaviour of the He rich CNONa model is the superposition of the separate effects of He and of NNa enhancement with CO depletion.
The models that are mixtures of the He rich CNONa model and the solar composition model are intermediate to the two model base models in rough proportion to the fraction of each model. 

\begin{figure}
\centering
 \includegraphics[width=240pt]{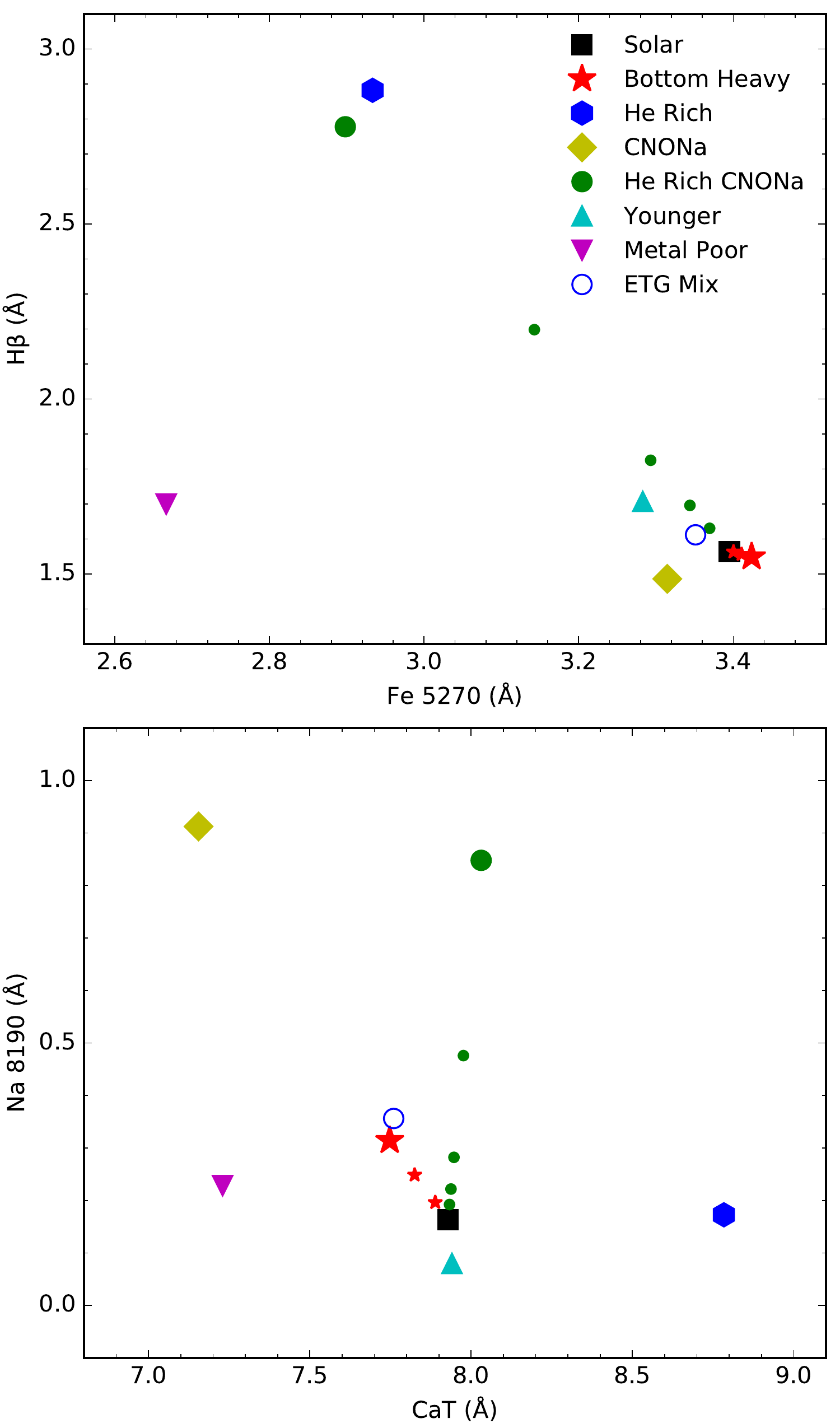}
 \caption{Line index strengths. The symbol shapes and colours are as in Figure \ref{Figure:colours}. \emph{Top:} Strength of H$\beta$ versus the Fe 5270 spectral index. The He rich models have significantly stronger H$\beta$ and weaker Fe 5270 index strengths with respect to the solar composition model making them appear like a younger population rather than a bottom heavy IMF which shows marginally stronger Fe 5270 strength. The CNONa model has weaker H$\beta$ and Fe 5270 indices. \emph{Bottom:} Strength of the Na 8190 index versus the CaT spectral index. The He rich solar model has stronger CaT values due to stronger Paschen lines while the CNONa model has stronger Na 8190 and weaker CaT strengths mimicking a bottom heavy IMF.}
 \label{Figure:age_metal_imf_indices}
\end{figure}

In the top panel of Figure \ref{Figure:age_metal_imf_indices} we show both the hydrogen H$\beta$ index and the iron Fe 5270 for our models which together have been commonly used to measure galaxy ages.
Going from the solar composition, Kroupa IMF to a bottom heavy IMF model with the same chemistry has little effect on these indices.
However, going from a solar composition model to the He rich model leads to a large increase in H$\beta$ and a decrease in Fe 5270 in a similar manner to a younger age.
The He rich CNONa model shows the same behaviour.

In the bottom panel of Figure \ref{Figure:age_metal_imf_indices} we show both the IMF sensitive Na 8190 and CaT indices.
While a bottom heavy IMF causes stronger Na 8190 and weaker CaT strengths, an increase in He leads to a stronger CaT index but almost no change in the Na 8190 index.
However, the He rich CNONa models show much stronger Na 8190 indices and little change in the CaT strength.
In Figure \ref{Figure:indices_2}, both the Mg indices and the TiO indices are stronger in the bottom heavy IMF models but weaker in the He rich or CNONa models.
Taken together these plots show that increased He can not explain the spectral feature evidence for a bottom heavy IMF.
A He rich population instead mimics the spectrum of a younger population.
While a enhanced NNa and depleted CO population can mimic some of the effects of a bottom heavy IMF (i.e. enhanced Na indices), it can not reproduce other indices (Mg indices, TiO bands). 

\section{Early-type galaxy mixture}\label{application}

Unsurprisingly, the ETG mixture behaves similarly to the He rich models and the CNONa models.
While the shape of the spectral energy distribution of ETG mixture model is similar to CNONa model redwards $\sim 4000$ \AA{}, bluewards of $\sim 4000$ \AA{} the shape of the spectral energy distribution is similar to the He rich CNONa model. This is most noticeable bluewards of 2500 \AA{} where the flux of the ETG mixture model is over 3 times higher than the solar composition model. The \textit{M}/\textit{L} of the ETG mixture model is similar to the 5\% He rich CNONa model as are the colours of the ETG mixture model. The increased He in our ETG mixture model leads to slightly stronger H indices in agreement with the other He rich models (Fig.~\ref{Figure:indices_1}). Like the CNONa model, the CH, Mg, Ca, TiO and Fe indices are weaker than the solar model while the CN and Na indices are stronger. In the H$\beta$-Fe 5270 plane (Fig.~\ref{Figure:age_metal_imf_indices}), the ETG mixture model follows the trend of the younger model rather than the bottom heavy IMF model. Thus ETG mixture model appears similar to the He rich models in this plane.

However, in the Na 8190-CaT plane (Fig.~\ref{Figure:age_metal_imf_indices}), the ETG mixture behaves similar to the CNONa model and mimics the appearance of the bottom heavy IMF model. While observations solely of the Na 8190 doublet and the CaT of the ETG mixture model would suggest a bottom heavy IMF, observations that cover a wider range of spectral features such as the Mg indices and the TiO bands would not support such an interpretation. Thus to disentangle chemical effects from IMF effects, we encourage the use of stellar population synthesis models such as \citet{Conroy12_counting} that allow for the abundances of a wide range of elements to be varied and for observations that cover a wide wavelength range.

In our toy model, the CO depleted, NNa enhanced stars change the mass weighted mean abundances of the population from solar to [C/Fe] = [O/Fe] $= -0.17$ and [N/Fe] = [Na/Fe] $= +0.28$. These changes are comparable to the abundance differences between the centre and one effective radii observed in massive ETGs by \citet{vanDokkum17}, suggesting that disrupted GCs can play important role in determining the mean abundances of the central regions of ETGs.

Disrupted GCs could also play a role in the origin of the UV-upturn.
Several authors \citep[e.g.][]{Dorman95, Yi97, Peng09} have proposed He rich stars as the origin of the UV-upturn and GCs provide a source of such He enhanced stars. 
The 5 \% of He enhanced stars in our ETG toy model is sufficient to raise the near-UV F225W luminosity by a factor of 2.
Hotter horizontal branch stars than those in our models ($\sim 12$ 000 K), for example from lower metallicity but still highly He enriched populations, produce even brighter far-UV and near-UV luminosities.
We note that massive ETGs show broad GC metallicity distributions \citep[e.g.][]{Peng06, Usher12} and that such galaxies contain GCs with bright UV luminosities \citep[e.g.][]{Sohn06, Rey09, Bellini15}.
Thus, disrupted GCs provide a plausible source for the UV bright stars responsible for the UV-upturn.

\section{Discussion}\label{Discussion}

In this section, we briefly discuss the assumptions of our models and the perspectives. 

We have assumed that ETGs are single age and single metallicity. For a fixed mean age, extended star formation histories lead to bluer colours as younger populations are both bluer and brighter than older populations. While ETGs do display a spread in ages, studies of their star formation histories show that the most massive ETGs have at least 90\% of their mass in place by 10 Gyr ago \citep[redshift of $z \sim 2$, e.g.,][]{Heavens04, Panter07, McDermid15}. The broad metallicity distribution present in ETGs \citep[e.g.,][]{Harris02, Bird10} leads to bluer colours in the ultraviolet than a narrow distribution with the same peak metallicity as lower metallicity populations are both brighter and bluer \citep{Conroy09,Tang14}. \medskip

The isochrones we use are not fully self-consistent since we use different database for different stages of stellar evolution. This is to due the lack of stellar evolution models with enhanced helium that both extend to low stellar masses ($\sim 0.1$ M$_{\sun}$) and fully cover post RGB evolution. \medskip

The models we present in this study do not account for blue stragglers (BSs), post-AGB stars, and binaries. BSs dominate the energy contribution in ultraviolet in our old and metal rich models without blue horizontal branch (HB) for a He normal population. If one compares to the blue HB formed from He rich stars (Fig.~\ref{Figure:isochrone}), it is expected that the low number of BSs usually present in GCs \citep{Piotto04} would lead only to a small contribution in the ultraviolet band. Thus, it is important to keep in mind that BSs will change the properties of our He normal population in the ultraviolet band. Post-AGB stars dominate the energy contribution only in the far-UV \citep[$\lesssim$ 2000~\AA{},][]{Chung17} at the ages and metallicities investigated in this paper. Finally \citet{HernandezPerez14} showed that interacting binaries might form extreme HB stars and in turn lead to strong emissions in the ultraviolet band,  providing an alternative solution to explain the UV-upturn phenomenon in ETGs. The binary fraction required then is at least 15\%, which is relatively similar to what is found in low-metallicity Galactic GCs \citep[see][]{Milone12}. Thus, BSs and post-AGB stars do not affect the conclusions of our study. However it would be interesting to include binaries in future studies and to couple them with He rich models in order to be more consistent when studying the UV contribution. \medskip 

As we have shown in this study, the He rich population has a non-negligible effect on mass of the stars still in the nuclear active phase but also the remnants when considered separately (for the extreme case $Y_\mathrm{ini} = 0.4$). However the effect becomes negligible when one looks at the total mass of the population. Thus, in studies where the mass of remnants or stars plays a crucial role, it is more consistent to take into account suitable initial-final mass relations and mass cut-off (minimum $M_\mathrm{ini}$ for remnants) for the He rich population. In addition, for this extreme He content, it would be interesting to know for stars which do not end their life as WDs if their final masses and mass boundaries change. For instance, as shown in this study and pointed out in \cite{Shingles15}, the He enhancement will increase the formation rate of NSs through electron-capture supernovae. \medskip

We have shown that from the theoretical point of view, the optical \textit{M}/\textit{L} of both He rich and He normal populations is similar. Then to observationally highlight the presence of a He rich population at work in ETGs through \textit{M}/\textit{L}, it is mandatory to focus on shorter wavelengths. The effects of the He rich population is already seen at $\sim$4000~\AA{} and $\sim$2600~\AA{} (if we compare to populations with $Y_\mathrm{ini} = 0.4$ and the ETG mixture respectively). Note that we have to be careful since a He rich population mimic the \textit{M}/\textit{L} of a He normal population at lower metallicity and/or younger age in U bands. \medskip  

\section{Conclusion}\label{Conclusion}

Massive ETGs display radial abundance gradients which do not correspond to standard chemical predictions and \textit{M}/\textit{L} variations which might be explained so far only with a debated change in the stellar IMF. Since multiple stellar populations of globular clusters display similar chemical trends and are expected to lead to \textit{M}/\textit{L} variations, we investigate in this paper if these multiple stellar populations could be the source of these patterns observed in their host galaxies. \medskip

We use a toy model to estimate the contribution of disrupted GCs to the centres of massive ETGs. We find that $\sim 35$\% of stars in a massive ETG should show the N, Na enhanced, C, O depleted chemistry of second population GC stars. This contribution of GC stars can explain the Na enriched, O and C depleted abundance pattern seen in the centres of massive ETGs \citep{vanDokkum17}. We also estimate that $\sim 5$\% of stars should show extreme He enhancement ($\Delta Y > 0.1$). \medskip

We have constructed single stellar population models with a range of He abundances, chemistries and IMFs. For each of our model of spectral energy distribution we calculate synthetic magnitudes and spectral indices. We consider the effects of helium abundance on stellar final masses allowing us to study the effects of helium on the mass-to-light ratio. \medskip

We find that the lower mass in nuclear burning stars in a He rich population is largely offset by the higher total mass locked into the remnants. As such, the lower \textit{M}/\textit{L} of a He rich population is due to its higher luminosity. Since changes in abundance pattern at fixed C+N+O have little effect on stellar evolution and since differences in the abundance pattern have little effect on broadband magnitudes, the N, Na enhanced, C, O depleted  models have quite similar \textit{M}/\textit{L} ratios to models with the same He abundance. \medskip

Both the colours and the spectral indices of He rich populations resemble a younger population rather than a bottom heavy IMF. However, a  N, Na enhanced, C, O depleted population can mimic the effects of a bottom heavy IMF on some IMF spectral indices (Na D, Na 8190, CaT) but not others (such as TiO bands). \medskip

We also find that the small contribution of He rich stars of our toy model (stars with $\Delta Y > 0.1$) has little effect on the spectral spectral energy distribution redwards of 4000 \AA{} but is enough to dramatically change the flux bluewards of 2600 \AA{} contributing to the UV-upturn phenomenon. \medskip

We conclude that He enhancement cannot mimic the observational signatures of a bottom heavy IMF as a He rich population resembles as younger population instead. However, we find disrupted GCs can supply a significant number of He, N and Na enhanced, C and O depleted stars to the field star population of massive ETGs. These stars can at least partially explain the UV-excess and the chemistry of the centres of massive ETGs. A similar conclusion was recently reached by \cite{Goudfrooij18} who studied the correlation between the UV-upturn strength and the specific frequency (the number of GCs per unit galaxy light) in early-type galaxies.\medskip

More detailed stellar population modelling is required to compare the predictions of this work with the observed spectra of the centres of massive ETGs. In a future work (Schiavon et al. in preparation) we will quantitatively explore the abundance gradients in ETGs and match them to the variations observed in globular clusters due to multiple populations.

\section*{Acknowledgements}
We warmly thank J. Pfeffer for useful discussions, and M. Salaris and R. Schiavon for careful reading and useful comments on the manuscript. W. Chantereau acknowledges funding from the Swiss National Science Foundation under grant P2GEP2\_171971 and thanks the International Space Science
Institute (ISSI, Bern, CH) for welcoming the activities of the Team 271 `Massive Star Clusters Across the Hubble Time'. C.U. and N.B. gratefully acknowledge financial support from the European Research Council (ERC-CoG-646928, Multi-Pop). N.B. gratefully acknowledge financial support from the Royal Society (University Research Fellowship).

This work made use of the Python packages \textsc{NumPy} \citep{numpy}, \textsc{Scipy} \citep{scipy} and \textsc{matplotlib} \citep{Matplotlib} as well as \textsc{Astropy}, a community-developed core Python package for astronomy \citep{astropy}.

\bibliographystyle{mnras}
\bibliography{bib}

\begin{thebibliography}{}
\makeatletter
\relax
\def\mn@urlcharsother{\let\do\@makeother \do\$\do\&\do\#\do\^\do\_\do\%\do\~}
\def\mn@doi{\begingroup\mn@urlcharsother \@ifnextchar [ {\mn@doi@}
  {\mn@doi@[]}}
\def\mn@doi@[#1]#2{\def\@tempa{#1}\ifx\@tempa\@empty \href
  {http://dx.doi.org/#2} {doi:#2}\else \href {http://dx.doi.org/#2} {#1}\fi
  \endgroup}
\def\mn@eprint#1#2{\mn@eprint@#1:#2::\@nil}
\def\mn@eprint@arXiv#1{\href {http://arxiv.org/abs/#1} {{\tt arXiv:#1}}}
\def\mn@eprint@dblp#1{\href {http://dblp.uni-trier.de/rec/bibtex/#1.xml}
  {dblp:#1}}
\def\mn@eprint@#1:#2:#3:#4\@nil{\def\@tempa {#1}\def\@tempb {#2}\def\@tempc
  {#3}\ifx \@tempc \@empty \let \@tempc \@tempb \let \@tempb \@tempa \fi \ifx
  \@tempb \@empty \def\@tempb {arXiv}\fi \@ifundefined
  {mn@eprint@\@tempb}{\@tempb:\@tempc}{\expandafter \expandafter \csname
  mn@eprint@\@tempb\endcsname \expandafter{\@tempc}}}

\bibitem[\protect\citeauthoryear{{Althaus}, {De Ger{\'o}nimo}, {C{\'o}rsico},
  {Torres}  \& {Garc{\'{\i}}a-Berro}}{{Althaus} et~al.}{2017}]{Althaus17}
{Althaus} L.~G.,  {De Ger{\'o}nimo} F.,  {C{\'o}rsico} A.,  {Torres} S.,
  {Garc{\'{\i}}a-Berro} E.,  2017, \mn@doi [\aap]
  {10.1051/0004-6361/201629909}, \href
  {http://cdsads.u-strasbg.fr/abs/2017A%26A...597A..67A} {597, A67}

\bibitem[\protect\citeauthoryear{{Asplund}}{{Asplund}}{2005}]{Asplund05}
{Asplund} M.,  2005, \mn@doi [\araa] {10.1146/annurev.astro.42.053102.134001},
  \href {http://adsabs.harvard.edu/abs/2005ARA%26A..43..481A} {43, 481}

\bibitem[\protect\citeauthoryear{{Asplund}, {Grevesse}, {Sauval}  \&
  {Scott}}{{Asplund} et~al.}{2009}]{Asplund09}
{Asplund} M.,  {Grevesse} N.,  {Sauval} A.~J.,   {Scott} P.,  2009, \mn@doi
  [\araa] {10.1146/annurev.astro.46.060407.145222}, \href
  {http://cdsads.u-strasbg.fr/abs/2009ARA%26A..47..481A} {47, 481}

\bibitem[\protect\citeauthoryear{{Astropy Collaboration} et~al.,}{{Astropy
  Collaboration} et~al.}{2013}]{astropy}
{Astropy Collaboration} et~al., 2013, \mn@doi [\aap]
  {10.1051/0004-6361/201322068}, \href
  {http://adsabs.harvard.edu/abs/2013A%26A...558A..33A} {558, A33}

\bibitem[\protect\citeauthoryear{{Baraffe}, {Chabrier}, {Allard}  \&
  {Hauschildt}}{{Baraffe} et~al.}{1998}]{Baraffe98}
{Baraffe} I.,  {Chabrier} G.,  {Allard} F.,   {Hauschildt} P.~H.,  1998, \aap,
  \href {http://adsabs.harvard.edu/abs/1998A%26A...337..403B} {337, 403}

\bibitem[\protect\citeauthoryear{{Bastian}}{{Bastian}}{2008}]{Bastian08}
{Bastian} N.,  2008, \mn@doi [\mnras] {10.1111/j.1365-2966.2008.13775.x}, \href
  {http://adsabs.harvard.edu/abs/2008MNRAS.390..759B} {390, 759}

\bibitem[\protect\citeauthoryear{{Bastian} \& {Lardo}}{{Bastian} \&
  {Lardo}}{2017}]{Bastian18}
{Bastian} N.,  {Lardo} C.,  2017, preprint, \href
  {http://adsabs.harvard.edu/abs/2017arXiv171201286B} {} (\mn@eprint {arXiv}
  {1712.01286})

\bibitem[\protect\citeauthoryear{{Bastian}, {Saglia}, {Goudfrooij},
  {Kissler-Patig}, {Maraston}, {Schweizer}  \& {Zoccali}}{{Bastian}
  et~al.}{2006}]{Bastian06}
{Bastian} N.,  {Saglia} R.~P.,  {Goudfrooij} P.,  {Kissler-Patig} M.,
  {Maraston} C.,  {Schweizer} F.,   {Zoccali} M.,  2006, \mn@doi [\aap]
  {10.1051/0004-6361:20054177}, \href
  {http://adsabs.harvard.edu/abs/2006A%26A...448..881B} {448, 881}

\bibitem[\protect\citeauthoryear{{Bastian} et~al.,}{{Bastian}
  et~al.}{2012}]{Bastian12}
{Bastian} N.,  et~al., 2012, \mn@doi [\mnras]
  {10.1111/j.1365-2966.2011.19909.x}, \href
  {http://adsabs.harvard.edu/abs/2012MNRAS.419.2606B} {419, 2606}

\bibitem[\protect\citeauthoryear{{Behr}}{{Behr}}{2003}]{Behr03}
{Behr} B.~B.,  2003, \mn@doi [\apjs] {10.1086/377509}, \href
  {http://adsabs.harvard.edu/abs/2003ApJS..149...67B} {149, 67}

\bibitem[\protect\citeauthoryear{{Belczynski}, {Bulik}, {Fryer}, {Ruiter},
  {Valsecchi}, {Vink}  \& {Hurley}}{{Belczynski} et~al.}{2010}]{Belczynski10}
{Belczynski} K.,  {Bulik} T.,  {Fryer} C.~L.,  {Ruiter} A.,  {Valsecchi} F.,
  {Vink} J.~S.,   {Hurley} J.~R.,  2010, \mn@doi [\apj]
  {10.1088/0004-637X/714/2/1217}, \href
  {http://cdsads.u-strasbg.fr/abs/2010ApJ...714.1217B} {714, 1217}

\bibitem[\protect\citeauthoryear{{Bellini} et~al.,}{{Bellini}
  et~al.}{2015}]{Bellini15}
{Bellini} A.,  et~al., 2015, \mn@doi [\apj] {10.1088/0004-637X/805/2/178},
  \href {http://adsabs.harvard.edu/abs/2015ApJ...805..178B} {805, 178}

\bibitem[\protect\citeauthoryear{{Bertelli}, {Girardi}, {Marigo}  \&
  {Nasi}}{{Bertelli} et~al.}{2008}]{Bertelli08}
{Bertelli} G.,  {Girardi} L.,  {Marigo} P.,   {Nasi} E.,  2008, \mn@doi [\aap]
  {10.1051/0004-6361:20079165}, \href
  {http://adsabs.harvard.edu/abs/2008A%26A...484..815B} {484, 815}

\bibitem[\protect\citeauthoryear{{Bird}, {Harris}, {Blakeslee}  \&
  {Flynn}}{{Bird} et~al.}{2010}]{Bird10}
{Bird} S.,  {Harris} W.~E.,  {Blakeslee} J.~P.,   {Flynn} C.,  2010, \mn@doi
  [\aap] {10.1051/0004-6361/201014876}, \href
  {http://adsabs.harvard.edu/abs/2010A%26A...524A..71B} {524, A71}

\bibitem[\protect\citeauthoryear{{Brown}, {Wallerstein}  \& {Oke}}{{Brown}
  et~al.}{1991}]{Brown91}
{Brown} J.~A.,  {Wallerstein} G.,   {Oke} J.~B.,  1991, \mn@doi [\aj]
  {10.1086/115798}, \href {http://cdsads.u-strasbg.fr/abs/1991AJ....101.1693B}
  {101, 1693}

\bibitem[\protect\citeauthoryear{{Brown}, {Ferguson}, {Davidsen}  \&
  {Dorman}}{{Brown} et~al.}{1997}]{Brown97}
{Brown} T.~M.,  {Ferguson} H.~C.,  {Davidsen} A.~F.,   {Dorman} B.,  1997,
  \mn@doi [\apj] {10.1086/304187}, \href
  {http://esoads.eso.org/abs/1997ApJ...482..685B} {482, 685}

\bibitem[\protect\citeauthoryear{{Brown}, {Bowers}, {Kimble}, {Sweigart}  \&
  {Ferguson}}{{Brown} et~al.}{2000}]{Brown00}
{Brown} T.~M.,  {Bowers} C.~W.,  {Kimble} R.~A.,  {Sweigart} A.~V.,
  {Ferguson} H.~C.,  2000, \mn@doi [\apj] {10.1086/308566}, \href
  {http://cdsads.u-strasbg.fr/abs/2000ApJ...532..308B} {532, 308}

\bibitem[\protect\citeauthoryear{{Bruzual} \& {Charlot}}{{Bruzual} \&
  {Charlot}}{2003}]{Bruzual03}
{Bruzual} G.,  {Charlot} S.,  2003, \mn@doi [\mnras]
  {10.1046/j.1365-8711.2003.06897.x}, \href
  {http://cdsads.u-strasbg.fr/abs/2003MNRAS.344.1000B} {344, 1000}

\bibitem[\protect\citeauthoryear{{Burstein}, {Bertola}, {Buson}, {Faber}  \&
  {Lauer}}{{Burstein} et~al.}{1988}]{Burstein88}
{Burstein} D.,  {Bertola} F.,  {Buson} L.~M.,  {Faber} S.~M.,   {Lauer} T.~R.,
  1988, \mn@doi [\apj] {10.1086/166304}, \href
  {http://cdsads.u-strasbg.fr/abs/1988ApJ...328..440B} {328, 440}

\bibitem[\protect\citeauthoryear{{Busso} et~al.,}{{Busso}
  et~al.}{2007}]{Busso07}
{Busso} G.,  et~al., 2007, \mn@doi [\aap] {10.1051/0004-6361:20077806}, \href
  {http://cdsads.u-strasbg.fr/abs/2007A%26A...474..105B} {474, 105}

\bibitem[\protect\citeauthoryear{{Cappellari} et~al.,}{{Cappellari}
  et~al.}{2012}]{Cappellari12}
{Cappellari} M.,  et~al., 2012, \mn@doi [\nat] {10.1038/nature10972}, \href
  {http://cdsads.u-strasbg.fr/abs/2012Natur.484..485C} {484, 485}

\bibitem[\protect\citeauthoryear{{Carretta}, {Cohen}, {Gratton}  \&
  {Behr}}{{Carretta} et~al.}{2001}]{Carretta01}
{Carretta} E.,  {Cohen} J.~G.,  {Gratton} R.~G.,   {Behr} B.~B.,  2001, \mn@doi
  [\aj] {10.1086/322116}, \href
  {http://adsabs.harvard.edu/abs/2001AJ....122.1469C} {122, 1469}

\bibitem[\protect\citeauthoryear{{Carretta} et~al.,}{{Carretta}
  et~al.}{2009a}]{Carretta09}
{Carretta} E.,  et~al., 2009a, \mn@doi [\aap] {10.1051/0004-6361/200912096},
  \href {http://cdsads.u-strasbg.fr/abs/2009A%26A...505..117C} {505, 117}

\bibitem[\protect\citeauthoryear{{Carretta}, {Bragaglia}, {Gratton}  \&
  {Lucatello}}{{Carretta} et~al.}{2009b}]{Carretta09_MgAl}
{Carretta} E.,  {Bragaglia} A.,  {Gratton} R.,   {Lucatello} S.,  2009b,
  \mn@doi [\aap] {10.1051/0004-6361/200912097}, \href
  {http://cdsads.u-strasbg.fr/abs/2009A%26A...505..139C} {505, 139}

\bibitem[\protect\citeauthoryear{{Carretta}, {Bragaglia}, {Gratton},
  {Recio-Blanco}, {Lucatello}, {D'Orazi}  \& {Cassisi}}{{Carretta}
  et~al.}{2010}]{Carretta10}
{Carretta} E.,  {Bragaglia} A.,  {Gratton} R.~G.,  {Recio-Blanco} A.,
  {Lucatello} S.,  {D'Orazi} V.,   {Cassisi} S.,  2010, \mn@doi [\aap]
  {10.1051/0004-6361/200913451}, \href
  {http://cdsads.u-strasbg.fr/abs/2010A%26A...516A..55C} {516, A55}

\bibitem[\protect\citeauthoryear{{Cassisi}, {Mucciarelli}, {Pietrinferni},
  {Salaris}  \& {Ferguson}}{{Cassisi} et~al.}{2013}]{Cassisi13}
{Cassisi} S.,  {Mucciarelli} A.,  {Pietrinferni} A.,  {Salaris} M.,
  {Ferguson} J.,  2013, \mn@doi [\aap] {10.1051/0004-6361/201321311}, \href
  {http://cdsads.u-strasbg.fr/abs/2013A%26A...554A..19C} {554, A19}

\bibitem[\protect\citeauthoryear{{Cassisi}, {Salaris}, {Pietrinferni}, {Vink}
  \& {Monelli}}{{Cassisi} et~al.}{2014}]{Cassisi14}
{Cassisi} S.,  {Salaris} M.,  {Pietrinferni} A.,  {Vink} J.~S.,   {Monelli} M.,
   2014, \mn@doi [\aap] {10.1051/0004-6361/201424540}, \href
  {http://cdsads.u-strasbg.fr/abs/2014A%26A...571A..81C} {571, A81}

\bibitem[\protect\citeauthoryear{{Catelan}, {Valcarce}  \&
  {Sweigart}}{{Catelan} et~al.}{2010}]{Catelan10}
{Catelan} M.,  {Valcarce} A.~A.~R.,   {Sweigart} A.~V.,  2010, in {de Grijs}
  R.,  {L{\'e}pine} J.~R.~D.,  eds,  IAU Symposium Vol. 266, Star Clusters:
  Basic Galactic Building Blocks Throughout Time and Space. pp 281--292
  (\mn@eprint {arXiv} {0910.1367}), \mn@doi{10.1017/S1743921309991153}

\bibitem[\protect\citeauthoryear{{Cenarro}, {Cardiel}, {Gorgas}, {Peletier},
  {Vazdekis}  \& {Prada}}{{Cenarro} et~al.}{2001}]{Cenarro01}
{Cenarro} A.~J.,  {Cardiel} N.,  {Gorgas} J.,  {Peletier} R.~F.,  {Vazdekis}
  A.,   {Prada} F.,  2001, \mn@doi [\mnras] {10.1046/j.1365-8711.2001.04688.x},
  \href {http://cdsads.u-strasbg.fr/abs/2001MNRAS.326..959C} {326, 959}

\bibitem[\protect\citeauthoryear{{Chabrier} \& {Baraffe}}{{Chabrier} \&
  {Baraffe}}{1997}]{Chabrier97}
{Chabrier} G.,  {Baraffe} I.,  1997, \aap, \href
  {http://adsabs.harvard.edu/abs/1997A%26A...327.1039C} {327, 1039}

\bibitem[\protect\citeauthoryear{{Chantereau}, {Charbonnel}  \&
  {Decressin}}{{Chantereau} et~al.}{2015}]{Chantereau15}
{Chantereau} W.,  {Charbonnel} C.,   {Decressin} T.,  2015, \mn@doi [\aap]
  {10.1051/0004-6361/201525929}, \href
  {http://cdsads.u-strasbg.fr/abs/2015A%26A...578A.117C} {578, A117}

\bibitem[\protect\citeauthoryear{{Chantereau}, {Charbonnel}  \&
  {Meynet}}{{Chantereau} et~al.}{2016}]{Chantereau16}
{Chantereau} W.,  {Charbonnel} C.,   {Meynet} G.,  2016, \mn@doi [\aap]
  {10.1051/0004-6361/201628418}, \href
  {http://cdsads.u-strasbg.fr/abs/2016A%26A...592A.111C} {592, A111}

\bibitem[\protect\citeauthoryear{{Chantereau}, {Charbonnel}  \&
  {Meynet}}{{Chantereau} et~al.}{2017}]{Chantereau17}
{Chantereau} W.,  {Charbonnel} C.,   {Meynet} G.,  2017, \mn@doi [\aap]
  {10.1051/0004-6361/201730537}, \href
  {http://cdsads.u-strasbg.fr/abs/2017A%26A...602A..13C} {602, A13}

\bibitem[\protect\citeauthoryear{{Charbonnel}}{{Charbonnel}}{2016}]{Charbonnel16_EES}
{Charbonnel} C.,  2016, in {Moraux} E.,  {Lebreton} Y.,   {Charbonnel} C.,
  eds,  EAS Publications Series Vol. 80, EAS Publications Series. pp 177--226
  (\mn@eprint {arXiv} {1611.08855}), \mn@doi{10.1051/eas/1680006}

\bibitem[\protect\citeauthoryear{{Charbonnel} \& {Chantereau}}{{Charbonnel} \&
  {Chantereau}}{2016}]{Charbonnel16}
{Charbonnel} C.,  {Chantereau} W.,  2016, \mn@doi [\aap]
  {10.1051/0004-6361/201527576}, \href
  {http://cdsads.u-strasbg.fr/abs/2016A%26A...586A..21C} {586, A21}

\bibitem[\protect\citeauthoryear{{Chung}, {Lee}, {Yoon}  \& {Lee}}{{Chung}
  et~al.}{2013}]{Chung13}
{Chung} C.,  {Lee} S.-Y.,  {Yoon} S.-J.,   {Lee} Y.-W.,  2013, \mn@doi [\apjl]
  {10.1088/2041-8205/769/1/L3}, \href
  {http://cdsads.u-strasbg.fr/abs/2013ApJ...769L...3C} {769, L3}

\bibitem[\protect\citeauthoryear{{Chung}, {Yoon}  \& {Lee}}{{Chung}
  et~al.}{2017}]{Chung17}
{Chung} C.,  {Yoon} S.-J.,   {Lee} Y.-W.,  2017, \mn@doi [\apj]
  {10.3847/1538-4357/aa6f19}, \href
  {http://cdsads.u-strasbg.fr/abs/2017ApJ...842...91C} {842, 91}

\bibitem[\protect\citeauthoryear{{Code} \& {Welch}}{{Code} \&
  {Welch}}{1979}]{Code79}
{Code} A.~D.,  {Welch} G.~A.,  1979, \mn@doi [\apj] {10.1086/156825}, \href
  {http://esoads.eso.org/abs/1979ApJ...228...95C} {228, 95}

\bibitem[\protect\citeauthoryear{{Coelho}, {Percival}  \& {Salaris}}{{Coelho}
  et~al.}{2011}]{Coehlo11}
{Coelho} P.,  {Percival} S.~M.,   {Salaris} M.,  2011, \mn@doi [\apj]
  {10.1088/0004-637X/734/1/72}, \href
  {http://cdsads.u-strasbg.fr/abs/2011ApJ...734...72C} {734, 72}

\bibitem[\protect\citeauthoryear{{Cohen} \& {Mel{\'e}ndez}}{{Cohen} \&
  {Mel{\'e}ndez}}{2005}]{Cohen05}
{Cohen} J.~G.,  {Mel{\'e}ndez} J.,  2005, \mn@doi [\aj] {10.1086/426369}, \href
  {http://cdsads.u-strasbg.fr/abs/2005AJ....129..303C} {129, 303}

\bibitem[\protect\citeauthoryear{{Conroy} \& {van Dokkum}}{{Conroy} \& {van
  Dokkum}}{2012a}]{Conroy12_counting}
{Conroy} C.,  {van Dokkum} P.,  2012a, \mn@doi [\apj]
  {10.1088/0004-637X/747/1/69}, \href
  {http://cdsads.u-strasbg.fr/abs/2012ApJ...747...69C} {747, 69}

\bibitem[\protect\citeauthoryear{{Conroy} \& {van Dokkum}}{{Conroy} \& {van
  Dokkum}}{2012b}]{Conroy12}
{Conroy} C.,  {van Dokkum} P.~G.,  2012b, \mn@doi [\apj]
  {10.1088/0004-637X/760/1/71}, \href
  {http://cdsads.u-strasbg.fr/abs/2012ApJ...760...71C} {760, 71}

\bibitem[\protect\citeauthoryear{{Conroy}, {Gunn}  \& {White}}{{Conroy}
  et~al.}{2009}]{Conroy09}
{Conroy} C.,  {Gunn} J.~E.,   {White} M.,  2009, \mn@doi [\apj]
  {10.1088/0004-637X/699/1/486}, \href
  {http://cdsads.u-strasbg.fr/abs/2009ApJ...699..486C} {699, 486}

\bibitem[\protect\citeauthoryear{{Conroy}, {Graves}  \& {van Dokkum}}{{Conroy}
  et~al.}{2014}]{Conroy14}
{Conroy} C.,  {Graves} G.~J.,   {van Dokkum} P.~G.,  2014, \mn@doi [\apj]
  {10.1088/0004-637X/780/1/33}, \href
  {http://esoads.eso.org/abs/2014ApJ...780...33C} {780, 33}

\bibitem[\protect\citeauthoryear{{Conroy}, {Villaume}, {van Dokkum}  \&
  {Lind}}{{Conroy} et~al.}{2018}]{Conroy18}
{Conroy} C.,  {Villaume} A.,  {van Dokkum} P.~G.,   {Lind} K.,  2018, \mn@doi
  [\apj] {10.3847/1538-4357/aaab49}, \href
  {http://adsabs.harvard.edu/abs/2018ApJ...854..139C} {854, 139}

\bibitem[\protect\citeauthoryear{{D'Antona}, {Caloi}  \& {Ventura}}{{D'Antona}
  et~al.}{2010}]{D'Antona10}
{D'Antona} F.,  {Caloi} V.,   {Ventura} P.,  2010, \mn@doi [\mnras]
  {10.1111/j.1365-2966.2010.16646.x}, \href
  {http://cdsads.u-strasbg.fr/abs/2010MNRAS.405.2295D} {405, 2295}

\bibitem[\protect\citeauthoryear{{Dalessandro}, {Schiavon}, {Rood}, {Ferraro},
  {Sohn}, {Lanzoni}  \& {O'Connell}}{{Dalessandro}
  et~al.}{2012}]{Dalessandro12}
{Dalessandro} E.,  {Schiavon} R.~P.,  {Rood} R.~T.,  {Ferraro} F.~R.,  {Sohn}
  S.~T.,  {Lanzoni} B.,   {O'Connell} R.~W.,  2012, \mn@doi [\aj]
  {10.1088/0004-6256/144/5/126}, \href
  {http://cdsads.u-strasbg.fr/abs/2012AJ....144..126D} {144, 126}

\bibitem[\protect\citeauthoryear{{Denisenkov} \& {Denisenkova}}{{Denisenkov} \&
  {Denisenkova}}{1990}]{Denisenkov90}
{Denisenkov} P.~A.,  {Denisenkova} S.~N.,  1990, Soviet Astronomy Letters,
  \href {http://esoads.eso.org/abs/1990SvAL...16..275D} {16, 275}

\bibitem[\protect\citeauthoryear{{Di Criscienzo}, {Tailo}, {Milone},
  {D'Antona}, {Ventura}, {Dotter}  \& {Brocato}}{{Di Criscienzo}
  et~al.}{2015}]{DiCriscienzo15}
{Di Criscienzo} M.,  {Tailo} M.,  {Milone} A.~P.,  {D'Antona} F.,  {Ventura}
  P.,  {Dotter} A.,   {Brocato} E.,  2015, \mn@doi [\mnras]
  {10.1093/mnras/stu2167}, \href
  {http://cdsads.u-strasbg.fr/abs/2015MNRAS.446.1469D} {446, 1469}

\bibitem[\protect\citeauthoryear{{Dickens}, {Croke}, {Cannon}  \&
  {Bell}}{{Dickens} et~al.}{1991}]{Dickens91}
{Dickens} R.~J.,  {Croke} B.~F.~W.,  {Cannon} R.~D.,   {Bell} R.~A.,  1991,
  \mn@doi [\nat] {10.1038/351212a0}, \href
  {http://cdsads.u-strasbg.fr/abs/1991Natur.351..212D} {351, 212}

\bibitem[\protect\citeauthoryear{{Donas} et~al.,}{{Donas}
  et~al.}{2007}]{Donas07}
{Donas} J.,  et~al., 2007, \mn@doi [\apjs] {10.1086/516643}, \href
  {http://cdsads.u-strasbg.fr/abs/2007ApJS..173..597D} {173, 597}

\bibitem[\protect\citeauthoryear{{Dorman}, {O'Connell}  \& {Rood}}{{Dorman}
  et~al.}{1995}]{Dorman95}
{Dorman} B.,  {O'Connell} R.~W.,   {Rood} R.~T.,  1995, \mn@doi [\apj]
  {10.1086/175428}, \href {http://esoads.eso.org/abs/1995ApJ...442..105D} {442,
  105}

\bibitem[\protect\citeauthoryear{{Dotter}, {Chaboyer}, {Jevremovi{\'c}},
  {Baron}, {Ferguson}, {Sarajedini}  \& {Anderson}}{{Dotter}
  et~al.}{2007}]{Dotter07}
{Dotter} A.,  {Chaboyer} B.,  {Jevremovi{\'c}} D.,  {Baron} E.,  {Ferguson}
  J.~W.,  {Sarajedini} A.,   {Anderson} J.,  2007, \mn@doi [\aj]
  {10.1086/517915}, \href {http://cdsads.u-strasbg.fr/abs/2007AJ....134..376D}
  {134, 376}

\bibitem[\protect\citeauthoryear{{Forbes}, {S{\'a}nchez-Bl{\'a}zquez}, {Phan},
  {Brodie}, {Strader}  \& {Spitler}}{{Forbes} et~al.}{2006}]{Forbes06}
{Forbes} D.~A.,  {S{\'a}nchez-Bl{\'a}zquez} P.,  {Phan} A.~T.~T.,  {Brodie}
  J.~P.,  {Strader} J.,   {Spitler} L.,  2006, \mn@doi [\mnras]
  {10.1111/j.1365-2966.2006.09763.x}, \href
  {http://adsabs.harvard.edu/abs/2006MNRAS.366.1230F} {366, 1230}

\bibitem[\protect\citeauthoryear{{Gallazzi}, {Charlot}, {Brinchmann}, {White}
  \& {Tremonti}}{{Gallazzi} et~al.}{2005}]{Gallazzi05}
{Gallazzi} A.,  {Charlot} S.,  {Brinchmann} J.,  {White} S.~D.~M.,   {Tremonti}
  C.~A.,  2005, \mn@doi [\mnras] {10.1111/j.1365-2966.2005.09321.x}, \href
  {http://cdsads.u-strasbg.fr/abs/2005MNRAS.362...41G} {362, 41}

\bibitem[\protect\citeauthoryear{{Gallazzi}, {Charlot}, {Brinchmann}  \&
  {White}}{{Gallazzi} et~al.}{2006}]{Gallazzi06}
{Gallazzi} A.,  {Charlot} S.,  {Brinchmann} J.,   {White} S.~D.~M.,  2006,
  \mn@doi [\mnras] {10.1111/j.1365-2966.2006.10548.x}, \href
  {http://adsabs.harvard.edu/abs/2006MNRAS.370.1106G} {370, 1106}

\bibitem[\protect\citeauthoryear{{Gieles}, {Larsen}, {Bastian}  \&
  {Stein}}{{Gieles} et~al.}{2006}]{Gieles06}
{Gieles} M.,  {Larsen} S.~S.,  {Bastian} N.,   {Stein} I.~T.,  2006, \mn@doi
  [\aap] {10.1051/0004-6361:20053589}, \href
  {http://adsabs.harvard.edu/abs/2006A%26A...450..129G} {450, 129}

\bibitem[\protect\citeauthoryear{{Goudfrooij}}{{Goudfrooij}}{2018}]{Goudfrooij18}
{Goudfrooij} P.,  2018, \mn@doi [\apj] {10.3847/1538-4357/aab553}, \href
  {http://adsabs.harvard.edu/abs/2018ApJ...857...16G} {857, 16}

\bibitem[\protect\citeauthoryear{{Goudfrooij}, {Gilmore}, {Whitmore}  \&
  {Schweizer}}{{Goudfrooij} et~al.}{2004}]{Goudfrooij04}
{Goudfrooij} P.,  {Gilmore} D.,  {Whitmore} B.~C.,   {Schweizer} F.,  2004,
  \mn@doi [\apjl] {10.1086/425071}, \href
  {http://adsabs.harvard.edu/abs/2004ApJ...613L.121G} {613, L121}

\bibitem[\protect\citeauthoryear{{Gratton}, {Carretta}  \&
  {Bragaglia}}{{Gratton} et~al.}{2012}]{Gratton12}
{Gratton} R.~G.,  {Carretta} E.,   {Bragaglia} A.,  2012, \mn@doi [\aapr]
  {10.1007/s00159-012-0050-3}, \href
  {http://cdsads.u-strasbg.fr/abs/2012A%26ARv..20...50G} {20, 50}

\bibitem[\protect\citeauthoryear{{Graves} \& {Schiavon}}{{Graves} \&
  {Schiavon}}{2008}]{Graves08}
{Graves} G.~J.,  {Schiavon} R.~P.,  2008, \mn@doi [\apjs] {10.1086/588097},
  \href {http://adsabs.harvard.edu/abs/2008ApJS..177..446G} {177, 446}

\bibitem[\protect\citeauthoryear{{Greggio} \& {Renzini}}{{Greggio} \&
  {Renzini}}{1990}]{Greggio90}
{Greggio} L.,  {Renzini} A.,  1990, \mn@doi [\apj] {10.1086/169384}, \href
  {http://cdsads.u-strasbg.fr/abs/1990ApJ...364...35G} {364, 35}

\bibitem[\protect\citeauthoryear{{Harris}}{{Harris}}{1996}]{Harris96}
{Harris} W.~E.,  1996, \mn@doi [\aj] {10.1086/118116}, \href
  {http://adsabs.harvard.edu/abs/1996AJ....112.1487H} {112, 1487}

\bibitem[\protect\citeauthoryear{{Harris} \& {Harris}}{{Harris} \&
  {Harris}}{2002}]{Harris02}
{Harris} W.~E.,  {Harris} G.~L.~H.,  2002, \mn@doi [\aj] {10.1086/340466},
  \href {http://adsabs.harvard.edu/abs/2002AJ....123.3108H} {123, 3108}

\bibitem[\protect\citeauthoryear{{Harris}, {Harris}  \& {Alessi}}{{Harris}
  et~al.}{2013}]{Harris13}
{Harris} W.~E.,  {Harris} G.~L.~H.,   {Alessi} M.,  2013, \mn@doi [\apj]
  {10.1088/0004-637X/772/2/82}, \href
  {http://adsabs.harvard.edu/abs/2013ApJ...772...82H} {772, 82}

\bibitem[\protect\citeauthoryear{{Heavens}, {Panter}, {Jimenez}  \&
  {Dunlop}}{{Heavens} et~al.}{2004}]{Heavens04}
{Heavens} A.,  {Panter} B.,  {Jimenez} R.,   {Dunlop} J.,  2004, \mn@doi [\nat]
  {10.1038/nature02474}, \href
  {http://adsabs.harvard.edu/abs/2004Natur.428..625H} {428, 625}

\bibitem[\protect\citeauthoryear{{Hern{\'a}ndez-P{\'e}rez} \&
  {Bruzual}}{{Hern{\'a}ndez-P{\'e}rez} \& {Bruzual}}{2014}]{HernandezPerez14}
{Hern{\'a}ndez-P{\'e}rez} F.,  {Bruzual} G.,  2014, \mn@doi [\mnras]
  {10.1093/mnras/stu1627}, \href
  {http://cdsads.u-strasbg.fr/abs/2014MNRAS.444.2571H} {444, 2571}

\bibitem[\protect\citeauthoryear{Hunter}{Hunter}{2007}]{Matplotlib}
Hunter J.~D.,  2007, \mn@doi [Computing In Science \& Engineering]
  {10.1109/MCSE.2007.55}, 9, 90

\bibitem[\protect\citeauthoryear{{Ivans}, {Sneden}, {Kraft}, {Suntzeff},
  {Smith}, {Langer}  \& {Fulbright}}{{Ivans} et~al.}{1999}]{Ivans99}
{Ivans} I.~I.,  {Sneden} C.,  {Kraft} R.~P.,  {Suntzeff} N.~B.,  {Smith} V.~V.,
   {Langer} G.~E.,   {Fulbright} J.~P.,  1999, \mn@doi [\aj] {10.1086/301017},
  \href {http://cdsads.u-strasbg.fr/abs/1999AJ....118.1273I} {118, 1273}

\bibitem[\protect\citeauthoryear{{Johansson}, {Thomas}  \&
  {Maraston}}{{Johansson} et~al.}{2012}]{Johansson12}
{Johansson} J.,  {Thomas} D.,   {Maraston} C.,  2012, \mn@doi [\mnras]
  {10.1111/j.1365-2966.2011.20316.x}, \href
  {http://adsabs.harvard.edu/abs/2012MNRAS.421.1908J} {421, 1908}

\bibitem[\protect\citeauthoryear{{Johnson}, {Rich}, {Kobayashi}, {Kunder}  \&
  {Koch}}{{Johnson} et~al.}{2014}]{Johnson14}
{Johnson} C.~I.,  {Rich} R.~M.,  {Kobayashi} C.,  {Kunder} A.,   {Koch} A.,
  2014, \mn@doi [\aj] {10.1088/0004-6256/148/4/67}, \href
  {http://adsabs.harvard.edu/abs/2014AJ....148...67J} {148, 67}

\bibitem[\protect\citeauthoryear{{Johnson} et~al.,}{{Johnson}
  et~al.}{2017}]{Johnson17}
{Johnson} L.~C.,  et~al., 2017, \mn@doi [\apj] {10.3847/1538-4357/aa6a1f},
  \href {http://adsabs.harvard.edu/abs/2017ApJ...839...78J} {839, 78}

\bibitem[\protect\citeauthoryear{Jones, Oliphant, Peterson  et~al.}{Jones
  et~al.}{2001}]{scipy}
Jones E.,  Oliphant T.,  Peterson P.,   et~al., 2001, {SciPy}: Open source
  scientific tools for {Python}, \url {http://www.scipy.org/}

\bibitem[\protect\citeauthoryear{{Kamath}, {Karakas}  \& {Wood}}{{Kamath}
  et~al.}{2012}]{Kamath12}
{Kamath} D.,  {Karakas} A.~I.,   {Wood} P.~R.,  2012, \mn@doi [\apj]
  {10.1088/0004-637X/746/1/20}, \href
  {http://adsabs.harvard.edu/abs/2012ApJ...746...20K} {746, 20}

\bibitem[\protect\citeauthoryear{{Karakas}}{{Karakas}}{2014}]{Karakas14}
{Karakas} A.~I.,  2014, \mn@doi [\mnras] {10.1093/mnras/stu1727}, \href
  {http://cdsads.u-strasbg.fr/abs/2014MNRAS.445..347K} {445, 347}

\bibitem[\protect\citeauthoryear{{King} et~al.,}{{King} et~al.}{2012}]{King12}
{King} I.~R.,  et~al., 2012, \mn@doi [\aj] {10.1088/0004-6256/144/1/5}, \href
  {http://cdsads.u-strasbg.fr/abs/2012AJ....144....5K} {144, 5}

\bibitem[\protect\citeauthoryear{{Kotulla}, {Fritze}, {Weilbacher}  \&
  {Anders}}{{Kotulla} et~al.}{2009}]{Kotulla09}
{Kotulla} R.,  {Fritze} U.,  {Weilbacher} P.,   {Anders} P.,  2009, \mn@doi
  [\mnras] {10.1111/j.1365-2966.2009.14717.x}, \href
  {http://cdsads.u-strasbg.fr/abs/2009MNRAS.396..462K} {396, 462}

\bibitem[\protect\citeauthoryear{{Kroupa}}{{Kroupa}}{2001}]{Kroupa01}
{Kroupa} P.,  2001, \mn@doi [\mnras] {10.1046/j.1365-8711.2001.04022.x}, \href
  {http://cdsads.u-strasbg.fr/abs/2001MNRAS.322..231K} {322, 231}

\bibitem[\protect\citeauthoryear{{Kruijssen}}{{Kruijssen}}{2014}]{Kruijssen14}
{Kruijssen} J.~M.~D.,  2014, \mn@doi [Classical and Quantum Gravity]
  {10.1088/0264-9381/31/24/244006}, \href
  {http://adsabs.harvard.edu/abs/2014CQGra..31x4006K} {31, 244006}

\bibitem[\protect\citeauthoryear{{Kruijssen}}{{Kruijssen}}{2015}]{Kruijssen15}
{Kruijssen} J.~M.~D.,  2015, \mn@doi [\mnras] {10.1093/mnras/stv2026}, \href
  {http://adsabs.harvard.edu/abs/2015MNRAS.454.1658K} {454, 1658}

\bibitem[\protect\citeauthoryear{{Kurucz}}{{Kurucz}}{1970}]{Kurucz70}
{Kurucz} R.~L.,  1970, SAO Special Report, \href
  {http://cdsads.u-strasbg.fr/abs/1970SAOSR.309.....K} {309}

\bibitem[\protect\citeauthoryear{{Kurucz}}{{Kurucz}}{2005}]{Kurucz05}
{Kurucz} R.~L.,  2005, Memorie della Societa Astronomica Italiana Supplementi,
  \href {http://cdsads.u-strasbg.fr/abs/2005MSAIS...8...14K} {8, 14}

\bibitem[\protect\citeauthoryear{{Kurucz} \& {Avrett}}{{Kurucz} \&
  {Avrett}}{1981}]{Kurucz81}
{Kurucz} R.~L.,  {Avrett} E.~H.,  1981, SAO Special Report, \href
  {http://cdsads.u-strasbg.fr/abs/1981SAOSR.391.....K} {391}

\bibitem[\protect\citeauthoryear{{Kurucz} \& {Furenlid}}{{Kurucz} \&
  {Furenlid}}{1979}]{Kurucz79}
{Kurucz} R.~L.,  {Furenlid} I.,  1979, SAO Special Report, \href
  {http://cdsads.u-strasbg.fr/abs/1979SAOSR.387.....K} {387}

\bibitem[\protect\citeauthoryear{{La Barbera}, {Ferreras}, {Vazdekis}, {de la
  Rosa}, {de Carvalho}, {Trevisan}, {Falc{\'o}n-Barroso}  \&
  {Ricciardelli}}{{La Barbera} et~al.}{2013}]{LaBarbera13}
{La Barbera} F.,  {Ferreras} I.,  {Vazdekis} A.,  {de la Rosa} I.~G.,  {de
  Carvalho} R.~R.,  {Trevisan} M.,  {Falc{\'o}n-Barroso} J.,   {Ricciardelli}
  E.,  2013, \mn@doi [\mnras] {10.1093/mnras/stt943}, \href
  {http://cdsads.u-strasbg.fr/abs/2013MNRAS.433.3017L} {433, 3017}

\bibitem[\protect\citeauthoryear{{La Barbera}, {Vazdekis}, {Ferreras},
  {Pasquali}, {Allende Prieto}, {R{\"o}ck}, {Aguado}  \& {Peletier}}{{La
  Barbera} et~al.}{2017}]{LaBarbera17}
{La Barbera} F.,  {Vazdekis} A.,  {Ferreras} I.,  {Pasquali} A.,  {Allende
  Prieto} C.,  {R{\"o}ck} B.,  {Aguado} D.~S.,   {Peletier} R.~F.,  2017,
  \mn@doi [\mnras] {10.1093/mnras/stw2407}, \href
  {http://cdsads.u-strasbg.fr/abs/2017MNRAS.464.3597L} {464, 3597}

\bibitem[\protect\citeauthoryear{{Lagioia} et~al.,}{{Lagioia}
  et~al.}{2014}]{Lagioia14}
{Lagioia} E.~P.,  et~al., 2014, \mn@doi [\apj] {10.1088/0004-637X/782/1/50},
  \href {http://adsabs.harvard.edu/abs/2014ApJ...782...50L} {782, 50}

\bibitem[\protect\citeauthoryear{{Larsen}}{{Larsen}}{2009}]{Larsen09}
{Larsen} S.~S.,  2009, \mn@doi [\aap] {10.1051/0004-6361:200811212}, \href
  {http://adsabs.harvard.edu/abs/2009A%26A...494..539L} {494, 539}

\bibitem[\protect\citeauthoryear{{Maraston}}{{Maraston}}{1998}]{Maraston98}
{Maraston} C.,  1998, \mn@doi [\mnras] {10.1046/j.1365-8711.1998.01947.x},
  \href {http://cdsads.u-strasbg.fr/abs/1998MNRAS.300..872M} {300, 872}

\bibitem[\protect\citeauthoryear{{Maraston}, {Greggio}, {Renzini}, {Ortolani},
  {Saglia}, {Puzia}  \& {Kissler-Patig}}{{Maraston} et~al.}{2003}]{Maraston03}
{Maraston} C.,  {Greggio} L.,  {Renzini} A.,  {Ortolani} S.,  {Saglia} R.~P.,
  {Puzia} T.~H.,   {Kissler-Patig} M.,  2003, \mn@doi [\aap]
  {10.1051/0004-6361:20021723}, \href
  {http://adsabs.harvard.edu/abs/2003A%26A...400..823M} {400, 823}

\bibitem[\protect\citeauthoryear{{Maraston}, {Bastian}, {Saglia},
  {Kissler-Patig}, {Schweizer}  \& {Goudfrooij}}{{Maraston}
  et~al.}{2004}]{Maraston04}
{Maraston} C.,  {Bastian} N.,  {Saglia} R.~P.,  {Kissler-Patig} M.,
  {Schweizer} F.,   {Goudfrooij} P.,  2004, \mn@doi [\aap]
  {10.1051/0004-6361:20031604}, \href
  {http://adsabs.harvard.edu/abs/2004A%26A...416..467M} {416, 467}

\bibitem[\protect\citeauthoryear{{Marigo}, {Girardi}, {Bressan}, {Groenewegen},
  {Silva}  \& {Granato}}{{Marigo} et~al.}{2008}]{Marigo08}
{Marigo} P.,  {Girardi} L.,  {Bressan} A.,  {Groenewegen} M.~A.~T.,  {Silva}
  L.,   {Granato} G.~L.,  2008, \mn@doi [\aap] {10.1051/0004-6361:20078467},
  \href {http://adsabs.harvard.edu/abs/2008A%26A...482..883M} {482, 883}

\bibitem[\protect\citeauthoryear{{Martell} \& {Grebel}}{{Martell} \&
  {Grebel}}{2010}]{Martell10}
{Martell} S.~L.,  {Grebel} E.~K.,  2010, \mn@doi [\aap]
  {10.1051/0004-6361/201014135}, \href
  {http://cdsads.u-strasbg.fr/abs/2010A%26A...519A..14M} {519, A14}

\bibitem[\protect\citeauthoryear{{Martell} et~al.,}{{Martell}
  et~al.}{2016}]{Martell16}
{Martell} S.~L.,  et~al., 2016, \mn@doi [\apj] {10.3847/0004-637X/825/2/146},
  \href {http://cdsads.u-strasbg.fr/abs/2016ApJ...825..146M} {825, 146}

\bibitem[\protect\citeauthoryear{{Martocchia} et~al.,}{{Martocchia}
  et~al.}{2017}]{Martocchia17}
{Martocchia} S.,  et~al., 2017, \mn@doi [\mnras] {10.1093/mnras/stx660}, \href
  {http://adsabs.harvard.edu/abs/2017MNRAS.468.3150M} {468, 3150}

\bibitem[\protect\citeauthoryear{{Martocchia} et~al.,}{{Martocchia}
  et~al.}{2018}]{Martocchia18}
{Martocchia} S.,  et~al., 2018, \mn@doi [\mnras] {10.1093/mnras/stx2556}, \href
  {http://adsabs.harvard.edu/abs/2018MNRAS.473.2688M} {473, 2688}

\bibitem[\protect\citeauthoryear{{Matteucci}}{{Matteucci}}{1994}]{Matteucci94}
{Matteucci} F.,  1994, \aap, \href
  {http://adsabs.harvard.edu/abs/1994A%26A...288...57M} {288, 57}

\bibitem[\protect\citeauthoryear{{McDermid} et~al.,}{{McDermid}
  et~al.}{2015}]{McDermid15}
{McDermid} R.~M.,  et~al., 2015, \mn@doi [\mnras] {10.1093/mnras/stv105}, \href
  {http://adsabs.harvard.edu/abs/2015MNRAS.448.3484M} {448, 3484}

\bibitem[\protect\citeauthoryear{{McLaughlin}}{{McLaughlin}}{1999}]{McLaughlin99}
{McLaughlin} D.~E.,  1999, \mn@doi [\aj] {10.1086/300836}, \href
  {http://adsabs.harvard.edu/abs/1999AJ....117.2398M} {117, 2398}

\bibitem[\protect\citeauthoryear{{Milone}}{{Milone}}{2015}]{Milone15_NGC6266}
{Milone} A.~P.,  2015, \mn@doi [\mnras] {10.1093/mnras/stu2198}, \href
  {http://adsabs.harvard.edu/abs/2015MNRAS.446.1672M} {446, 1672}

\bibitem[\protect\citeauthoryear{{Milone} et~al.,}{{Milone}
  et~al.}{2012}]{Milone12}
{Milone} A.~P.,  et~al., 2012, \mn@doi [\aap] {10.1051/0004-6361/201016384},
  \href {http://cdsads.u-strasbg.fr/abs/2012A%26A...540A..16M} {540, A16}

\bibitem[\protect\citeauthoryear{{Milone} et~al.,}{{Milone}
  et~al.}{2015}]{Milone15}
{Milone} A.~P.,  et~al., 2015, \mn@doi [\apj] {10.1088/0004-637X/808/1/51},
  \href {http://cdsads.u-strasbg.fr/abs/2015ApJ...808...51M} {808, 51}

\bibitem[\protect\citeauthoryear{{Milone} et~al.,}{{Milone}
  et~al.}{2017}]{Milone17}
{Milone} A.~P.,  et~al., 2017, \mn@doi [\mnras] {10.1093/mnras/stw2531}, \href
  {http://adsabs.harvard.edu/abs/2017MNRAS.464.3636M} {464, 3636}

\bibitem[\protect\citeauthoryear{{Norris} \& {Da Costa}}{{Norris} \& {Da
  Costa}}{1995}]{Norris95}
{Norris} J.~E.,  {Da Costa} G.~S.,  1995, \mn@doi [\apj] {10.1086/175909},
  \href {http://cdsads.u-strasbg.fr/abs/1995ApJ...447..680N} {447, 680}

\bibitem[\protect\citeauthoryear{{O'Connell}}{{O'Connell}}{1999}]{OConnell99}
{O'Connell} R.~W.,  1999, \mn@doi [\araa] {10.1146/annurev.astro.37.1.603},
  \href {http://cdsads.u-strasbg.fr/abs/1999ARA%26A..37..603O} {37, 603}

\bibitem[\protect\citeauthoryear{{Pace}, {Recio-Blanco}, {Piotto}  \&
  {Momany}}{{Pace} et~al.}{2006}]{Pace06}
{Pace} G.,  {Recio-Blanco} A.,  {Piotto} G.,   {Momany} Y.,  2006, \mn@doi
  [\aap] {10.1051/0004-6361:20054593}, \href
  {http://adsabs.harvard.edu/abs/2006A%26A...452..493P} {452, 493}

\bibitem[\protect\citeauthoryear{{Panter}, {Jimenez}, {Heavens}  \&
  {Charlot}}{{Panter} et~al.}{2007}]{Panter07}
{Panter} B.,  {Jimenez} R.,  {Heavens} A.~F.,   {Charlot} S.,  2007, \mn@doi
  [\mnras] {10.1111/j.1365-2966.2007.11909.x}, \href
  {http://adsabs.harvard.edu/abs/2007MNRAS.378.1550P} {378, 1550}

\bibitem[\protect\citeauthoryear{{Peng} \& {Nagai}}{{Peng} \&
  {Nagai}}{2009}]{Peng09}
{Peng} F.,  {Nagai} D.,  2009, \mn@doi [\apjl] {10.1088/0004-637X/705/1/L58},
  \href {http://adsabs.harvard.edu/abs/2009ApJ...705L..58P} {705, L58}

\bibitem[\protect\citeauthoryear{{Peng} et~al.,}{{Peng} et~al.}{2006}]{Peng06}
{Peng} E.~W.,  et~al., 2006, \mn@doi [\apj] {10.1086/498210}, \href
  {http://adsabs.harvard.edu/abs/2006ApJ...639...95P} {639, 95}

\bibitem[\protect\citeauthoryear{{Peng} et~al.,}{{Peng} et~al.}{2008}]{Peng08}
{Peng} E.~W.,  et~al., 2008, \mn@doi [\apj] {10.1086/587951}, \href
  {http://adsabs.harvard.edu/abs/2008ApJ...681..197P} {681, 197}

\bibitem[\protect\citeauthoryear{{Pfeffer}, {Kruijssen}, {Crain}  \&
  {Bastian}}{{Pfeffer} et~al.}{2017}]{Pfeffer17}
{Pfeffer} J.,  {Kruijssen} J.~M.~D.,  {Crain} R.~A.,   {Bastian} N.,  2017,
  preprint, \href {http://adsabs.harvard.edu/abs/2017arXiv171200019P} {}
  (\mn@eprint {arXiv} {1712.00019})

\bibitem[\protect\citeauthoryear{{Pietrinferni}, {Cassisi}, {Salaris},
  {Percival}  \& {Ferguson}}{{Pietrinferni} et~al.}{2009}]{Pietrinferni09}
{Pietrinferni} A.,  {Cassisi} S.,  {Salaris} M.,  {Percival} S.,   {Ferguson}
  J.~W.,  2009, \mn@doi [\apj] {10.1088/0004-637X/697/1/275}, \href
  {http://cdsads.u-strasbg.fr/abs/2009ApJ...697..275P} {697, 275}

\bibitem[\protect\citeauthoryear{{Piotto}}{{Piotto}}{2009}]{Piotto09}
{Piotto} G.,  2009, in {Mamajek} E.~E.,  {Soderblom} D.~R.,   {Wyse} R.~F.~G.,
  eds,  IAU Symposium Vol. 258, The Ages of Stars. pp 233--244,
  \mn@doi{10.1017/S1743921309031883}

\bibitem[\protect\citeauthoryear{{Piotto} et~al.,}{{Piotto}
  et~al.}{2004}]{Piotto04}
{Piotto} G.,  et~al., 2004, \mn@doi [\apjl] {10.1086/383617}, \href
  {http://cdsads.u-strasbg.fr/abs/2004ApJ...604L.109P} {604, L109}

\bibitem[\protect\citeauthoryear{{Ram{\'{\i}}rez}, {Mel{\'e}ndez}  \&
  {Chanam{\'e}}}{{Ram{\'{\i}}rez} et~al.}{2012}]{Ramirez12}
{Ram{\'{\i}}rez} I.,  {Mel{\'e}ndez} J.,   {Chanam{\'e}} J.,  2012, \mn@doi
  [\apj] {10.1088/0004-637X/757/2/164}, \href
  {http://cdsads.u-strasbg.fr/abs/2012ApJ...757..164R} {757, 164}

\bibitem[\protect\citeauthoryear{{Renzini} \& {Ciotti}}{{Renzini} \&
  {Ciotti}}{1993}]{Renzini93}
{Renzini} A.,  {Ciotti} L.,  1993, \mn@doi [\apjl] {10.1086/187068}, \href
  {http://cdsads.u-strasbg.fr/abs/1993ApJ...416L..49R} {416, L49}

\bibitem[\protect\citeauthoryear{{Rey} et~al.,}{{Rey} et~al.}{2009}]{Rey09}
{Rey} S.-C.,  et~al., 2009, \mn@doi [\apjl] {10.1088/0004-637X/700/1/L11},
  \href {http://adsabs.harvard.edu/abs/2009ApJ...700L..11R} {700, L11}

\bibitem[\protect\citeauthoryear{{Salaris}, {Weiss}, {Ferguson}  \&
  {Fusilier}}{{Salaris} et~al.}{2006}]{Salaris06}
{Salaris} M.,  {Weiss} A.,  {Ferguson} J.~W.,   {Fusilier} D.~J.,  2006,
  \mn@doi [\apj] {10.1086/504520}, \href
  {http://cdsads.u-strasbg.fr/abs/2006ApJ...645.1131S} {645, 1131}

\bibitem[\protect\citeauthoryear{{Salpeter}}{{Salpeter}}{1955}]{Salpeter55}
{Salpeter} E.~E.,  1955, \mn@doi [\apj] {10.1086/145971}, \href
  {http://adsabs.harvard.edu/abs/1955ApJ...121..161S} {121, 161}

\bibitem[\protect\citeauthoryear{{Sarzi}, {Spiniello}, {La Barbera},
  {Krajnovi{\'c}}  \& {van den Bosch}}{{Sarzi} et~al.}{2017}]{Sarzi17}
{Sarzi} M.,  {Spiniello} C.,  {La Barbera} F.,  {Krajnovi{\'c}} D.,   {van den
  Bosch} R.,  2017, preprint, \href
  {http://adsabs.harvard.edu/abs/2017arXiv171108980S} {} (\mn@eprint {arXiv}
  {1711.08980})

\bibitem[\protect\citeauthoryear{{Sbordone}, {Salaris}, {Weiss}  \&
  {Cassisi}}{{Sbordone} et~al.}{2011}]{Sbordone11}
{Sbordone} L.,  {Salaris} M.,  {Weiss} A.,   {Cassisi} S.,  2011, \mn@doi
  [\aap] {10.1051/0004-6361/201116714}, \href
  {http://cdsads.u-strasbg.fr/abs/2011A%26A...534A...9S} {534, A9}

\bibitem[\protect\citeauthoryear{{Schiavon}}{{Schiavon}}{2007}]{Schiavon07}
{Schiavon} R.~P.,  2007, \mn@doi [\apjs] {10.1086/511753}, \href
  {http://esoads.eso.org/abs/2007ApJS..171..146S} {171, 146}

\bibitem[\protect\citeauthoryear{{Schiavon}, {Rose}, {Courteau}  \&
  {MacArthur}}{{Schiavon} et~al.}{2004}]{Schiavon04}
{Schiavon} R.~P.,  {Rose} J.~A.,  {Courteau} S.,   {MacArthur} L.~A.,  2004,
  \mn@doi [\apjl] {10.1086/422251}, \href
  {http://cdsads.u-strasbg.fr/abs/2004ApJ...608L..33S} {608, L33}

\bibitem[\protect\citeauthoryear{{Schiavon}, {Caldwell}, {Conroy}, {Graves},
  {Strader}, {MacArthur}, {Courteau}  \& {Harding}}{{Schiavon}
  et~al.}{2013}]{Schiavon13}
{Schiavon} R.~P.,  {Caldwell} N.,  {Conroy} C.,  {Graves} G.~J.,  {Strader} J.,
   {MacArthur} L.~A.,  {Courteau} S.,   {Harding} P.,  2013, \mn@doi [\apjl]
  {10.1088/2041-8205/776/1/L7}, \href
  {http://adsabs.harvard.edu/abs/2013ApJ...776L...7S} {776, L7}

\bibitem[\protect\citeauthoryear{{Schiavon} et~al.,}{{Schiavon}
  et~al.}{2017a}]{Schiavon17}
{Schiavon} R.~P.,  et~al., 2017a, \mn@doi [\mnras] {10.1093/mnras/stw2162},
  \href {http://esoads.eso.org/abs/2017MNRAS.465..501S} {465, 501}

\bibitem[\protect\citeauthoryear{{Schiavon} et~al.,}{{Schiavon}
  et~al.}{2017b}]{Schiavon17_inner_gcs}
{Schiavon} R.~P.,  et~al., 2017b, \mn@doi [\mnras] {10.1093/mnras/stw3093},
  \href {http://adsabs.harvard.edu/abs/2017MNRAS.466.1010S} {466, 1010}

\bibitem[\protect\citeauthoryear{{Serven}, {Worthey}  \& {Briley}}{{Serven}
  et~al.}{2005}]{Serven05}
{Serven} J.,  {Worthey} G.,   {Briley} M.~M.,  2005, \mn@doi [\apj]
  {10.1086/430400}, \href {http://cdsads.u-strasbg.fr/abs/2005ApJ...627..754S}
  {627, 754}

\bibitem[\protect\citeauthoryear{{Shingles}, {Doherty}, {Karakas},
  {Stancliffe}, {Lattanzio}  \& {Lugaro}}{{Shingles} et~al.}{2015}]{Shingles15}
{Shingles} L.~J.,  {Doherty} C.~L.,  {Karakas} A.~I.,  {Stancliffe} R.~J.,
  {Lattanzio} J.~C.,   {Lugaro} M.,  2015, \mn@doi [\mnras]
  {10.1093/mnras/stv1489}, \href
  {http://cdsads.u-strasbg.fr/abs/2015MNRAS.452.2804S} {452, 2804}

\bibitem[\protect\citeauthoryear{{Short}, {Young}  \& {Layden}}{{Short}
  et~al.}{2015}]{Short15}
{Short} C.~I.,  {Young} M.~E.,   {Layden} N.,  2015, \mn@doi [\apj]
  {10.1088/0004-637X/810/1/76}, \href
  {http://adsabs.harvard.edu/abs/2015ApJ...810...76S} {810, 76}

\bibitem[\protect\citeauthoryear{{Smith}, {Shetrone}, {Bell}, {Churchill}  \&
  {Briley}}{{Smith} et~al.}{1996}]{Smith96}
{Smith} G.~H.,  {Shetrone} M.~D.,  {Bell} R.~A.,  {Churchill} C.~W.,   {Briley}
  M.~M.,  1996, \mn@doi [\aj] {10.1086/118119}, \href
  {http://cdsads.u-strasbg.fr/abs/1996AJ....112.1511S} {112, 1511}

\bibitem[\protect\citeauthoryear{{Sohn}, {O'Connell}, {Kundu}, {Landsman},
  {Burstein}, {Bohlin}, {Frogel}  \& {Rose}}{{Sohn} et~al.}{2006}]{Sohn06}
{Sohn} S.~T.,  {O'Connell} R.~W.,  {Kundu} A.,  {Landsman} W.~B.,  {Burstein}
  D.,  {Bohlin} R.~C.,  {Frogel} J.~A.,   {Rose} J.~A.,  2006, \mn@doi [\aj]
  {10.1086/499039}, \href {http://cdsads.u-strasbg.fr/abs/2006AJ....131..866S}
  {131, 866}

\bibitem[\protect\citeauthoryear{{Spiesman}}{{Spiesman}}{1992}]{Spiesman92}
{Spiesman} W.~J.,  1992, \mn@doi [\apjl] {10.1086/186555}, \href
  {http://cdsads.u-strasbg.fr/abs/1992ApJ...397L.103S} {397, L103}

\bibitem[\protect\citeauthoryear{{Strader} et~al.,}{{Strader}
  et~al.}{2013}]{Strader13}
{Strader} J.,  et~al., 2013, \mn@doi [\apjl] {10.1088/2041-8205/775/1/L6},
  \href {http://esoads.eso.org/abs/2013ApJ...775L...6S} {775, L6}

\bibitem[\protect\citeauthoryear{{Tailo}, {Di Criscienzo}, {D'Antona}, {Caloi}
  \& {Ventura}}{{Tailo} et~al.}{2016}]{Tailo16}
{Tailo} M.,  {Di Criscienzo} M.,  {D'Antona} F.,  {Caloi} V.,   {Ventura} P.,
  2016, \mn@doi [\mnras] {10.1093/mnras/stw319}, \href
  {http://cdsads.u-strasbg.fr/abs/2016MNRAS.457.4525T} {457, 4525}

\bibitem[\protect\citeauthoryear{{Tang}, {Worthey}  \& {Davis}}{{Tang}
  et~al.}{2014}]{Tang14}
{Tang} B.,  {Worthey} G.,   {Davis} A.~B.,  2014, \mn@doi [\mnras]
  {10.1093/mnras/stu1867}, \href
  {http://cdsads.u-strasbg.fr/abs/2014MNRAS.445.1538T} {445, 1538}

\bibitem[\protect\citeauthoryear{{Tang} et~al.,}{{Tang} et~al.}{2017}]{Tang17}
{Tang} B.,  et~al., 2017, \mn@doi [\mnras] {10.1093/mnras/stw2739}, \href
  {http://adsabs.harvard.edu/abs/2017MNRAS.465...19T} {465, 19}

\bibitem[\protect\citeauthoryear{{Thomas}, {Maraston}  \& {Bender}}{{Thomas}
  et~al.}{2003}]{Thomas03}
{Thomas} D.,  {Maraston} C.,   {Bender} R.,  2003, \mn@doi [\mnras]
  {10.1046/j.1365-8711.2003.06248.x}, \href
  {http://cdsads.u-strasbg.fr/abs/2003MNRAS.339..897T} {339, 897}

\bibitem[\protect\citeauthoryear{{Thomas}, {Maraston}, {Bender}  \& {Mendes de
  Oliveira}}{{Thomas} et~al.}{2005}]{Thomas05}
{Thomas} D.,  {Maraston} C.,  {Bender} R.,   {Mendes de Oliveira} C.,  2005,
  \mn@doi [\apj] {10.1086/426932}, \href
  {http://adsabs.harvard.edu/abs/2005ApJ...621..673T} {621, 673}

\bibitem[\protect\citeauthoryear{{Trager}, {Faber}, {Worthey}  \&
  {Gonz{\'a}lez}}{{Trager} et~al.}{2000}]{Trager00}
{Trager} S.~C.,  {Faber} S.~M.,  {Worthey} G.,   {Gonz{\'a}lez} J.~J.,  2000,
  \mn@doi [\aj] {10.1086/301442}, \href
  {http://adsabs.harvard.edu/abs/2000AJ....120..165T} {120, 165}

\bibitem[\protect\citeauthoryear{{Usher} et~al.,}{{Usher}
  et~al.}{2012}]{Usher12}
{Usher} C.,  et~al., 2012, \mn@doi [\mnras] {10.1111/j.1365-2966.2012.21801.x},
  \href {http://adsabs.harvard.edu/abs/2012MNRAS.426.1475U} {426, 1475}

\bibitem[\protect\citeauthoryear{{Usher} et~al.,}{{Usher}
  et~al.}{2017}]{Usher17}
{Usher} C.,  et~al., 2017, \mn@doi [\mnras] {10.1093/mnras/stx713}, \href
  {http://adsabs.harvard.edu/abs/2017MNRAS.468.3828U} {468, 3828}

\bibitem[\protect\citeauthoryear{{Valcarce}, {Catelan}  \&
  {Sweigart}}{{Valcarce} et~al.}{2012}]{Valcarce12}
{Valcarce} A.~A.~R.,  {Catelan} M.,   {Sweigart} A.~V.,  2012, \mn@doi [\aap]
  {10.1051/0004-6361/201219510}, \href
  {http://cdsads.u-strasbg.fr/abs/2012A%26A...547A...5V} {547, A5}

\bibitem[\protect\citeauthoryear{{Wachter}, {Schr{\"o}der}, {Winters}, {Arndt}
  \& {Sedlmayr}}{{Wachter} et~al.}{2002}]{Wachter02}
{Wachter} A.,  {Schr{\"o}der} K.-P.,  {Winters} J.~M.,  {Arndt} T.~U.,
  {Sedlmayr} E.,  2002, \mn@doi [\aap] {10.1051/0004-6361:20020022}, \href
  {http://adsabs.harvard.edu/abs/2002A%26A...384..452W} {384, 452}

\bibitem[\protect\citeauthoryear{{Wedemeyer}, {Kucinskas}, {Klevas}  \&
  {Ludwig}}{{Wedemeyer} et~al.}{2017}]{Wedemeyer17}
{Wedemeyer} S.,  {Kucinskas} A.,  {Klevas} J.,   {Ludwig} H.-G.,  2017,
  preprint, \href {http://adsabs.harvard.edu/abs/2017arXiv170509641W} {}
  (\mn@eprint {arXiv} {1705.09641})

\bibitem[\protect\citeauthoryear{{Whitmore}, {Zhang}, {Leitherer}, {Fall},
  {Schweizer}  \& {Miller}}{{Whitmore} et~al.}{1999}]{Whitmore99}
{Whitmore} B.~C.,  {Zhang} Q.,  {Leitherer} C.,  {Fall} S.~M.,  {Schweizer} F.,
    {Miller} B.~W.,  1999, \mn@doi [\aj] {10.1086/301041}, \href
  {http://adsabs.harvard.edu/abs/1999AJ....118.1551W} {118, 1551}

\bibitem[\protect\citeauthoryear{{Worthey}}{{Worthey}}{1994}]{Worthey94_comp}
{Worthey} G.,  1994, \mn@doi [\apjs] {10.1086/192096}, \href
  {http://cdsads.u-strasbg.fr/abs/1994ApJS...95..107W} {95, 107}

\bibitem[\protect\citeauthoryear{{Worthey} \& {Ottaviani}}{{Worthey} \&
  {Ottaviani}}{1997}]{Worthey97}
{Worthey} G.,  {Ottaviani} D.~L.,  1997, \mn@doi [\apjs] {10.1086/313021},
  \href {http://cdsads.u-strasbg.fr/abs/1997ApJS..111..377W} {111, 377}

\bibitem[\protect\citeauthoryear{{Worthey}, {Faber}  \& {Gonzalez}}{{Worthey}
  et~al.}{1992}]{Worthey92}
{Worthey} G.,  {Faber} S.~M.,   {Gonzalez} J.~J.,  1992, \mn@doi [\apj]
  {10.1086/171836}, \href {http://adsabs.harvard.edu/abs/1992ApJ...398...69W}
  {398, 69}

\bibitem[\protect\citeauthoryear{{Worthey}, {Faber}, {Gonzalez}  \&
  {Burstein}}{{Worthey} et~al.}{1994}]{Worthey94}
{Worthey} G.,  {Faber} S.~M.,  {Gonzalez} J.~J.,   {Burstein} D.,  1994,
  \mn@doi [\apjs] {10.1086/192087}, \href
  {http://cdsads.u-strasbg.fr/abs/1994ApJS...94..687W} {94, 687}

\bibitem[\protect\citeauthoryear{{Worthey}, {Tang}  \& {Serven}}{{Worthey}
  et~al.}{2014}]{Worthey14}
{Worthey} G.,  {Tang} B.,   {Serven} J.,  2014, \mn@doi [\apj]
  {10.1088/0004-637X/783/1/20}, \href
  {http://adsabs.harvard.edu/abs/2014ApJ...783...20W} {783, 20}

\bibitem[\protect\citeauthoryear{{Yi}, {Demarque}  \& {Oemler}}{{Yi}
  et~al.}{1997}]{Yi97}
{Yi} S.,  {Demarque} P.,   {Oemler} Jr. A.,  1997, \mn@doi [\apj]
  {10.1086/304498}, \href {http://adsabs.harvard.edu/abs/1997ApJ...486..201Y}
  {486, 201}

\bibitem[\protect\citeauthoryear{{Yi}, {Lee}, {Sheen}, {Jeong}, {Suh}  \&
  {Oh}}{{Yi} et~al.}{2011}]{Yi11}
{Yi} S.~K.,  {Lee} J.,  {Sheen} Y.-K.,  {Jeong} H.,  {Suh} H.,   {Oh} K.,
  2011, \mn@doi [\apjs] {10.1088/0067-0049/195/2/22}, \href
  {http://cdsads.u-strasbg.fr/abs/2011ApJS..195...22Y} {195, 22}

\bibitem[\protect\citeauthoryear{{Yong}, {Grundahl}  \& {Norris}}{{Yong}
  et~al.}{2015}]{Yong15}
{Yong} D.,  {Grundahl} F.,   {Norris} J.~E.,  2015, \mn@doi [\mnras]
  {10.1093/mnras/stu2334}, \href
  {http://cdsads.u-strasbg.fr/abs/2015MNRAS.446.3319Y} {446, 3319}

\bibitem[\protect\citeauthoryear{{Zaritsky}, {Gil de Paz}  \&
  {Bouquin}}{{Zaritsky} et~al.}{2015}]{Zaritsky15}
{Zaritsky} D.,  {Gil de Paz} A.,   {Bouquin} A.~Y.~K.,  2015, \mn@doi [\mnras]
  {10.1093/mnras/stu2245}, \href
  {http://cdsads.u-strasbg.fr/abs/2015MNRAS.446.2030Z} {446, 2030}

\bibitem[\protect\citeauthoryear{{Zoccali}, {Renzini}, {Ortolani}, {Bica}  \&
  {Barbuy}}{{Zoccali} et~al.}{2001}]{Zoccali01}
{Zoccali} M.,  {Renzini} A.,  {Ortolani} S.,  {Bica} E.,   {Barbuy} B.,  2001,
  \mn@doi [\aj] {10.1086/320411}, \href
  {http://adsabs.harvard.edu/abs/2001AJ....121.2638Z} {121, 2638}

\bibitem[\protect\citeauthoryear{{Zoccali} et~al.,}{{Zoccali}
  et~al.}{2004}]{Zoccali04}
{Zoccali} M.,  et~al., 2004, \mn@doi [\aap] {10.1051/0004-6361:20041014}, \href
  {http://adsabs.harvard.edu/abs/2004A%26A...423..507Z} {423, 507}

\bibitem[\protect\citeauthoryear{{van Dokkum} \& {Conroy}}{{van Dokkum} \&
  {Conroy}}{2010}]{vanDokkum10}
{van Dokkum} P.~G.,  {Conroy} C.,  2010, \mn@doi [\nat] {10.1038/nature09578},
  \href {http://adsabs.harvard.edu/abs/2010Natur.468..940V} {468, 940}

\bibitem[\protect\citeauthoryear{{van Dokkum} \& {Conroy}}{{van Dokkum} \&
  {Conroy}}{2014}]{vanDokkum14}
{van Dokkum} P.~G.,  {Conroy} C.,  2014, \mn@doi [\apj]
  {10.1088/0004-637X/797/1/56}, \href
  {http://adsabs.harvard.edu/abs/2014ApJ...797...56V} {797, 56}

\bibitem[\protect\citeauthoryear{{van Dokkum}, {Conroy}, {Villaume}, {Brodie}
  \& {Romanowsky}}{{van Dokkum} et~al.}{2017}]{vanDokkum17}
{van Dokkum} P.,  {Conroy} C.,  {Villaume} A.,  {Brodie} J.,   {Romanowsky}
  A.~J.,  2017, \mn@doi [\apj] {10.3847/1538-4357/aa7135}, \href
  {http://adsabs.harvard.edu/abs/2017ApJ...841...68V} {841, 68}

\bibitem[\protect\citeauthoryear{van~der Walt, Colbert  \& Varoquaux}{van~der
  Walt et~al.}{2011}]{numpy}
van~der Walt S.,  Colbert S.~C.,   Varoquaux G.,  2011, \mn@doi [Computing in
  Science Engineering] {10.1109/MCSE.2011.37}, 13, 22

\makeatother
\end{thebibliography}

\appendix
\section{A comparison with the Conroy \& van Dokkum (2012) models}
\label{appendix_conroy}
To assess the reliability of our stellar population models, we compared the colours, spectral indices and \textit{M}/\textit{L} predicted by our models with those of \citet{Conroy12_counting}.
We downloaded version 1.2 of the \citeauthor{Conroy12_counting} spectral energy distributions and after smoothing them to match the velocity dispersion of our models, we performed the same magnitude and spectral index measurements as we did for our models (Section \ref{mags_indices}).
To calculate mass-to-light ratios for the Conroy models we used their initial-to-final mass relations.
We are able to make a comparison with equivalent models for our base solar composition, Milky Way-like IMF model, our bottom heavy IMF ($\alpha = -3$) model, our Salpeter IMF model and our young model.
We can not compare our lower metallicity model with the Conroy models as only models where a single element is varied are publicly available.  
Likewise, we cannot directly compare our CNONa models to any of the Conroy models with variations of a single element due to the complex interdependence of the abundance of different atomic and molecular species on each other.
However, a visual comparison of our Figures \ref{Figure:indices_1} and \ref{Figure:indices_2} with the figures in \citet{Conroy12_counting} do generally show qualitative agreement for the effects of changing the dominant element of an index.
We note that we vary the abundances of each of C, N, O and Na by larger amounts than do \citet{Conroy12_counting} and that the effects of [N/Fe] dominates the effects of [C/Fe] on the CN$_{1}$ index.

\begin{figure}
\centering
 \includegraphics[width=240pt]{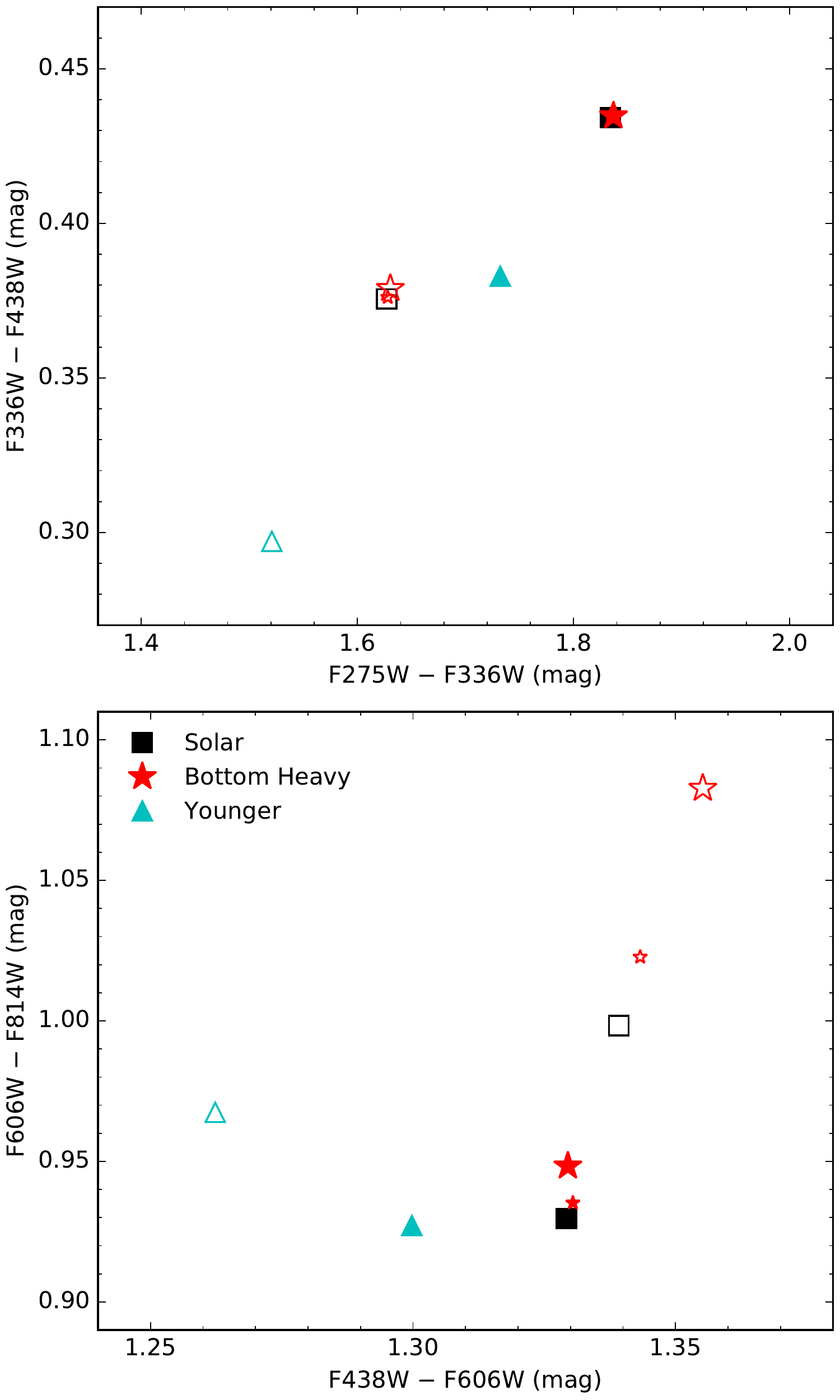}
 \caption{Comparison of the predicted colours for our models (solid points) with the predictions of the models of \citet{Conroy12_counting} (open points).
 As in Figure \ref{Figure:colours}, the black squares are models with  solar abundances and a Milky Way like IMF, the red stars are solar composition models with bottom heavy IMFs and the cyan triangles are the younger (8.9 Gyr) models.
 While the colours plotted are the same as in Figure \ref{Figure:colours}, the axes ranges are different.
 Although the colours of the Conroy models are bluer in the bluer bands (top) and redder in the redder bands (bottom) than our models, the effects of a younger age or a bottom heavy IMF are broadly similar for both models although a bottom heavy IMF has a much larger effect on the (F606W - F814W) colour of the Conroy models than for our models.}
 \label{Figure:colours_conroy}
\end{figure}

\begin{figure}
\centering
 \includegraphics[width=240pt]{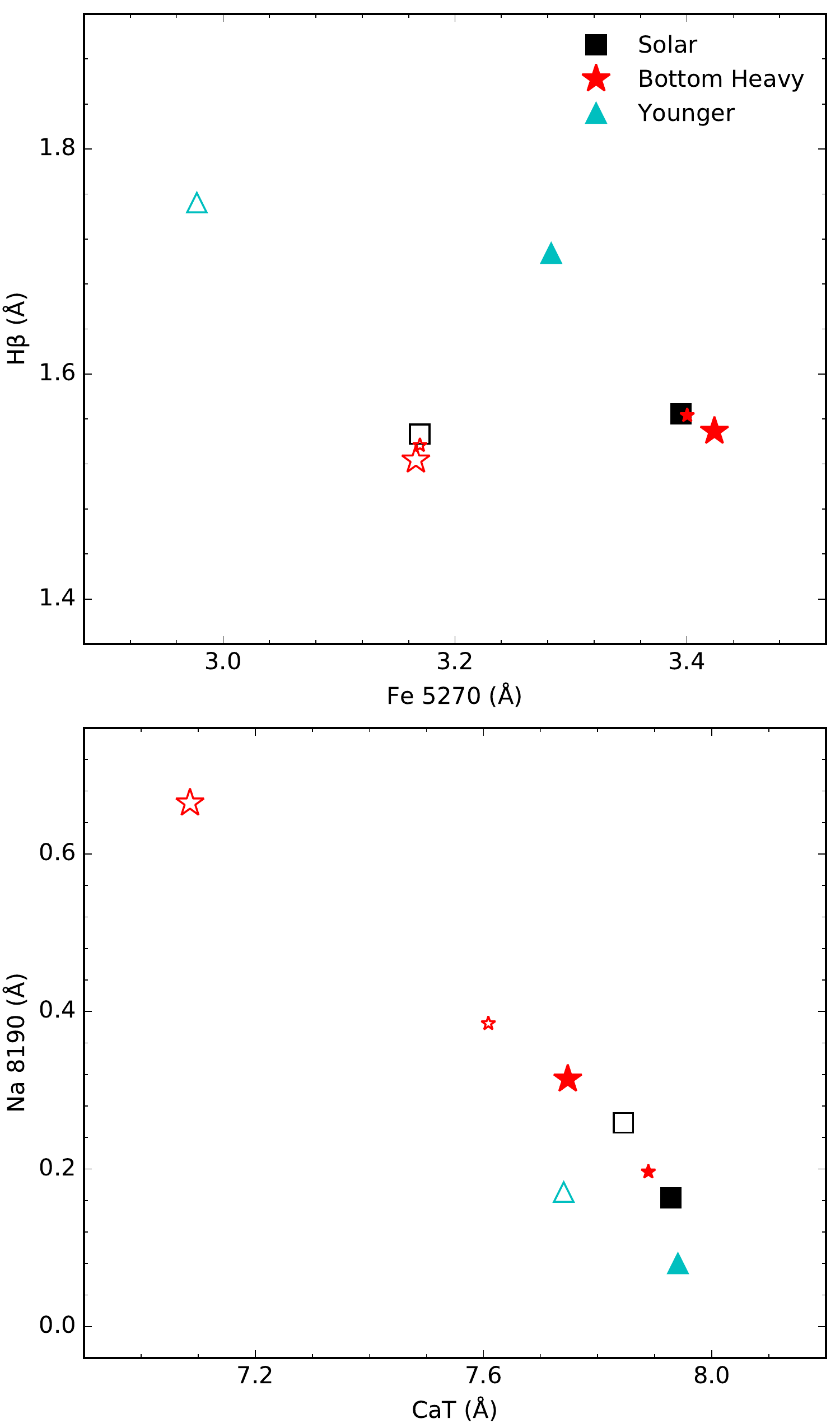}
 \caption{Comparison of the predicted spectral indices for our models (solid points) with the predictions of the models of \citet{Conroy12_counting} (open points).
 The colours and shapes of the points are the same as in Figure \ref{Figure:colours_conroy}.
  While the colours plotted are the same as in Figure \ref{Figure:age_metal_imf_indices}, the axes ranges are different.
  While there are differences in index values between our models and the Conroy models, both sets of models show similar behaviour although the Conroy models show larger IMF effects.}
 \label{Figure:age_metal_imf_indices_conroy}
\end{figure}

\begin{figure}
\centering
 \includegraphics[width=240pt]{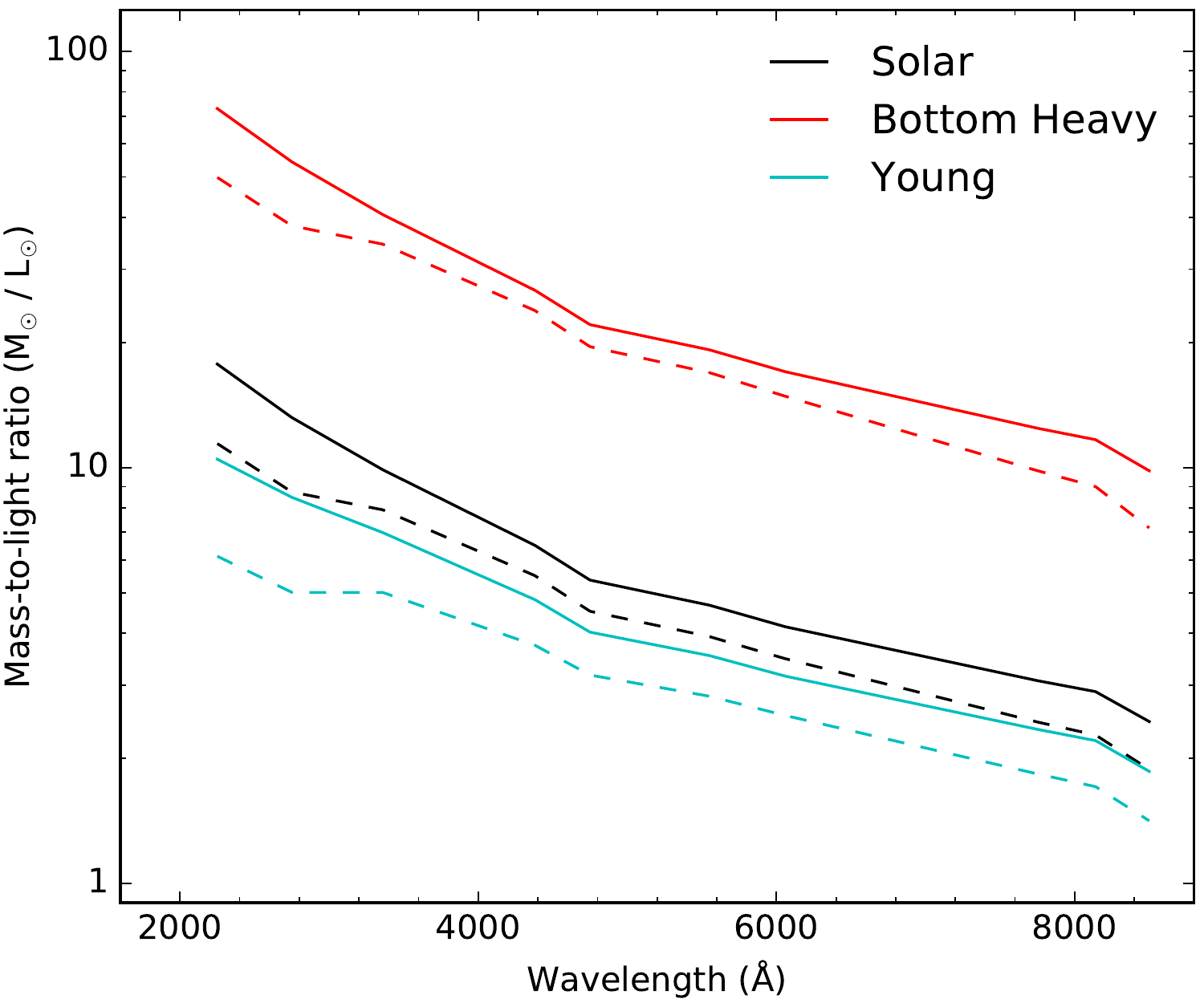}
 \caption{Comparison of the predicted \textit{M}/\textit{L} for our models (solid lines) with the predictions of the models of \citet{Conroy12_counting} (dashed lines).
 The colours of the lines are the same as in Figure \ref{Figure:colours_conroy}.
 Other than a 0.1 dex offset, there is good agreement between the two sets of models.}
 \label{Figure:conroy_m_l}
\end{figure}

We plot a comparison of the colours predicted by the two sets of models in Figure \ref{Figure:colours_conroy}, a comparison of a subset of the spectral indices in Figure \ref{Figure:age_metal_imf_indices_conroy} and a comparison of the predicted \textit{M}/\textit{L} in Figure \ref{Figure:conroy_m_l}.
While we see do zeropoint offsets between our models and those of \citet{Conroy12_counting}, we generally see similar qualitative behaviour with age and the IMF for both sets of models. 
Both sets of models predict bluer colours for younger ages although there are differences in the size of the colour effect for the redder colours.
While both sets of models predict little effect of the IMF on the bluer colours, the \citet{Conroy12_counting} models predict a much larger effect of a bottom heavy IMF on redder colours in the bottom of Figure \ref{Figure:colours_conroy}.
In H$\beta$-Fe 5270 space, both sets of models display similar behaviour with age (stronger H$\beta$, weaker Fe 5270 indices) and with a bottom heavy IMF (little effect on either H$\beta$ or Fe 5270).
In Na 8190-CaT space both models predict stronger Na 8190 indices and weaker CaT indices for a bottom heavy IMF compared to a Kroupa IMF although the \citet{Conroy12_counting} models predict a much larger effect.
Both models predict weaker Na 8190 indices at younger ages, but the \citet{Conroy12_counting} models predict a slightly weaker CaT at younger ages while our models predict no effect of age on the CaT.
Modulo a 0.1 dex offset between our models and the Conroy models, we see excellent agreement in the \textit{M}/\textit{L} predictions.

The differences between our models and those of \citet{Conroy12_counting} are not surprising given that we use synthetic spectra where the Conroy models use empirical spectral libraries supplemented by synthetic spectra.
Differences in the sizes of the effect of changing IMF between the two sets of models are likely due to lower mass stars. 
The lowest mass stellar model we compute is a 3200 K 0.12 M$_{\odot}$ star - the lowest mass point in the \citet{Dotter07} isochrones - while the Conroy models include empirical spectra of stars as cool as 2200 K and as low mass as 0.08 M$_{\odot}$.
Although both sets of models are based on similar isochrones for stars more massive than 0.17 M$_{\odot}$ (\citealt{Dotter07} isochrones for the MS and RGB; Padova isochrones \citealt{Marigo08, Bertelli08} for the post RGB evolution), the Conroy models utilize the Lyon isochrones \citep{Chabrier97, Baraffe98} for the lowest mass stars.
As our models are intended to allow us to explore the differential effects of various abundance patterns and IMF and not to replicate observations, we consider the similar qualitative behaviour of our models to those of \citet{Conroy12_counting} to be sufficient for our purposes.

\section{A comparison with the metal rich globular clusters NGC 6528 and NGC 6553}
\label{appendix_waggs}
To test our models we compared our predicted indices with those measured on the observed spectra of the metal rich Milky Way GCs NGC 6528 and NGC 6553.
Both NGC 6528 and NGC 6553 are old \citep[11 to 13 Gyr, e.g.][]{Zoccali01, Lagioia14} and have slightly subsolar metallicities ([Fe/H] $= -0.11$ and $-0.18$ respectively, 2010 edition of the \citealt{Harris96} catalogue).
These GCs show slight $\alpha$-element enhancements and the light element abundance anti-correlations characteristic of multiple populations in GCs \citep[e.g.][]{Carretta01, Zoccali04, Johnson14, Schiavon17_inner_gcs}.
We use publicly available spectra from the WAGGS project \citep{Usher17} which cover the wavelength range of 3300 to 9050 \AA{} at a resolution of $R \sim 6800$.
As was done for the \citeauthor{Conroy12_counting} models, we smoothed the observed spectra to match the velocity dispersion of our models and we performed the same spectral index measurements as we did for our models (Section \ref{mags_indices}).

\begin{figure*}
   \centering
   \includegraphics[width=504pt]{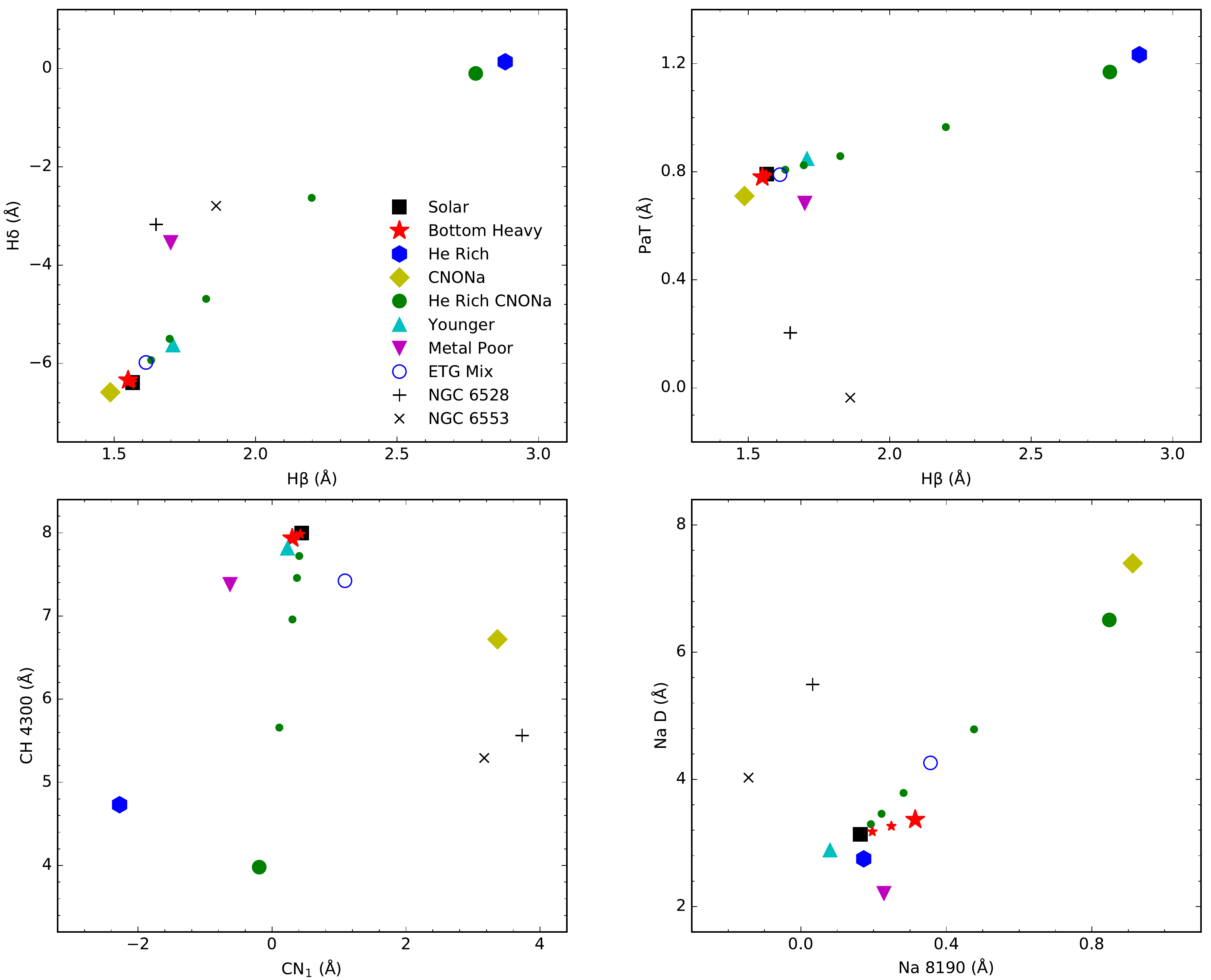}   
    \caption{Comparison of model line index strengths with observations.
    The model symbol shapes and colours are as in Figure \ref{Figure:colours} while the GCs NGC 6528 and NGC 6553 are represented by a black plus and x markers respectively.
    \emph{Top left:} Strength of the H$\delta$ line index versus the strength of the H$\beta$ line index.
    \emph{Top right:} Strength of PaT Paschen line index versus the strength of the H$\beta$ line index.
    \emph{Bottom left:} Strength of the CH 4300 molecular index versus the strength of the CN$_{1}$ molecular index.
    \emph{Bottom right:} the Na D and Na 8190 line indices.
The Balmer indices are consistent with the subsolar [Fe/H] observed GCs.
The observed C dominated CH 4300 index is weaker and the N dominated CN$_{1}$ index is stronger than our solar abundance models but are broadly consistent with the C depleted, N enhanced model.
The stronger Na D indices than our solar model are consistent with both GCs being Na enhanced.
Both the PaT and the Na 8190 indices are affected by TiO bands and the weak observed indices are likely due to the effects of strong TiO absorption on their continuum passbands.}
    \label{Figure:indices_1_observed}
\end{figure*}

\begin{figure*}
   \centering
   \includegraphics[width=504pt]{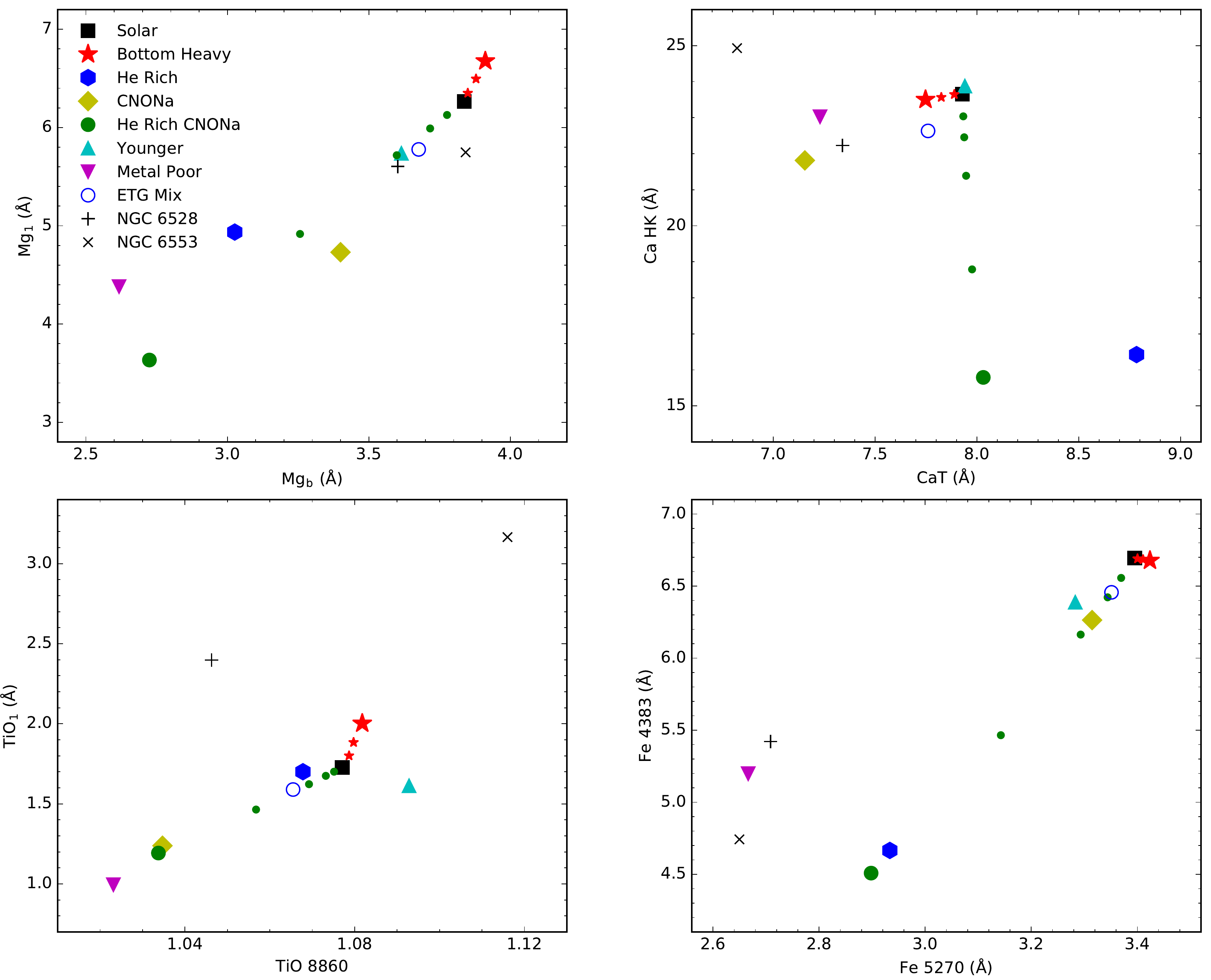}   
    \caption{Comparison of model line index strengths with observations.
    The symbol shapes and colours are as in Figure \ref{Figure:colours}.
    \emph{Top left:} Mg$_{1}$ MgH molecular band strength versus Mg$_{b}$ line index.
    \emph{Top right:} Strength of the Ca H and K lines versus the strength of the calcium triplet (CaT).
    \emph{Bottom left:} The TiO$_{1}$ and TiO 8860 TiO molecular indices.
    \emph{Bottom right:} The Fe 4383 and Fe 5270 indices.
The Mg and Fe indices are all consistent with the subsolar [Fe/H] and slightly supersolar [Mg/Fe] observed GCs.
Given the large observational uncertainty on the Ca H \& K lines, the models are consistent with the observations of the near solar [Ca/H] GCs.
The models do not predict the observed TiO bands nor do the TiO band strengths of the two GCs agree with one another.
The strong observed TiO bands weaken the observed CaT strengths.}
    \label{Figure:indices_2_observed}
\end{figure*}

We plot a comparison of our model predictions with the indices measured on the observed spectra of NGC 6528 and NGC 6553 in Figures \ref{Figure:indices_1_observed} and \ref{Figure:indices_2_observed}.
Given the similar ages, metallicities and chemistries of the two observed GCs, the differences in measured indices provides a rough measure of the uncertainty of the index measurements.
For the Balmer indices and the Fe indices, the observations lie close to our metal poor model ([Fe/H]=$-0.3$) which is unsurprising given their slightly subsolar [Fe/H] abundances.
The observed Mg indices are slightly weaker than our base solar metallicity model which is consistent with the GCs being having a subsolar [Fe/H] but slightly supersolar [Mg/Fe].
Given the large observational uncertainties on the Ca H and K lines, our models are broadly consistent with near solar [Ca/H] GCs.
The observed CH 4300 bands are weaker than our solar metallicity model but the CN$_{1}$ bands are stronger than the model.
This is consistent with both the C-N anti-correlation observed in these GCs and the effects of the first dredge-up which are not included in our models.
Our C depleted, N enhanced models predict the observed CN$_{1}$ strengths but under predicts the CH 4300 abundances.
Likewise, we measure stronger Na D strengths than predicted by our solar abundance model but are consistent with enhanced [Na/Fe] ratios.
We note that interstellar Na D absorption can increase the observed Na D, especially for GCs with significant foreground extinction such as NGC 6528 and NGC 6553.

Perhaps unsurprisingly, our models do not predict the observed TiO band strengths with the spectra of the two GCs showing wildly different TiO strengths.
Observationally, the TiO band strengths are the spectral features most effected by stochastic effects as they are mostly produced by the brightest, coolest giants \citet{vanDokkum14, Usher17}.
Our spectral synthesis is least reliable for such cool, low surface gravity stars. 
The strong TiO bands affect a number of the spectral features we consider in the red.
The red continuum passband for the Na 8190 index overlaps with a TiO bandhead, so the stronger observed TiO absorption leads to a weaker Na 8190 strength.
The wavelength region covered by the CaT and PaT indices is also affected by TiO bands with stronger TiO bands weakly lowering the observe CaT strength but strongly lowering the PaT strength. Due to the sensitivity of some of the indices on TiO, we do include these bands in our comparison.

Given the statistical and systematic uncertainties in both the observed spectra and in the literature metallicities and abundances, and the limitations of our modelling, we consider our models to be in agreement with observations.
We remind the reader that our models were designed to allow us to study the differential effects of various chemical mixtures and IMFs and were not intended to replicate observations in detail.

\bsp	
\label{lastpage}
\end{document}